\documentclass[amsmath,amssymb,aps,prd,nofootinbib,onecolumn,longbibliography,notitlepage]{revtex4-1}

\usepackage{graphicx}
\usepackage{hyperref}
\usepackage{xcolor}

\usepackage[inline]{enumitem}

\usepackage{amsmath}
\usepackage{amsfonts}
\usepackage{relsize}

\newcommand{\araa}{Annu. Rev. Astron. Astrophys.}
\newcommand{\apjl}{Astrophys. J. Lett.}
\newcommand{\procspie}{Proc. SPIE}
\newcommand{\mnras}{Mon. Not. Royal Astron.}
\newcommand{\aap}{Astron. Astrophys.}

\usepackage[inline]{enumitem}
\usepackage{comment}
\usepackage{multirow}

\newcommand{\rhonuc}{\ensuremath{\rho_\mathrm{nuc}}}
\newcommand{\rhonucCGS}{\ensuremath{2.8\times10^{14}\, \mathrm{g}/\mathrm{cm}^3}}

\newcommand{\EOS}{EOS}

\newcommand{\Bayes}{\ensuremath{B^\mathcal{A}_\mathcal{I}}}

\newcommand{\mmaxPOSTun}{\ensuremath{2.09^{+0.37}_{-0.16}}}

\newcommand{\mmaxPOSTmod}{\ensuremath{2.04^{+0.22}_{-0.002}}}

\newcommand{\pressureunits}{\ensuremath{\mathrm{dyn}/\mathrm{cm}^2}}

\newcommand{\gwphenomunloPRIORpressureattworhonuc}{\ensuremath{1.09^{ +8.8}_{-1.1}\times10^{34}}} 

\newcommand{\gwphenomunloPOSTpressureattworhonuc}{\ensuremath{1.35^{+1.8}_{-1.2}\times10^{34}}}  
\newcommand{\gwphenomunloPOSTpressureatsixrhonuc}{\ensuremath{8.86^{+4.3}_{-5.9}\times10^{35}}}  

\newcommand{\gwphenommodloPRIORpressureattworhonuc}{\ensuremath{4.96^{+1.3}_{-2.9}\times10^{34}}} 

\newcommand{\gwphenommodloPOSTpressureattworhonuc}{\ensuremath{4.73^{+1.4}_{-2.5}\times10^{34}}}  
\newcommand{\gwphenommodloPOSTpressureatsixrhonuc}{\ensuremath{7.55^{+2.0}_{-3.2}\times10^{35}}}  

\newcommand{\bayesgwphenomlo}{\ensuremath{1.12 \pm 0.06}} 

\newcommand{\gwphenomunloPOSTmone}{\ensuremath{1.46^{+0.16}_{-0.09}}}  
\newcommand{\gwphenomunloPOSTmtwo}{\ensuremath{1.26^{+0.10}_{-0.10}}}  
\newcommand{\gwphenomunloPOSTlone}{\ensuremath{110^{+278}_{ -81}}}     
\newcommand{\gwphenomunloPOSTltwo}{\ensuremath{361^{+529}_{-252}}}     
\newcommand{\gwphenomunloPOSTlamtilde}{\ensuremath{210^{+383}_{-113}}} 

\newcommand{\gwphenomunloPOSTlamonepointfour}{\ensuremath{160^{+448}_{-113}}}  

\newcommand{\gwphenommodloPOSTmone}{\ensuremath{1.41^{+0.15}_{-0.04}}}  
\newcommand{\gwphenommodloPOSTmtwo}{\ensuremath{1.32^{+0.04}_{-0.13}}}  
\newcommand{\gwphenommodloPOSTlone}{\ensuremath{426^{+214}_{-229}}}     
\newcommand{\gwphenommodloPOSTltwo}{\ensuremath{772^{+641}_{-212}}}     
\newcommand{\gwphenommodloPOSTlamtilde}{\ensuremath{631^{+164}_{-122}}} 

\newcommand{\gwphenommodloPOSTlamonepointfour}{\ensuremath{556^{+163}_{-172}}}  

\newcommand{\bayesinjonehfour}{\ensuremath{0.26 \pm 0.01}} 

\newcommand{\injoneunhfourPOSTmone}{\ensuremath{1.57^{+0.12}_{-0.22}}}  
\newcommand{\injoneunhfourPOSTmtwo}{\ensuremath{1.15^{+0.18}_{-0.07}}}  
\newcommand{\injoneunhfourPOSTlone}{\ensuremath{ 341^{+504}_{-265}}}    
\newcommand{\injoneunhfourPOSTltwo}{\ensuremath{1200^{+1440}_{-488}}}   
\newcommand{\injoneunhfourPOSTlamtilde}{\ensuremath{836^{+337}_{-366}}} 

\newcommand{\injonemodhfourPOSTmone}{\ensuremath{1.60^{+0.10}_{-0.23}}}  
\newcommand{\injonemodhfourPOSTmtwo}{\ensuremath{1.14^{+0.17}_{-0.07}}}  
\newcommand{\injonemodhfourPOSTlone}{\ensuremath{ 231^{ +413}_{-109}}}   
\newcommand{\injonemodhfourPOSTltwo}{\ensuremath{1152^{+1340}_{-407}}}   
\newcommand{\injonemodhfourPOSTlamtilde}{\ensuremath{740^{+167}_{-164}}} 

\newcommand{\bayesinjonemsone}{\ensuremath{22.6 \pm 0.7}} 

\newcommand{\injoneunmsonePOSTmone}{\ensuremath{1.40^{+0.27}_{-0.04}}}   
\newcommand{\injoneunmsonePOSTmtwo}{\ensuremath{1.30^{+0.05}_{-0.20}}}   
\newcommand{\injoneunmsonePOSTlone}{\ensuremath{ 611^{+979}_{-276}}}     
\newcommand{\injoneunmsonePOSTltwo}{\ensuremath{2350^{+2030}_{-897}}}    
\newcommand{\injoneunmsonePOSTlamtilde}{\ensuremath{1530^{+387}_{-455}}} 

\newcommand{\injonemodmsonePOSTmone}{\ensuremath{1.72^{+0.03}_{-0.27}}}  
\newcommand{\injonemodmsonePOSTmtwo}{\ensuremath{1.08^{+0.18}_{-0.03}}}  
\newcommand{\injonemodmsonePOSTlone}{\ensuremath{ 210^{+472}_{  -70}}}   
\newcommand{\injonemodmsonePOSTltwo}{\ensuremath{3190^{+279}_{-2070}}}   
\newcommand{\injonemodmsonePOSTlamtilde}{\ensuremath{890^{+177}_{-112}}} 

\newcommand{\bayesinjonesly}{\ensuremath{0.44 \pm 0.01}} 

\newcommand{\injoneunslyPOSTmone}{\ensuremath{1.41^{+0.27}_{-0.05}}}  
\newcommand{\injoneunslyPOSTmtwo}{\ensuremath{1.22^{+0.11}_{-0.13}}}  
\newcommand{\injoneunslyPOSTlone}{\ensuremath{155^{+363}_{-121}}}     
\newcommand{\injoneunslyPOSTltwo}{\ensuremath{792^{+842}_{-563}}}     
\newcommand{\injoneunslyPOSTlamtilde}{\ensuremath{463^{+277}_{-320}}} 

\newcommand{\injonemodslyPOSTmone}{\ensuremath{1.40^{+0.24}_{-0.05}}}  
\newcommand{\injonemodslyPOSTmtwo}{\ensuremath{1.30^{+0.05}_{-0.18}}}  
\newcommand{\injonemodslyPOSTlone}{\ensuremath{316^{ +328}_{-192}}}    
\newcommand{\injonemodslyPOSTltwo}{\ensuremath{862^{+1080}_{-281}}}    
\newcommand{\injonemodslyPOSTlamtilde}{\ensuremath{667^{+147}_{-144}}} 

\newcommand{\bayesinjfourhfour}{\ensuremath{0.30 \pm 0.01}} 

\newcommand{\injfourunhfourPOSTmone}{\ensuremath{1.39^{+0.22}_{-0.04}}}  
\newcommand{\injfourunhfourPOSTmtwo}{\ensuremath{1.31^{+0.04}_{-0.17}}}  
\newcommand{\injfourunhfourPOSTlone}{\ensuremath{341^{ +527}_{-257}}}    
\newcommand{\injfourunhfourPOSTltwo}{\ensuremath{942^{+1240}_{-452}}}    
\newcommand{\injfourunhfourPOSTlamtilde}{\ensuremath{721^{+461}_{-386}}} 

\newcommand{\injfourmodhfourPOSTmone}{\ensuremath{1.39^{+0.23}_{-0.04}}}  
\newcommand{\injfourmodhfourPOSTmtwo}{\ensuremath{1.31^{+0.04}_{-0.18}}}  
\newcommand{\injfourmodhfourPOSTlone}{\ensuremath{306^{ +440}_{-121}}}    
\newcommand{\injfourmodhfourPOSTltwo}{\ensuremath{942^{+1080}_{-276}}}    
\newcommand{\injfourmodhfourPOSTlamtilde}{\ensuremath{728^{+178}_{-162}}} 

\newcommand{\bayesinjfourmsone}{\ensuremath{0.85 \pm 0.02}} 

\newcommand{\injfourunmsonePOSTmone}{\ensuremath{1.40^{+0.20}_{-0.05}}}   
\newcommand{\injfourunmsonePOSTmtwo}{\ensuremath{1.29^{+0.06}_{-0.15}}}   
\newcommand{\injfourunmsonePOSTlone}{\ensuremath{ 636^{ +637}_{ -546}}}   
\newcommand{\injfourunmsonePOSTltwo}{\ensuremath{1670^{+1330}_{-1028}}}   
\newcommand{\injfourunmsonePOSTlamtilde}{\ensuremath{1200^{+541}_{-686}}} 

\newcommand{\injfourmodmsonePOSTmone}{\ensuremath{1.40^{+0.23}_{-0.05}}}  
\newcommand{\injfourmodmsonePOSTmtwo}{\ensuremath{1.29^{+0.05}_{-0.17}}}  
\newcommand{\injfourmodmsonePOSTlone}{\ensuremath{ 406^{ +345}_{-226}}}   
\newcommand{\injfourmodmsonePOSTltwo}{\ensuremath{1030^{+1210}_{-329}}}   
\newcommand{\injfourmodmsonePOSTlamtilde}{\ensuremath{776^{+192}_{-199}}} 

\newcommand{\bayesinjfoursly}{\ensuremath{1.40 \pm 0.03}} 

\newcommand{\injfourunslyPOSTmone}{\ensuremath{1.40^{+0.17}_{-0.05}}}  
\newcommand{\injfourunslyPOSTmtwo}{\ensuremath{1.29^{+0.06}_{-0.13}}}  
\newcommand{\injfourunslyPOSTlone}{\ensuremath{140^{+278}_{-109}}}     
\newcommand{\injfourunslyPOSTltwo}{\ensuremath{341^{+581}_{-217}}}     
\newcommand{\injfourunslyPOSTlamtilde}{\ensuremath{247^{+332}_{-162}}} 

\newcommand{\injfourmodslyPOSTmone}{\ensuremath{1.41^{+0.13}_{-0.06}}}  
\newcommand{\injfourmodslyPOSTmtwo}{\ensuremath{1.29^{+0.06}_{-0.10}}}  
\newcommand{\injfourmodslyPOSTlone}{\ensuremath{431^{+210}_{-202}}}     
\newcommand{\injfourmodslyPOSTltwo}{\ensuremath{902^{+547}_{-304}}}     
\newcommand{\injfourmodslyPOSTlamtilde}{\ensuremath{655^{+161}_{-130}}} 

\newcommand{\bayesinjfivehfour}{\ensuremath{0.44 \pm 0.01}} 

\newcommand{\injfiveunhfourPOSTmone}{\ensuremath{1.58^{+0.16}_{-0.16}}}  
\newcommand{\injfiveunhfourPOSTmtwo}{\ensuremath{1.14^{+0.16}_{-0.07}}}  
\newcommand{\injfiveunhfourPOSTlone}{\ensuremath{ 251^{ +605}_{ -193}}}  
\newcommand{\injfiveunhfourPOSTltwo}{\ensuremath{2010^{+1409}_{-1130}}}  
\newcommand{\injfiveunhfourPOSTlamtilde}{\ensuremath{998^{+376}_{-515}}} 

\newcommand{\injfivemodhfourPOSTmone}{\ensuremath{1.63^{+0.12}_{-0.17}}}  
\newcommand{\injfivemodhfourPOSTmtwo}{\ensuremath{1.13^{+0.14}_{-0.06}}}  
\newcommand{\injfivemodhfourPOSTlone}{\ensuremath{ 210^{+308}_{  -94}}}   
\newcommand{\injfivemodhfourPOSTltwo}{\ensuremath{1960^{+713}_{-1050}}}   
\newcommand{\injfivemodhfourPOSTlamtilde}{\ensuremath{697^{+190}_{-154}}} 

\newcommand{\bayesinjfivemsone}{\ensuremath{1.97 \pm 0.06}} 

\newcommand{\injfiveunmsonePOSTmone}{\ensuremath{1.62^{+0.13}_{-0.20}}}   
\newcommand{\injfiveunmsonePOSTmtwo}{\ensuremath{1.14^{+0.17}_{-0.07}}}   
\newcommand{\injfiveunmsonePOSTlone}{\ensuremath{ 361^{ +725}_{ -278}}}   
\newcommand{\injfiveunmsonePOSTltwo}{\ensuremath{2550^{+1620}_{-1470}}}   
\newcommand{\injfiveunmsonePOSTlamtilde}{\ensuremath{1210^{+458}_{-474}}} 

\newcommand{\injfivemodmsonePOSTmone}{\ensuremath{1.67^{+0.09}_{-0.19}}}  
\newcommand{\injfivemodmsonePOSTmtwo}{\ensuremath{1.13^{+0.13}_{-0.06}}}  
\newcommand{\injfivemodmsonePOSTlone}{\ensuremath{ 246^{ +194}_{-121}}}   
\newcommand{\injfivemodmsonePOSTltwo}{\ensuremath{1830^{+1200}_{-895}}}   
\newcommand{\injfivemodmsonePOSTlamtilde}{\ensuremath{782^{+191}_{-229}}} 

\newcommand{\bayesinjfivesly}{\ensuremath{0.53 \pm 0.01}} 

\newcommand{\injfiveunslyPOSTmone}{\ensuremath{1.62^{+0.10}_{-0.24}}}  
\newcommand{\injfiveunslyPOSTmtwo}{\ensuremath{1.15^{+0.17}_{-0.08}}}  
\newcommand{\injfiveunslyPOSTlone}{\ensuremath{110^{ +333}_{ -92}}}    
\newcommand{\injfiveunslyPOSTltwo}{\ensuremath{641^{+1020}_{-480}}}    
\newcommand{\injfiveunslyPOSTlamtilde}{\ensuremath{337^{+359}_{-230}}} 

\newcommand{\injfivemodslyPOSTmone}{\ensuremath{1.46^{+0.25}_{-0.09}}}  
\newcommand{\injfivemodslyPOSTmtwo}{\ensuremath{1.17^{+0.16}_{-0.09}}}  
\newcommand{\injfivemodslyPOSTlone}{\ensuremath{ 231^{ +331}_{-121}}}   
\newcommand{\injfivemodslyPOSTltwo}{\ensuremath{1040^{+1030}_{-458}}}   
\newcommand{\injfivemodslyPOSTlamtilde}{\ensuremath{637^{+133}_{-139}}} 

\begin{document}

\title{ Non-parametric inference of the neutron star equation of state \\ from gravitational wave observations
}

\author{Philippe Landry}
\email{landryp@uchicago.edu}
\affiliation{Enrico Fermi Institute and Kavli Institute for Cosmological Physics, The University of Chicago, 5640 South Ellis Avenue, Chicago, Illinois, 60637, USA}
\author{Reed Essick}
\email{reedessick@kicp.uchicago.edu}
\affiliation{Kavli Institute for Cosmological Physics, The University of Chicago, 5640 South Ellis Avenue, Chicago, Illinois, 60637, USA}

\date{\today}

\begin{abstract}
We develop a non-parametric method for inferring the universal neutron star (NS) equation of state (\EOS) from gravitational wave (GW) observations. 
Many different possible realizations of the \EOS~are generated with a Gaussian process conditioned on a set of nuclear-theoretic models.
These synthetic {\EOS}s are causal and thermodynamically stable by construction, span a broad region of the pressure-density plane, and can be selected to satisfy astrophysical constraints on the NS mass.
Associating every synthetic \EOS~with a pair of component masses $M_{1,2}$ and calculating the corresponding tidal deformabilities $\Lambda_{1,2}$, we perform Monte Carlo integration over the GW likelihood for $M_{1,2}$ and $\Lambda_{1,2}$ to directly infer a posterior process for the NS \EOS.
We first demonstrate that the method can accurately recover an injected GW signal, and subsequently use it to analyze data from GW170817, finding a canonical deformability of $\Lambda_{1.4} = \gwphenomunloPOSTlamonepointfour$ and $p(2\rho_{\mathrm{nuc}})=\gwphenomunloPOSTpressureattworhonuc$ \pressureunits~for the pressure at twice the nuclear saturation density at 90\% confidence, in agreement with previous studies, when assuming a loose \EOS~prior.
With a prior more tightly constrained to resemble the theoretical \EOS~models, we recover $\Lambda_{1.4} = \gwphenommodloPOSTlamonepointfour$ and $p(2\rho_{\mathrm{nuc}})=\gwphenommodloPOSTpressureattworhonuc$ \pressureunits.
We further infer the maximum NS mass supported by the EOS to be $M_\mathrm{max}=\mmaxPOSTun$ (\mmaxPOSTmod) $M_\odot$ with the loose (tight) prior.
The Bayes factor between the two priors is $\Bayes \simeq 1.12$, implying that neither is strongly preferred by the data and suggesting that constraints on the \EOS~from GW170817 alone may be relatively prior-dominated.
\end{abstract}

\maketitle

\section{Introduction}\label{sec:intro}

Determining the neutron star (NS) equation of state (\EOS) is a major unsolved problem in nuclear astrophysics. 
The NS \EOS~is the zero-temperature pressure-density relation that arises from the star's nuclear microphysics, encoding its composition and internal structure, and determining its macroscopic properties.
Because young NSs cool rapidly via neutrino emission and quickly reach $\beta$-equilibrium, thermal and dissipative corrections to the EOS are negligible, and NS matter is typically modeled as a perfect fluid \cite{Rot_rel_stars}.
The \EOS~for the low-density NS crust is well known \cite{Lattimer_nsmass}, but the conditions in the dense NS core are so extreme that laboratory experiments are unable to probe its constituent supranuclear matter. 
Consequently, astrophysical observations of NSs offer the best opportunity for constraining the unknown core \EOS. 

Because a given candidate \EOS~prescribes a unique mass-radius relation \cite{Lindblom_mr}, simultaneous mass and radius measurements have been attempted for a number of NSs \cite{Lattimer_eos}. 
However, obtaining an accurate radius measurement, which involves modeling the star's thermal X-ray emission, is notoriously challenging \cite{Ozel}. 
Likewise, simultaneous pulsar mass and moment of inertia measurements have been proposed as a means of constraining the NS \EOS~via the mass-moment of inertia relation \cite{Morrison,Lattimer_moi}. 
Existing radio observatories are, however, unable to measure the relativistic periastron advance of any known binary pulsar with sufficient precision to infer the stellar moment of inertia \cite{Kramer}. 
Although future moment of inertia measurements from next-generation radio telescopes, such as the Square Kilometre Array \cite{SKA}, and forthcoming radius measurements from the NICER soft X-ray observatory \cite{NICER} are expected to make significant contributions to the study of ultra-dense matter, at present the most informative constraints on the NS \EOS~come from gravitational-wave (GW) astronomy. 
Indeed, Advanced LIGO's \cite{LIGO} and Virgo's \cite{Virgo} GW measurement of NS tidal deformability in GW170817 \cite{LVC_GW170817}, the loud signal of a binary NS merger, produced the most discriminating constraints on the supranuclear \EOS~to date \cite{De,LVC_eos}.

The tidal forces that arise in NS binaries during inspiral deform the stars away from their spherical equilibrium shape. 
The mass quadrupole moments they acquire draw energy from the orbit and add to the binary's gravitational radiation, enhancing its GW luminosity and leading to a slight acceleration of the coalescence.
This manifests as a phase shift in the waveform relative to the merger of point-particles \cite{Kochanek,Lai,Flanagan}, which is measurable with existing GW detectors \cite{Hinderer,Read_nsmatter,DelPozzo,Wade,Damour}.
Its magnitude depends on the size of the induced stellar quadrupoles, measured by the tidal deformabilities $\Lambda_{1,2}$ of the two NSs.
$\Lambda$ is an \EOS-dependent, dimensionless function of the NS mass $M$ that correlates with the pressure gradients inside the star---i.e.~with the stiffness of the \EOS. 
A stiff EOS has large pressure gradients that support bigger, more diffuse stars with large $\Lambda$, while a soft EOS has small pressure gradients that yield more compact stars with small $\Lambda$.
Thus, as with the mass-radius or mass-moment of inertia relation, simultaneous measurements can pick out the preferred $M$-$\Lambda$ relation, which is in one-to-one correspondence with the NS \EOS.

The first GW constraints on NS tidal deformability were reported in Ref.~\cite{LVC_GW170817}. 
The parameters $\Lambda_{1,2}$ were included in the waveform model and inferred, along with the masses $M_{1,2}$ and the other source properties, from GW170817 strain data \cite{Veitch}.
The resulting posterior distribution over the tidal deformabilities and masses established an upper bound on the chirp deformability
\begin{equation} \label{binarydef}
    \tilde{\Lambda} = \frac{16}{13} \frac{(M_1+12M_2){M_1}^4 \Lambda_1 + (M_2 + 12M_1){M_2}^4\Lambda_2}{(M_1 + M_2)^5} ,
\end{equation}
the particular mass-weighted average of $\Lambda_{1,2}$ that appears in the waveform at lowest post-Newtonian order \cite{Flanagan,Favata}. 
Assuming that GW170817's components rotated slowly \cite{Hannam,Damour,Landry_pulsar}, with dimensionless spin \mbox{$\chi_{1,2} := c S_{1,2}/G {M_{1,2}}^2 \leq 0.05$}, Ref.~\cite{LVC_GW170817} placed a 90\%-credible upper bound $\tilde{\Lambda} \leq 800$.\footnote{Unless otherwise specified, quoted credible regions refer to 90\% confidence throughout the paper.}
The upper limit on the deformability of a canonical $1.4\, M_\odot$ NS was also found to be $\Lambda_{1.4} := \Lambda(1.4 \, M_{\odot}) \leq 800$.
This was refined to $\tilde{\Lambda} = 300^{+420}_{-230}$ in a subsequent analysis \cite{LVC_properties}, which made use of improved waveform models and included lower frequencies within the GW data.
For comparison, the widely used candidate EOS \texttt{sly} \cite{Douchin}, among the softer theoretical models, has $\Lambda_{1.4} \approx 290$, while \texttt{ms1b} \cite{Muller}, one of the stiffest, has $\Lambda_{1.4} \approx 1220$.
Overall, these constraints favor a relatively soft NS \EOS, and bounds from Ref.~\cite{LVC_properties} rule out several candidate {\EOS}s at 90\% confidence. 

Nonetheless, the analyses of Refs.~\cite{LVC_GW170817,LVC_properties} neglected important correlations between $\Lambda_1$ and $\Lambda_2$ and did not attempt to translate the tidal constraints into a direct inference of the NS \EOS. 
Because a NS's composition is dictated by universal nuclear many-body physics, every NS is thought to share a common \EOS. 
Their individual properties can thus be parameterized by the stellar mass alone.
In particular, since all NSs conform to the same $M$-$\Lambda$ relation, the waveform parameters $\Lambda_{1,2}$ are not truly independent. 
For example, with the standard convention $M_2 \leq M_1$, $\Lambda_2 \geq \Lambda_1$ for any physically realistic \EOS, as $\Lambda(M)$ decreases monotonically.
Moreover, the correlation between $\Lambda_{1,2}$ has been shown to be approximately universal \cite{Yagi_BiLoveLett,Yagi_BiLove}; knowledge of $\Lambda_1$ and the binary mass ratio is sufficient to determine $\Lambda_2$ with high fidelity, regardless of the \EOS. 
Deviations from the universal relation are at most $\sim 10\%$, and are substantially smaller for nearly equal-mass systems. 
Incorporating these correlations, known as \emph{binary Love relations}, into a prior on $\Lambda_{1,2}$ can significantly improve the recovery of tidal information from GWs.
An injection study performed in Ref.~\cite{Chatziioannou} demonstrated that the area of the 90\%-credible region of the $\Lambda_1$-$\Lambda_2$ posterior is reduced by a factor of two or more.

The intrinsic correlations captured by the binary Love relations were taken into account in a pair of reanalyses of GW170817. 
Ref.~\cite{LVC_eos} made use of the \EOS-insensitive, mass-ratio dependent binary Love relations from Refs.~\cite{Yagi_BiLoveLett,Yagi_BiLove} to compute $\Lambda_2$ as a function of the other parameters in the waveform model. 
Ref.~\cite{De} used a closely related approximation, $\Lambda_2 = (M_1/M_2)^6 \Lambda_1$, for the same purpose. 
With the inclusion of these physical restrictions on the tidal deformabilities, both works tightened the tidal constraints inferred from GW170817.
Ref.~\cite{De} reported $\tilde{\Lambda} = 222^{+420}_{-138}$ for a uniform component-mass prior, while Ref.~\cite{LVC_eos} found $\Lambda_{1.4} = 190^{+390}_{-120}$.
Both studies employed the low-spin prior described above, and they prefer a marginally softer \EOS~than was originally inferred.

Ref.~\cite{LVC_eos} also sought to constrain the NS \EOS~directly, instead of working exclusively with the tidal deformabilities. 
It adopted a spectral parameterization for the \EOS~\cite{Lindblom_review,Lindblom_spectral,Lindblom_stellarproblem,Lindblom_stellarproblem2} following methodology originally developed for piecewise polytropes \cite{Read_pwpoly,Lackey,Raaijmakers,Carney}.
The spectral coefficients were sampled in place of $\Lambda_{1,2}$ in the waveform, and the most probable \EOS~was then reconstructed, with error bars, from the first four spectral coefficients. 
This \emph{spectral method} ultimately produced a \emph{process} in the pressure-density plane, in addition to a joint posterior on $\Lambda_{1,2}$; just as a distribution gives the relative probability of finite sets of variates, a process gives the relative probability of functional degrees of freedom.
Although the $\Lambda_{1,2}$ posterior can be used for Bayesian hypothesis ranking of different candidate {\EOS}s \cite{DelPozzo,Agathos,Abdelsalhin}, the EOS process is inherently more informative. 
The discrete set of nuclear theory models tested in a hypothesis ranking scheme may not capture the full range of possibilities for the \EOS, and reducing each model to an overall evidence score may make it difficult to deduce preferences for specific fine-grained features of the \EOS. 
In contrast, Ref.~\cite{LVC_eos} was able to report, e.g., the recovered pressure at twice and six times the nuclear saturation density ($\rhonuc=\rhonucCGS$) independently of existing nuclear-theoretic models with $\sim 60\%$ uncertainty.

However, parametric {\EOS}~inference has limitations of its own. 
The reduction of a complicated pressure-density function to a small number of parameters necessarily leads to modeling errors~\cite{Carney}. 
This problem is especially acute for sharp features in the pressure-density relation, such as first-order phase transitions, and a four-parameter spectral representation has been shown to model such discontinuous behavior rather poorly \cite{Lindblom_spectral}. 
Indeed, it performs no better in this regard than a three-segment piecewise polytrope parameterization, whose issues with phase transitions are widely acknowledged \cite{Lackey,Lindblom_spectral}.
While it may be possible to design an alternate parameterization specially adapted to such phase transitions, given the extremely varied phenomenology of hybrid {\EOS}s (e.g.~\cite{Han,hybrid1,hybrid2,hybrid3}; see Ref.~\cite{Haensel} for an overview), it is unlikely that any single parametric model will be able to faithfully represent the full range of \EOS~variability with only a handful of parameters.
Accordingly, a \emph{non-parametric} representation of the \EOS~may be better suited to the task of inferring the internal structure of NSs from astrophysical observations.

The faithfulness of a non-parametric representation of a function $f$ scales with the amount of underlying knowledge of $f$---the way, e.g., the appropriate number of bins to use in a histogram grows as more data are observed---instead of the number of parameters chosen in a parametric model. 
Non-parametric representations typically involve a set of \emph{hyperparameters} that control allowed types of functional behavior---like the number and placement of bins in a histogram---but the resulting freedom in $f$ is much larger than that afforded by a parametric model.
This is because hyperparameters describe the correlations between a function's values rather than the values themselves.
Hence, the uncertainty in the non-parametric representation decreases as observations of the function accrue, while systematic errors in the parametric representation are fixed by the degree to which the model is suited to $f$. 
Given that information about the unknown \EOS~will accumulate through successive GW observations, a non-parametric approach is well-adapted to the so-called relativistic inverse stellar structure problem.

In this paper, we introduce a non-parametric method for directly inferring the NS \EOS~from GW data. 
Rather than prescribing a functional form for the relation between the star's pressure $p$ and its total energy density $\mu$, as was done in Ref.~\cite{LVC_eos}, we use Gaussian process regression (GPR) to generate a large number of possible realizations $\mu^{(a)}(p)$ of the EOS, labeled by $a = 1,2,...,N$. These \emph{synthetic {\EOS}s} span the full range of stiffnesses, core pressures and other characteristics consistent with thermodynamic stability, causality, astrophysical observations of NSs, and candidate models $\mu^{(\alpha)}(p)$, $\alpha = 1,2,...,n$, from nuclear theory.\footnote{
We index tabulated candidate {\EOS}s with Greek letters $(\alpha), (\beta), (\gamma), ...$ and synthetic {\EOS}s with Latin letters $(a), (b), (c), ...$ .}
We then associate each synthetic \EOS~$\mu^{(a)}(p)$ with NS masses $M_{1,2}^{(a)}$ drawn from a prior distribution and calculate the corresponding tidal deformabilities $\Lambda_{1,2}^{(a)}$.
A Gaussian kernel density estimate (KDE) approximates the marginal conditional likelihood $\mathcal{L}^{(a)} := \mathcal{L}(d|M_1^{(a)}, M_2^{(a)}, \Lambda_1^{(a)}, \Lambda_2^{(a)}; \mathcal{H})$ obtained from standard parameter estimation~\cite{Veitch}.
By assigning $\mathcal{L}^{(a)}$ to $\mu^{(a)}(p)$, we establish a statistical map between observables ($M_{1,2}$, $\Lambda_{1,2}$) and the pressure-density plane.
In the limit of many samples ($N \rightarrow \infty$), we obtain a smooth \emph{EOS posterior process}.
This inference scheme is similar to what was used in Ref.~\cite{Lackey} in the context of piecewise polytropes, but our approach adopts a different representation of the EOS prior and employs different techniques for obtaining the posterior.
Ref.~\cite{Most2018} employs integration techniques similar to ours in order to bound macroscopic NS observables, but adopts an \textit{ad hoc} \EOS~prior and does not weight their prior draws by the full likelihood.

Besides possessing systematics that are completely independent from those for parametric methods, a non-parametric approach to \EOS~inference has several attractive features.
Instead of requiring a separate analysis of GW strain data with an alternate waveform model, our method operates on the standard GW $M_{1,2}$-$\Lambda_{1,2}$ likelihoods produced by parameter estimation.
Moreover, a GPR representation of the EOS naturally comes equipped with smooth error estimates for the function, unlike simple interpolation or a spectral construction.
Most importantly, such a representation provides immense flexibility when selecting the \EOS~prior; by conditioning the GPR on different sets of candidate EOSs, priors specialized to different classes of theoretical models can easily be specified. 
With these, one could investigate, e.g., the prevalence of hyperonic degrees of freedom or first-order phase transitions in the EOS by calculating the relative posterior support for the relevant priors.
Astrophysical priors on NS masses and radii can also be accommodated by the GPR, and physical constraints on the \EOS---like causality, thermodynamic stability, and the binary Love relations---are automatically incorporated.
Furthermore, conditioning the priors on tabulated {\EOS}s may make our prior beliefs more transparent compared to \textit{ad hoc} choices for parameters which may not have immediate physical interpretations. 
The flexibility and transparency afforded by a non-parametric representation of the \EOS~may become increasingly important as more binary NS systems are observed, reducing statistical uncertainty to levels comparable to systematic model uncertainties~\cite{ObservingScenarios}.

Leveraging these advantages, we develop two non-parametric priors with Gaussian processes conditioned on a fiducial set of tabulated candidate {\EOS}s.
Our \emph{model-agnostic prior} attempts to span the full range of physically plausible {\EOS}s~and is only loosely informed by the theoretical models, while our \emph{model-informed prior} is more tightly constrained by the candidate {\EOS}s.
Applying both priors to a series of simulated signals, we confirm that our inference scheme behaves as expected.
Performing our analysis on GW170817, we infer $\tilde{\Lambda}=\gwphenomunloPOSTlamtilde$ (\gwphenommodloPOSTlamtilde) and $\Lambda_{1.4}=\gwphenomunloPOSTlamonepointfour$ (\gwphenommodloPOSTlamonepointfour) with the model-agnostic (model-informed) prior.
We also infer the pressures \mbox{$p(2\rhonuc)=\gwphenomunloPOSTpressureattworhonuc$ (\gwphenommodloPOSTpressureattworhonuc) \pressureunits}~and $p(6\rhonuc)=\gwphenomunloPOSTpressureatsixrhonuc$ (\gwphenommodloPOSTpressureatsixrhonuc) \pressureunits.
Both the model-agnostic and model-informed results are consistent with previous analyses of GW170817 \cite{De,LVC_eos}.
Additionally, we extrapolate the correlations between mid-range and high densities in the synthetic EOSs to place the first constraints on the maximum NS mass derived from GW data alone.
Our analysis with the model-agnostic (model-informed) prior yields $M_\mathrm{max}=\mmaxPOSTun$ (\mmaxPOSTmod) $M_\odot$.
Computing the Bayes factor $\Bayes$ between the two priors, we find that neither is strongly preferred by the data: $\Bayes = \bayesgwphenomlo$ (point estimate and 1-$\sigma$ uncertainty).
This suggests that constraints on the \EOS~and tidal deformabilities from GW170817 may be relatively prior-dominated, and modeling systematics associated with the parameterization of prior beliefs may therefore be important.

We detail our development of Gaussian processes in Sec.~\ref{sec:nonparam} and describe our inference scheme in Sec.~\ref{sec:bayes}.
Sec.~\ref{sec:method} reports the specific priors used in this analysis.
We apply them to simulated signals in Sec.~\ref{sec:injections} and to real GW170817 data in Sec.~\ref{sec:gw170817}.
We present our conclusions in Sec.~\ref{sec:conclusion}.

\section{Non-parametric representation of the equation of state} \label{sec:nonparam}

Our non-parametric approach to GW-based EOS inference relies heavily on Gaussian processes (GPs). We therefore begin by reviewing GPs in general, before explaining our specific implementation for causal, thermodynamically stable {\EOS}s.
GPs have been applied to many problems in GW astronomy, such as the modeling of gravitational waveforms~\cite{Doctor2017,Huerta} and electromagnetic counterpart light curves~\cite{Coughlin2018}, the optimization of parameter estimation strategies~\cite{MooreGair,Moore,Lange}, and hierarchical population inference \cite{Taylor,VitaleFarr}, but they have not previously been used in the relativistic inverse stellar structure problem.
Furthermore, many of these applications only use GPs to interpolate between a sparse sample of known functions, marginalizing over interpolation uncertainty, rather than employing a GP directly as a Bayesian prior.
We use GPs for both purposes.
A helpful general reference about GPs can be found in Ref.~\cite{GPR}.

\subsection{Gaussian processes} \label{sec:gp}

GPs provide an extremely flexible and compact representation of the uncertainty in a function's values (ordinates).
A GP treats a real-valued function $f$ as an element of an infinite-dimensional vector space, with the correlations between its ordinates $f_i := f(x_i) \in (-\infty, \infty)$ modeled as a multivariate Gaussian distribution $\mathcal{N}$ with mean $\left<f_i\right>$ and covariance $\mathrm{Cov}(f_i, f_j)$.
The joint distribution on the function's ordinates, conditioned on their corresponding abscissae $x_i$, is thus

\begin{equation} \label{firstdistr}
    f_i \, | \, x_i \sim \mathcal{N}\Big( \left<f_i\right>, \mathrm{Cov}(f_i, f_j) \Big) .
\end{equation}
The elements of the covariance matrix are determined by a covariance kernel $K$ such that \mbox{$K_{ij} := K(x_i, x_j) = \mathrm{Cov}(f_i, f_j)$}.
Typically, $K$ is chosen such that the covariance is larger for abscissae that are closer together, effectively constraining $f$ to be smooth over a length scale set by the kernel.
More generally, $K$ determines the preferred functional behavior of realizations drawn from the GP. 
Each realization $f$ of the GP is therefore a sequence of correlated random variables, but all such functions nonetheless share some general features.

A standard choice for $K$ is the \textit{squared-exponential kernel}
\begin{equation} \label{sqexpK}
    K_\mathrm{se}(x, x') = \sigma^2 \exp\left( -\frac{(x-x')^2}{2l^2} \right) ,
\end{equation}
in which the hyperparameters $\sigma$ and $l$ determine the functions' behavior: $\sigma$ sets the overall strength of the correlations, and $l$ governs the length scale over which they occur.
Other common choices are the Ornstein-Uhlenbeck and Mort\'{e}n kernels (see Sec. 4.2 of Ref.~\cite{GPR} for a more complete list).
We choose the squared-exponential because it is analytic and adequately captures the observed variations within our data.
Hence, our GP is built from this kernel, along with variants of the white-noise kernel
\begin{equation} \label{wn}
K_\mathrm{wn}(x, x') = \sigma_{\mathrm{wn}}^2(x) \delta(x-x'),
\end{equation}
which models measurement uncertainty at a specific point.

One can determine conditional distributions for an arbitrary set of ordinates $f_i$ given a set $\{ f_{j^\ast} \}$ of known values via
\begin{equation}\label{eqn:condition}
    P\left( f_i | x_i, \{ f_{j^\ast}, x_{j^\ast} \} \right) = \frac{P(f_i, \{f_{j^\ast}\} | x_i, \{ x_{j^*}\} )}{P(\{f_{j^\ast}\}|\{x_{j^\ast}\})} ,
\end{equation}
the definition of a conditional probability.
Here and throughout this work, starred indices (e.g.~$x_{i^*}$) indicate abscissae upon which we condition the GP, as opposed to generic abscissae (e.g.~$x_i$) at which the ordinates are unknown. 
We also adopt the Einstein summation convention throughout.
Applying Eq.~\eqref{eqn:condition} to a GP yields
\begin{equation}\label{eqn:a post gpr}
   f_i \ | \ x_i, \{ f_{j^\ast}, x_{j^\ast} \} \ \sim \ \mathcal{N} \left( \left<f_i\right> + K_{i k^\ast}\left(K^{-1}\right)_{k^\ast j^\ast}\left(f_{j^\ast} - \left<f_{j^\ast}\right>\right), K_{ij} - K_{i m^\ast} \left( K^{-1} \right)_{m^\ast n^\ast} K_{n^\ast j} \right) .
\end{equation}
As the set $\{ f_{j^\ast} \}$ of known values is expanded, the conditioned processes from which the ordinates are drawn become tighter, reducing the variance between different realizations $f$ of the GP. 
That is, the uncertainty in the representation decreases as knowledge of the function accrues.

Furthermore, conditioned processes can be inferred simultaneously for $f_i$ and its first derivative $\partial_i f := df/dx|_{x=x_i}$ via a self-consistent joint distribution (see Secs. 4.1.1 and 9.4 of Ref.~\cite{GPR} for a derivation)
\begin{equation}\label{eqn:default process}
    \left. \begin{bmatrix} f_i \\ \partial_i f \end{bmatrix} \ \right| \ x_i \ \sim \ \mathcal{N} \left( \begin{bmatrix} \left<f_i\right> \\ \left<\partial_i f\right> \end{bmatrix} , \begin{bmatrix} K_{ij} & \partial_j K_{i \centerdot} \\ \partial_i K_{\centerdot j} & \partial_i \partial_j K_{\centerdot \centerdot} \end{bmatrix} \right) .
\end{equation}
Here we introduce the notation $\partial_i K_{\centerdot j} := \partial_x K(x, x_j)|_{x=x_i}$, $\partial_j K_{i \centerdot} := \partial_{x'} K(x_i, x')|_{x'=x_j}$, and \linebreak $\partial_i \partial_j K_{\centerdot \centerdot} := \partial_{x} \partial_{x'} K(x, x')|_{x=x_i, x'=x_j}$.
Applying Eq.~\eqref{eqn:condition}, the conditioned distribution becomes
\begin{equation} \label{eq:a post gpr2}
    \left. \begin{bmatrix} f_i \\ \partial_i f \end{bmatrix} \  \right| \ x_i, \{f_{j^\ast}, x_{j^\ast}\} \ \sim \ \mathcal{N} \Bigg( \mathrm{E}\left(\begin{bmatrix} f_i \\ \partial_i f \end{bmatrix}\right),\ \text{Cov}\left(\begin{bmatrix} f_i \\ \partial_i f \end{bmatrix}, \begin{bmatrix} f_j \\ \partial_j f \end{bmatrix}\right) \Bigg) ,
\end{equation}
where
\begin{align}
    \mathrm{E}\left(\begin{bmatrix} f_i \\ \partial_i f \end{bmatrix} \right) & = \begin{bmatrix} \left<f_i\right> \\ \left<\partial_i f\right> \end{bmatrix} + \begin{bmatrix} K_{i j^\ast} \\ \partial_i K_{\centerdot j^\ast} \end{bmatrix} \left(K^{-1}\right)_{j^\ast k^\ast} \left( f_{k^\ast} - \left<f_{k^\ast}\right> \right) , \label{expectation} \\
    \text{Cov}\left(\begin{bmatrix} f_i \\ \partial_i f \end{bmatrix}, \begin{bmatrix} f_j \\ \partial_j f \end{bmatrix} \right) & = \begin{bmatrix} K_{ij} & \partial_j K_{i \centerdot} \\ \partial_i K_{\centerdot j} & \partial_i \partial_j K_{\centerdot \centerdot} \end{bmatrix} - \begin{bmatrix} K_{i j^\ast} \\ \partial_i K_{\centerdot j^\ast} \end{bmatrix} \left( K^{-1} \right)_{j^\ast k^\ast} \begin{bmatrix} K_{k^\ast j} & \partial_j K_{k^\ast \centerdot} \end{bmatrix} \label{cov} ,
\end{align} 
given known values $\left\{ f_{j^\ast} \right \}$ and corresponding abscissae $\left\{ x_{j^\ast} \right\}$.
Contingent on our choice of $K$ and hyperparameters, this conditioned process allows us to make probabilistic statements about $f_i$ and $\partial_i f$ at arbitrary abscissae based on observations of ordinates in other regions of the function's domain.
In other words, Eq.~\eqref{eq:a post gpr2} generates a function over its whole domain from a finite set of known values.
GPR is more sophisticated than simple interpolation, as it automatically comes equipped with estimates of the uncertainty in the interpolated values.

In our application, we use GPR to model uncertainty in the NS \EOS.
Viewing the GP as a generator of random functions with (specifiable) common overall behavior, we harness GPR to produce many random synthetic {\EOS}s that resemble candidate {\EOS}s from nuclear theory.
We find that the squared-exponential covariance suitably reproduces the variability observed in the nuclear-theoretic candidate {\EOS}s when the hyperparameters are appropriately selected. In principle, one could choose the hyperparameters by optimizing the likelihood $P(\{ f_{j^\ast} \}|\{ x_{j^\ast} \}, \sigma, l)$ of observing the known data $f_{j^\ast}$ at $x_{j^\ast}$ with respect to ($\sigma$, $l$); or, better still, marginalize over the hyperparameters according to this likelihood and a prior $P(\sigma, l)$ so that
\begin{equation} \label{mixture}
    P(f_i|x_i, \{ f_{j^\ast}, x_{j^\ast} \}) = \int d\sigma dl \, P(f_i|x_i, \{f_{j^\ast}, x_{j^\ast} \}; \sigma, l)\, P(\{ f_{j^\ast} \}|\{ x_{j^\ast} \}; \sigma, l)\, P(\sigma, l) , 
\end{equation}
in effect defining a mixture model of many GPs.
However, in our implementation we simply set the hyperparameters to approximate their maximum-likelihood values.
This manual selection of hyperparameters does not influence our results, and exploring refinements of these choices is left to future work.

\subsection{Gaussian process model for the equation of state} \label{sec:model}

The NS \EOS~is a function $\mu(p)$ that relates the star's pressure $p$ to its total energy density $\mu$, the sum of rest-mass energy density $\rho$ and internal energy density $\epsilon$. 
The rest-mass and total energy densities are related by the first law of thermodynamics,

\begin{equation} \label{1stlaw}
d\mu =  \frac{(\mu + p)}{\rho}  d\rho .
\end{equation}
Since the precise functional relationship between $\mu$ and $p$ is uncertain, but partial knowledge of the nuclear microphysics places some restrictions on its behavior, a GP conditioned on candidate models from nuclear theory is especially well-suited to modeling \EOS~uncertainty.
In particular, individual realizations of the GP serve as samples from the set of physically viable {\EOS}s.

The simplest way to construct a non-parametric representation of the EOS would be to apply GPR directly in the $p$-$\mu$ plane. 
However, the resulting GP would formally support {\EOS}s~with $\mu < 0$, since its range is necessarily the whole real line. 
Moreover, by default, GPR on $\mu(p)$ would include thermodynamically unstable or acausal {\EOS}s---i.e.~functions with $dp/d\mu < 0$ or superluminal sound speeds $c_s= \sqrt{dp/d\mu} > c$.
To incorporate causality, thermodynamic stability and positivity of the total energy automatically, we instead build a GP over
\begin{equation}\label{phi}
    \phi = \log\left(c^2\frac{d \mu}{d p} - 1 \right).
\end{equation}
The auxiliary variable $\phi$, first introduced in Ref.~\cite{Lindblom_spectral}, can take any value along the real line, and it naturally incorporates the desired physical constraints because $\phi \in \mathbb{R}$ corresponds to $0 \leq dp/d\mu \leq c^2$.
Positivity of $p$ then ensures $c_s \leq c$ and $\mu \geq 0$.
The spectral approach to \EOS~inference \cite{Lindblom_stellarproblem,Lindblom_stellarproblem2} employed in Ref.~\cite{LVC_eos} also starts with a similar transformation, but goes on to decompose the \EOS~onto a small set of basis functions.
In contrast, we assign a prior process to $\phi = \phi(\log p)$ via GPR conditioned on tabulated {\EOS}s.
The process over $\phi$ can easily be translated to a process over $\mu$.
Since the map from $\phi$ to $\mu$ is nonlinear, Gaussian uncertainty in $\phi$ will, however, generally not correspond to Gaussian uncertainty in $\mu$.

We condition the GP for $\phi(\log p)$ on a training set of $n$ candidate {\EOS}s $\mu^{(\alpha)}(p)$ from the literature.
The data for the $\alpha^{\text{th}}$ candidate \EOS~constitute a function $\log \mu^{(\alpha)}(\log p_{j^*}^{(\alpha)})$ with ordinates $\log \mu^{(\alpha)}_{j*}$ and abscissae $\log p_{j^*}^{(\alpha)}$.\footnote{Candidate \EOS~data tabulated as $p$ vs.~$\rho$ can also be accommodated by transforming the rest-mass energy density to total energy density via Eq.~\eqref{1stlaw}.} 
As the amount of $\{\log\mu^{(\alpha)}_{j*}, \log p_{j^*}^{(\alpha)} \}$ data available may vary from candidate \EOS~to candidate \EOS, and the relative weight assigned to each model in the training set is proportional to the number of data points included, we resample each $\mu^{(\alpha)}(p)$ to $s$ points so that the GP for $\phi(\log p)$ is conditioned equally on every input \EOS.
This need not be the case---one could formulate a mixture model of GPs (see, e.g., Refs.~\cite{NguyenTuong2008, Meier2014}) to establish a weighted training set of candidate EOSs---but for simplicity we assume equal weights here.
This resampling could be performed with, e.g., linear interpolation, but we instead use GPR to obtain an estimate of the uncertainty associated with the interpolation.
Thus, we construct a GP representation of $\phi^{(\alpha)}(\log p)$ for each candidate \EOS~in the training set, such that one realization of the $\alpha^{\text{th}}$ GP is an $s$-fold list of ordinates $\{ \phi_{i}^{(\alpha)} \}$ at evenly spaced points in $\log p$.
In this way, the maps from $\mu$ to $\phi$ for the tabulated EOSs---which come without uncertainties---are effectively equipped with error bars.
The GPs for $\phi^{(\alpha)}(\log p)$ are subordinate to, and used as input for, the overarching GP for $\phi(\log p)$. We next describe the construction of the GPs for $\phi^{(\alpha)}(\log p)$ in some detail, before addressing the GP for $\phi(\log p)$ itself.

For every candidate \EOS~$\mu^{(\alpha)}(p)$, we first fit $\log \mathbf{\mu}^{(\alpha)}$ with a low-order polynomial in $\log p$ and construct a GP for the residuals.
The resulting joint distribution on $\log \mu^{(\alpha)}$, its first derivative $\partial_i \log \mu^{(\alpha)} := d\log\mu^{(\alpha)}/d\log p \, |_{p=p_i}$ and the tabulated data $\{\log \mu^{(\alpha)}_{j^\ast}\}$ is
\begin{equation}\label{eqn:gp appendix 1}
    \left. \begin{bmatrix} \log\mu_i^{(\alpha)} \\ \partial_i \log\mu^{(\alpha)} \\ \log\mu_{j^\ast}^{(\alpha)} \end{bmatrix} \ \right| \ \log p_i, \{\log p^{(\alpha)}_{j^\ast}\} \ \sim \ \mathcal{N}\left( \begin{bmatrix} \log\hat{\mu}^{(\alpha)}_i \\ \partial_i \log\hat{\mu}^{(\alpha)} \\ \log\hat{\mu}^{(\alpha)}_{j^\ast} \end{bmatrix},\ \begin{bmatrix} K^{(\alpha)}_{ij} & \partial_j K^{(\alpha)}_{i\centerdot} & K^{(\alpha)}_{i j^\ast} \\ \partial_i K^{(\alpha)}_{\centerdot j} & \partial_i \partial_j K^{(\alpha)}_{\centerdot\centerdot} & \partial_i K^{(\alpha)}_{\centerdot j^\ast} \\ K^{(\alpha)}_{i^\ast j} & \partial_j K^{(\alpha)}_{i^\ast \centerdot} & K^{(\alpha)}_{i^\ast j^\ast} \end{bmatrix} \right) .
\end{equation}
Here we take the values $\{ \log \hat{\mu}_i^{(\alpha)}, \partial_i \log \hat{\mu}^{(\alpha)}, \log \hat{\mu}^{(\alpha)}_{j^*} \}$ from the low-order polynomial fit as the mean in the process, and model correlations among the residuals around this fit.
While we assume a squared-exponential covariance kernel for all tabulated {\EOS}s, we choose hyperparameters $(\sigma^{(\alpha)}, l^{(\alpha)})$ separately for each $K^{(\alpha)}_{ij}$.

\begin{figure} 
    \includegraphics[width=0.49\textwidth, clip=True, trim=0.15cm 0.30cm 0.30cm 0.30cm]{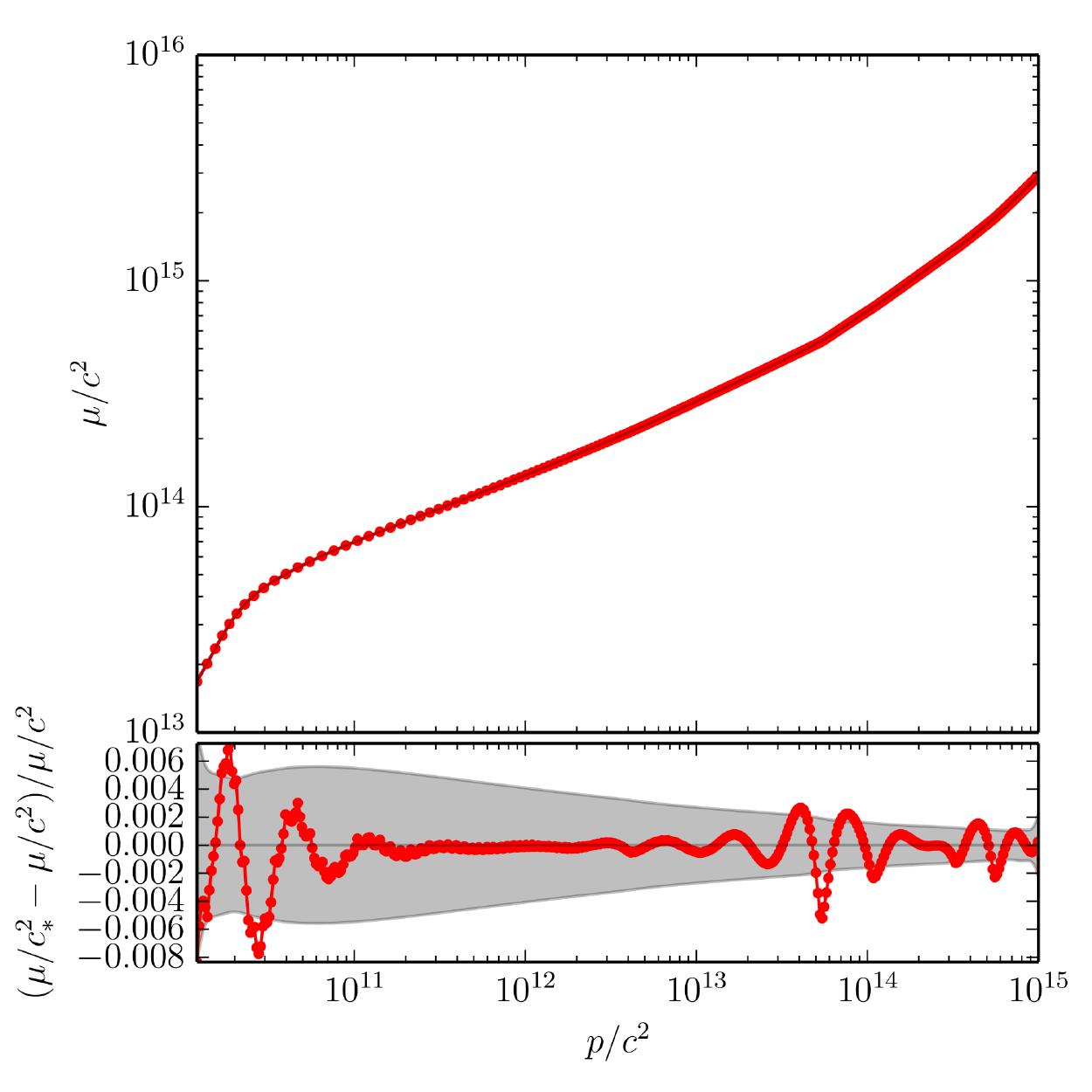}
    \includegraphics[width=0.49\textwidth, clip=True, trim=0.15cm 0.30cm 0.30cm 0.30cm]{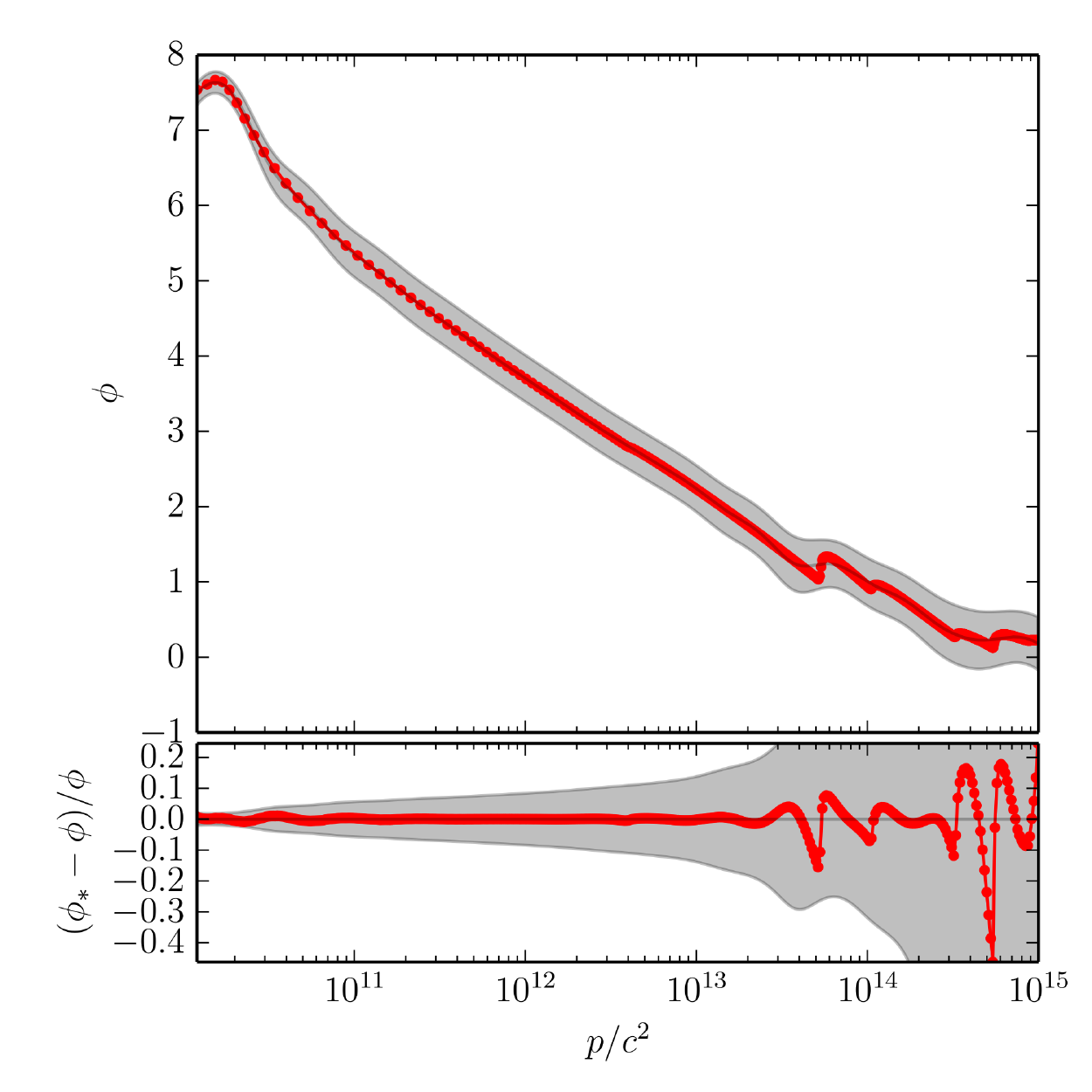}
    \caption{
        A demonstration of our GP mapping from $\mu^{(\texttt{h4})}$ to $\phi^{(\texttt{h4})}$ using a 3$^\mathrm{rd}$-order polynomial fit and a squared-exponential kernel with $\sigma^{(\texttt{h4})}=0.1$ and $l^{(\texttt{h4})}=0.6$ (see Eq.~\eqref{eqn:gp appendix 1}).
        Irregularly sampled tabulated values and estimates obtained from numeric differentiation (\textit{red}) are resampled to a regular grid in $\log p$ via GPR, along with residuals and associated error estimates (\textit{black}).
        We note that the residuals between tabulated values and our interpolation of $\mu^{(\alpha)}$ are $\lesssim 1\%$ at high pressures, resulting in $\lesssim10\%$ relative uncertainty in $\phi$.
    }
    \label{fig:gpr resamp}
\end{figure}

Conditioning the joint distribution on the tabulated ordinates $\{ \log\mu^{(\alpha)}_{j^\ast} \}$ via Eq.~\eqref{eqn:condition}, we obtain
\begin{equation}\label{eqn:gp appendix 2}
    \left. \begin{bmatrix} \log\mu_i^{(\alpha)} \\ \partial_i \log\mu^{(\alpha)} \end{bmatrix} \ \right| \ \log p_i, \left\{\log\mu^{(\alpha)}_{j^\ast}, \log p^{(\alpha)}_{j^\ast}\right\} \ \sim \ \mathcal{N}\Bigg(\mathrm{E}^{(\alpha)}\left(\begin{bmatrix} \log\mu_i \\ \partial_i \log\mu \end{bmatrix}\right), \ \mathrm{Cov}^{(\alpha)}\left(\begin{bmatrix} \log\mu_i \\ \partial_i \log\mu \end{bmatrix}, \begin{bmatrix} \log\mu_j \\ \partial_j \log\mu \end{bmatrix} \right) \Bigg) ,
\end{equation}
where the explicit expressions for the expectation value and conditioned covariance follow from Eqs.~\eqref{expectation} and \eqref{cov}.
The corresponding process for $\phi^{(\alpha)}$ is calculated from Eq.~\eqref{phi} as
\begin{equation} \label{phigp}
    \phi_i^{(\alpha)} \ | \ \log p_i, \left\{ \log\mu^{(\alpha)}_{j^\ast}, \log p^{(\mathrm{\alpha})}_{j^\ast} \right\} \ \sim \ \mathcal{N}\left( \mathrm{E}^{(\alpha)}(\phi_i),\ \mathrm{Cov}^{(\alpha)}\left(\phi_i, \phi_j\right)\right) ,
\end{equation}
with
\begin{equation}\label{eqn:phi mean}
    \mathrm{E}^{(\alpha)}(\phi_i) = \log\left( \mathrm{E}^{(\alpha)}\left(\partial_i \log \mu\right) \left(\frac{e^{\mathrm{E}^{(\alpha)}\left(\log \mu_i\right)}}{p_i}\right) c^2 - 1\right) .
\end{equation}
We approximate the covariance matrix $\mathrm{Cov}^{(\alpha)}(\phi_i, \phi_j)$ through a first order Taylor expansion for $\phi_i$ in terms of $\log\mu_i$ and $\partial_i \log\mu$, i.e.~
\begin{equation}\label{eqn:mu to phi}
    \phi_i(\log \mu_i, \partial_i \log \mu) \approx \phi_i + \left.\frac{\delta \phi}{\delta \log\mu}\right|_i \delta \log\mu_i + \left.\frac{\delta \phi}{\delta (\partial \log\mu)}\right|_i \delta (\partial_i \log\mu) ,
\end{equation}
so that
\begin{align}\label{eqn:mu to phi cov}
    \mathrm{Cov}\left(\phi_i,\phi_j\right) = &   \left(\left. \frac{\delta \phi}{\delta \log\mu} \right|_i\right) \left(\left. \frac{\delta \phi}{\delta \log\mu} \right|_j\right)                   \mathrm{Cov}\left(\log\mu_i, \log\mu_j\right) \nonumber \\
                                             & + \left(\left. \frac{\delta \phi}{\delta (\partial \log\mu)} \right|_i\right) \left( \left. \frac{\delta\phi}{\delta \log\mu} \right|_j\right)          \mathrm{Cov}\left(\partial_i \log \mu, \log\mu_j\right) \nonumber \\
                                             & + \left(\left. \frac{\delta \phi}{\delta \log\mu} \right|_i\right) \left(\left. \frac{\delta \phi}{\delta (\partial \log\mu)} \right|_j\right)          \mathrm{Cov}\left(\log \mu_i, \partial_j \log \mu\right) \nonumber \\
                                             & + \left(\left. \frac{\delta \phi}{\delta (\partial \log\mu)} \right|_i\right) \left(\left. \frac{\delta \phi}{\delta (\partial \log\mu)} \right|_j\right) \mathrm{Cov}\left(\partial_i \log \mu, \partial_j \log\mu\right) .
\end{align}
This allows us to translate $\log \mu^{(\alpha)}(\log p^{(\alpha)}_{j^\ast})$ to the auxiliary variable $\phi^{(\alpha)}(\log p_i)$, interpolated at an arbitrary set of abscissae, while keeping track of the uncertainty associated with the interpolation.
We refer to this as \emph{resampling} the tabulated {\EOS}s, because we take irregularly spaced tabulated data for $\mu^{(\alpha)}(p)$ and transform it into a process for $\phi^{(\alpha)}(\log p)$ from which we can extract regularly sampled data. 
Fig.~\ref{fig:gpr resamp} shows the $\phi^{(\alpha)}(\log p)$ process we construct for one of the candidate {\EOS}s described in Sec.~\ref{sec:method}.

The overarching GP for $\phi(\log p)$ is subsequently conditioned on the collection of processes $\phi^{(\alpha)}(\log p)$ for the individual candidate {\EOS}s.
We again use a squared-exponential kernel, but supplement it with a white-noise kernel (cf.~Eq.~\eqref{wn})
\begin{equation} \label{wncov}
    K_{ij}^{(\mathrm{EOS})} = \delta_{ij} \left. \frac{{\sigma_n}^2}{n} \sum\limits_{\alpha}\left(\mathrm{E}^{(\alpha)}\left(\phi_i\right) - \mathrm{E}^{(\mathrm{\EOS})}(\phi_i)\right)^2 \quad \right| \quad \mathrm{E}^{(\mathrm{\EOS})}(\phi_i) = \frac{1}{n}\sum_\beta \mathrm{E}^{(\beta)}\left(\phi_i\right)
\end{equation}
scaled by the observed variance between the $n$ tabulated {\EOS}s at the pressures $p_i$.
In this way, we not only model the covariance between $\phi$ at different pressures, but also the spread in candidate {\EOS}s~at each abscissa.
The scaling hyperparameter $\sigma_n$ in Eq.~\eqref{wncov} allows us to specify how closely we wish the synthetic {\EOS}s to follow the tabulated ones; small values enforce the conditioned process to closely follow the average of the tabulated {\EOS}s.
For simplicity, we condition based on the interpolated ordinates $\phi^{(\alpha)}_{i^\ast}$ at the same regularly sampled set of pressures $\log p_{i^\ast}$ regardless of the EOS, though this is not strictly necessary.

\begin{figure}
    \includegraphics[width=0.49\textwidth, clip=True, trim=0.30cm 0.30cm 0.35cm 0.45cm]{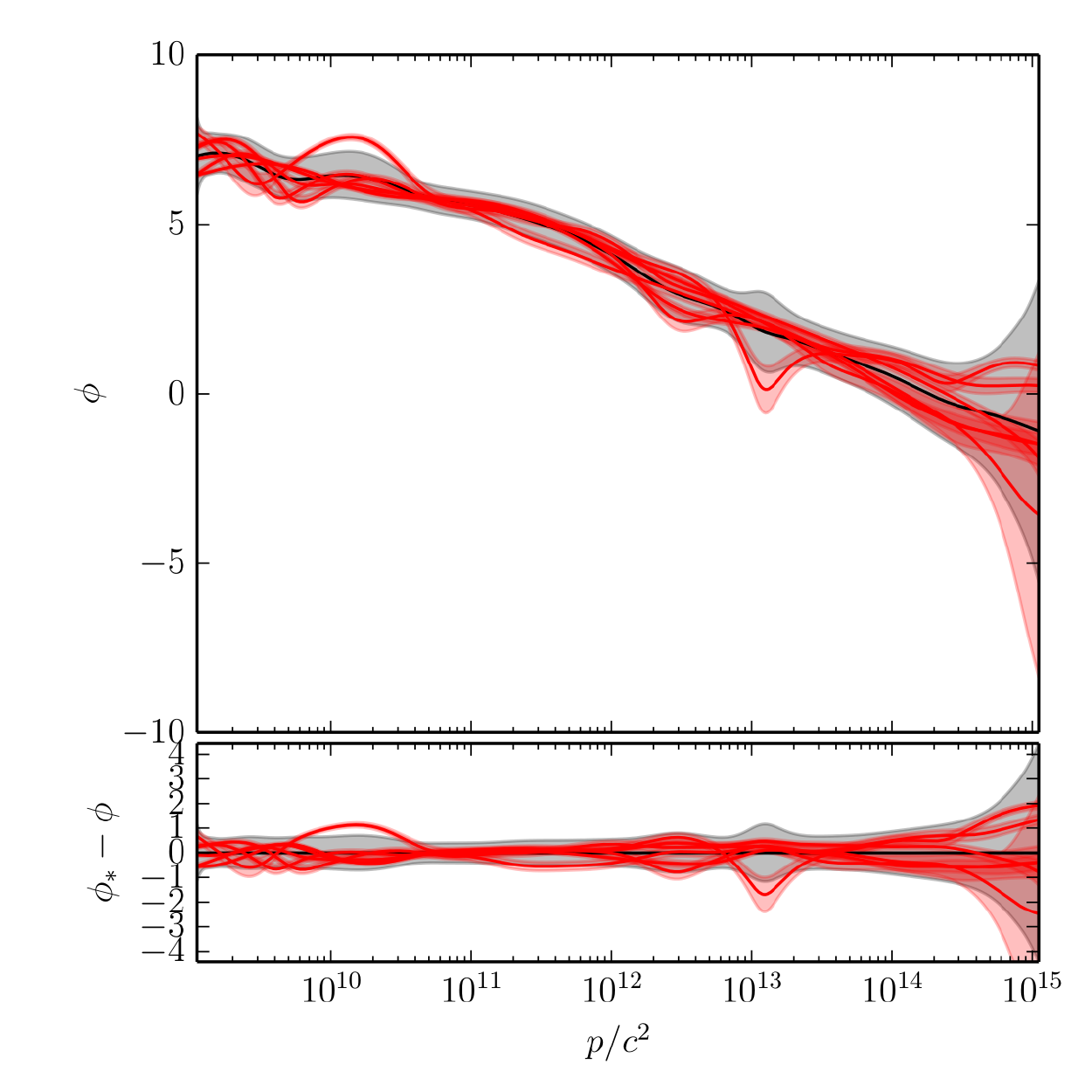}
    \includegraphics[width=0.49\textwidth, clip=True, trim=0.30cm 0.30cm 0.35cm 0.45cm]{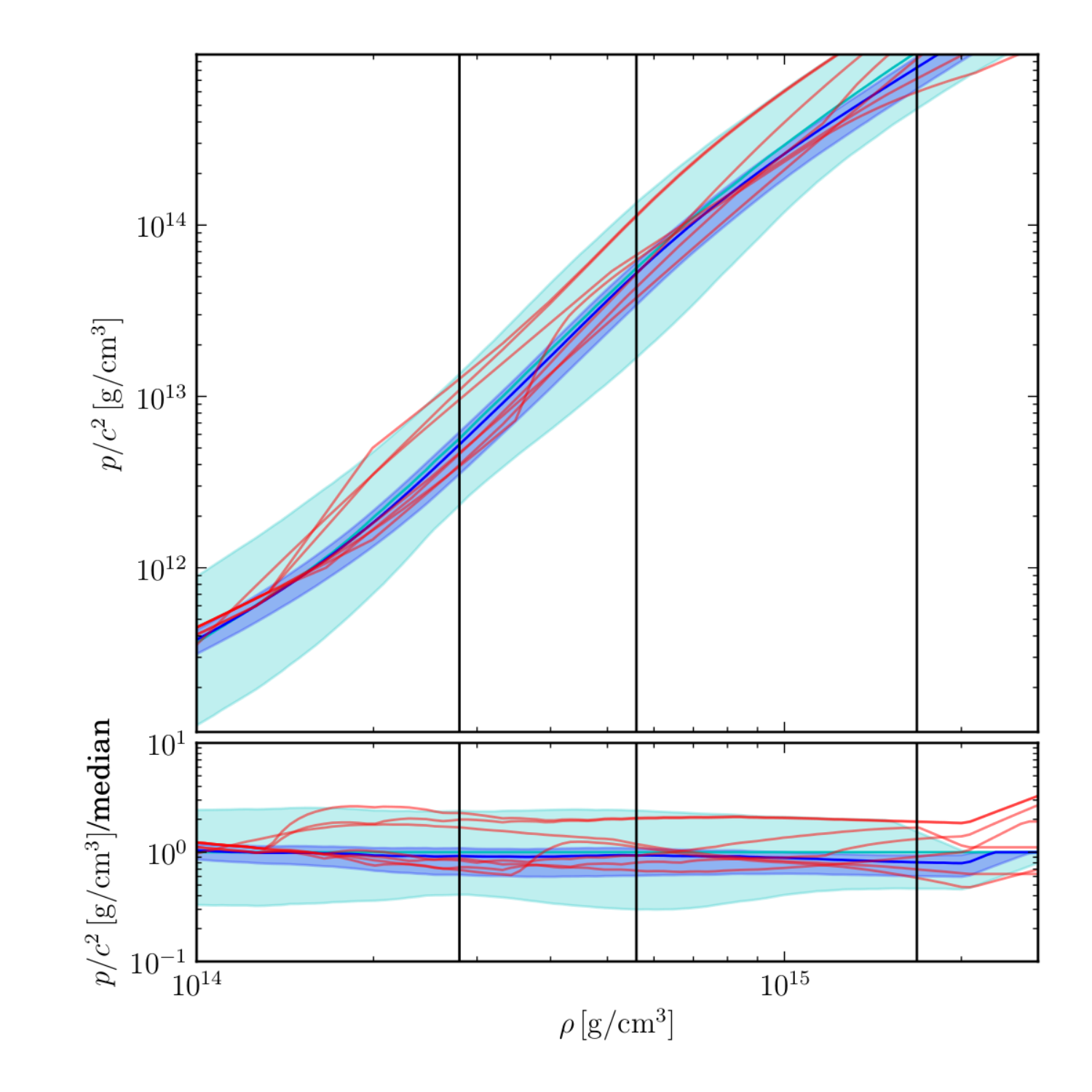}
    \caption{
        Inferred processes for several $\phi^{(\alpha)}(\log p)$ (\textit{red}) and an example overarching GP $\phi(\log p)$ (\textit{black}) conditioned on all of them (\textit{left}) along with the same reference {\EOS}s and the model-agnostic (\textit{light blue}, loosely informed by nuclear theory) and model-informed (\textit{dark blue}, more tightly constrained by nuclear theory) priors used in this analysis (\textit{right}, see Sec.~\ref{sec:method}).
        Shaded regions correspond to 1-$\sigma$ confidence regions.
        We see that the model-agnostic prior comfortably contains all the reference {\EOS}s, while the model-informed prior more closely follows a group of tabulated \EOS.
    }
    \label{fig:gpr draw}
\end{figure}

Equipped with processes $\phi^{(\alpha)}(\log p)$ for each tabulated \EOS, and the white-noise kernel approximating modeling uncertainty at each pressure, we construct the overarching GP for $\phi(\log p)$ conditioned on all the tabulated {\EOS}s.
The joint distribution on $\phi(\log p)$ and the subordinate $\phi^{(\alpha)}(\log p)$ is thus
\begin{equation}
    \left. \begin{bmatrix} \phi_k \\ \phi^{(\alpha=1)}_{i^\ast} \\ \phi^{(\alpha=2)}_{i^\ast} \\ \vdots \\ \phi^{(\alpha=n)}_{i^\ast} \end{bmatrix} \ \right| \ \log p_k, \log p_{i^\ast} \ \sim \ \mathcal{N}\left( \mathrm{E}, \mathrm{Cov} \right)
\end{equation}
with
\begin{equation} \label{E}
    \mathrm{E} = \begin{bmatrix} \hat{\mathrm{E}}^{(\mathrm{\EOS})}(\phi_k) \\ \hat{\mathrm{E}}^{(\mathrm{\EOS})}(\phi_{i^\ast}) \\ \hat{\mathrm{E}}^{(\mathrm{\EOS})}(\phi_{i^\ast}) \\ \vdots \\ \hat{\mathrm{E}}^{(\mathrm{\EOS})}(\phi_{i^\ast}) \end{bmatrix} ,
\end{equation}
where $\hat{\mathrm{E}}^{(\mathrm{\EOS})}(\phi_{i^\ast})$ is a low-order polynomial fit to $\mathrm{E}^{(\mathrm{\EOS})}(\phi_{i^\ast})$ in $\log p$, and
\begin{equation} \label{Cov}
    \mathrm{Cov} = \begin{bmatrix} K_{ij} & K_{i j^\ast} & K_{i j^\ast} & \cdots & K_{i j^\ast} \\ K_{i^\ast j} & K_{i^\ast j^\ast} + \mathrm{Cov}^{(\alpha=1)}_{i^\ast j^\ast} + K^{(\mathrm{\EOS})}_{i^\ast j^\ast} & K_{i^\ast j^\ast} & \cdots & K_{i^\ast j^\ast} \\ K_{i^\ast j} & K_{i^\ast j^\ast} & K_{i^\ast j^\ast} + \mathrm{Cov}^{(\alpha=2)}_{i^\ast j^\ast} + K^{(\mathrm{\EOS})}_{i^\ast j^{\ast}} & \cdots & K_{i^\ast j^\ast} \\ \vdots & \vdots & \vdots & \ddots & \vdots \\ K_{i^\ast j} & K_{i^\ast j^\ast} & K_{i^\ast j^\ast} & \cdots & K_{i^\ast j^\ast} + \mathrm{Cov}^{(\alpha=n)}_{i^\ast j^\ast} + K^{(\mathrm{\EOS})}_{i^\ast j^\ast} \end{bmatrix} .
\end{equation}
The covariance matrix for this GP representation of the NS \EOS~is composed of several terms:
\begin{enumerate}[label=\emph{\alph*})]
    \item{block diagonal contributions $\mathrm{Cov}^{(\alpha)}$ from the conditioned uncertainty in $\phi^{(\alpha)}_{i}$ for each candidate \EOS, as in Eq.~\eqref{phigp};}
    \item{diagonal contributions $K^{(\mathrm{\EOS})}$ from the white-noise kernel \eqref{wncov}, encoding modeling uncertainty in the ordinates $\{\phi_{j^*}^{(\alpha)}\}$ from the candidate {\EOS}s at each pressure $p_{j^*}$; and}
    \item{a squared-exponential covariance matrix $K$, as in Eq.~\eqref{sqexpK}, describing the correlations between all ordinates $\phi_{i}$.}
\end{enumerate}
Specific choices for the squared-exponential kernel hyperparameters ($\sigma$, $l$) and the white-noise kernel scaling factor $\sigma_n$ complete the model.
Conditioning this process on the means $\mathrm{E}^{(\alpha)}$ for the tabulated {\EOS}s yields 
\begin{equation}\label{final}
\phi_i|\log p_i, \{\mathrm{E}^{(\alpha=1)}(\phi_{j^\ast}), \mathrm{E}^{(\alpha=2)}(\phi_{j^\ast}), \cdots, \mathrm{E}^{(\alpha=n)}(\phi_{j^\ast}), \log p_{j^\ast} \} \sim \mathcal{N} \left(\tilde{\text{E}} , \widetilde{\text{Cov}} \right) ,
\end{equation}
which approximates the desired process $\phi_i|\log p_i, \{\log\mu^{(\alpha=1)}_{j^\ast}, \log p_{j^\ast}^{(\alpha=1)}\}, \{\log\mu^{(\alpha=2)}_{k^\ast}, \log p_{k^\ast}^{(\alpha=2)}\}, \cdots, \linebreak \{\log\mu^{(\alpha=n)}_{l^\ast}, \log p_{l^\ast}^{(\alpha=n)}\}$ by conditioning on interpolated values $E^{(\alpha)}$ with associated uncertainties instead of conditioning directly on the tabulated EOS data.
This is illustrated in Fig.~\ref{fig:gpr draw}.
Concrete expressions for the final conditioned expectation value and covariance follow from Eq.~\eqref{eqn:a post gpr}.

The GP defined by Eq.~\eqref{final} models $\phi(\log p)$, rather than $\mu(p)$.
Nonetheless, for each realization $\phi^{(a)}(\log p)$ of the process, we invert Eq.~\eqref{phi} and numerically integrate $\partial \mu^{(a)}/\partial p = (1 + e^{\phi^{(a)}})/c^2$ to obtain $\mu^{(a)}(p)$, which is suitable for calculating macroscopic NS properties.
The rest-mass density can be obtained from $\mu^{(a)}$ via Eq.~\eqref{1stlaw}, allowing us to plot the EOS in conventional $p^{(a)}(\rho)$ form.
The individual realizations of the {\EOS} are stitched onto a model for the low-density crust \EOS, which is well known below nuclear saturation density; we use a piecewise polytrope implementation of \texttt{sly}~\cite{Read_pwpoly}.
In this way, we obtain a \emph{unified \EOS} valid at all densities up to a specified maximum value. The distribution of functions $\mu^{(a)}(p)$ calculated in this manner defines the \emph{EOS prior process} $P(\mu(p)|\{\mu^{(\alpha)}_{j^\ast}\}, \vec{\sigma})$.
Each \emph{synthetic {\EOS}} drawn from this process is a random function that resembles the input tabulated {\EOS}s both qualitatively and quantitatively. 
The choice of hyperparameters $\vec{\sigma} := \left[ \sigma,\, l,\, \sigma_n,\, \sigma^{(\alpha=1)},\, l^{(\alpha=1)}, \sigma^{(\alpha=2)},\, l^{(\alpha=2)}, ..., \sigma^{(\alpha=n)},\, l^{(\alpha=n)} \right]$ for the GP provides the freedom to self-consistently explore a wide swath of the space of possible {\EOS}s, or to mimic to the behavior observed in the candidate {\EOS}s more closely.

The process for $\mu(p)$, conditioned on the full training set of candidate {\EOS}s, generates a large set of new proposals for the pressure-density relation without resorting to a parameterization.
Astrophysical constraints on the \EOS, such as the requirement that it support observed NS masses, are applied by directly checking, e.g., the maximum mass for each synthetic \EOS.
Any synthetic {\EOS} that violates the constraint is discarded from the prior.
Similar astrophysical cutoffs could incorporate future information about NS radii and moments of inertia from electromagnetic astronomy.

\section{Bayesian inference of the equation of state via Monte Carlo integration} \label{sec:bayes}

GP representations of the uncertainty in NS internal structure play a fundamental role in our non-parametric inference of the \EOS. 
The candidate {\EOS}s upon which the GP is conditioned, as well as the choice of the hyperparameters $\vec{\sigma}$, determine the characteristics of the \EOS~prior process $P(\mu(p)|\{\mu^{(\alpha)}_{j^\ast}\}, \vec{\sigma})$. 
In particular, the degree to which elements $\mu^{(a)}(p) \sim P(\mu(p)|\{\mu^{(\alpha)}_{j^\ast}\}, \vec{\sigma})$ of the prior process resemble the input {\EOS}s is controlled by the hyperparameters $\sigma$, $l$ and $\sigma_n$. 
Here we show how the GW likelihood from parameter estimation is used in conjunction with the prior to produce an \emph{\EOS~posterior process}.

The posterior process for the \EOS~is ultimately derived from GW detector data ($d$), which is sensitive to macroscopic NS observables like masses and tidal deformabilities.
Source properties are recovered from $d$ via Bayesian inference for the parameters of a prescribed waveform model.
The waveform parameters include the masses $M_{1,2}$, the tidal deformabilities $\Lambda_{1,2}$, and other source properties and phenomenological parameters denoted as $\vartheta$.
The priors chosen for $M_{1,2}$ and $\vartheta$, along with the underlying description of the waveform, constitute the Bayesian model $\mathcal{H}$ for the data.

Strictly speaking, the waveform parameters $M_{1,2}$ and $\Lambda_{1,2}$ are not completely independent, as a specification of $M$ and the \EOS~is sufficient to determine $\Lambda$. 
Indeed, an \EOS~$\mu(p)$ predicts a unique $M$-$\Lambda$ relation, which we denote as $\Lambda(M; \mu)$. 
This relation is calculated by first integrating the Tolman-Oppenheimer-Volkoff (TOV) equations \cite{Tolman,Oppenheimer} and the field equation governing the quadrupolar tidal perturbation \cite{Landry_surficial} simultaneously from the center of the NS to its surface for a choice of central density $\rho_c$. 
These equations of stellar structure determine the mass, radius and tidal deformability of the NS.
A sequence of stable neutron stars is then constructed by repeating the integration for a sequence of central densities up to $\rho_c^{\text{max}}$, the central density for which the mass reaches a maximum $M_{\text{max}}$ and beyond which the star becomes unstable \cite{Harrison}.
The resulting tidal deformability and mass sequence constitutes $\Lambda(M;\mu)$.
Integration of the equations of stellar structure also gives us access to the $M$-$\rho_c$ relation $M(\rho_c;\mu)$, which includes $M_{\text{max}}$ information.

The \EOS~posterior process $P(\mu|d; \{\mu^{(\alpha)}_{j^\ast}\}, \vec{\sigma}, \mathcal{H})$ is obtained from the GW data through an application of Bayes' theorem, which states
\begin{equation} \label{bayesinf}
    P(\mu|d; \{\mu^{(\alpha)}_{j^\ast}\}, \vec{\sigma}, \mathcal{H}) \propto P(d|\mu; \mathcal{H})\, P(\mu|\{\mu^{(\alpha)}_{j^\ast}\}, \vec{\sigma}) .
\end{equation}
$P(d|\mu_i;\mathcal{H})$ is the likelihood of the data given an \EOS~$\mu(p)$ and the model $\mathcal{H}$.
Marginalizing separately over the waveform parameters of interest $(M_{1,2},\Lambda_{1,2})$ and the nuisance parameters $\vartheta$, it can be written as
\begin{equation}
    P(d|\mu; \mathcal{H}) = \int dM_1 dM_2\, P(M_1, M_2|\mathcal{H}) \left[ \int d\vartheta P(\vartheta|\mathcal{H}) P\Big(d|\vartheta, M_1, M_2, \Lambda_1(M_1; \mu), \Lambda_2(M_2; \mu)\Big)\right] .
\end{equation}
The integral in square brackets is the marginal conditional likelihood $\mathcal{L}(d| M_1, M_2, \Lambda_1(M_1; \mu), \Lambda_2(M_2; \mu); \mathcal{H})$ for the observed GW data given a pair of masses and tidal deformabilities.
In terms of this likelihood, Eq.~\eqref{bayesinf} reads
\begin{align} \label{distr}
    P(\mu_i|d; \{\mu^{(\alpha)}_{j^\ast}\}, \vec{\sigma}, \mathcal{H}) \propto& \, P(\mu_i|\{\mu^{(\alpha)}_{j^\ast}\}, \vec{\sigma}) \nonumber \\ &\times \int d\Lambda_1 d\Lambda_2 dM_1 dM_2\, P(M_1, M_2|\mathcal{H}) \delta(\Lambda_1 - \Lambda(M_1; \mu_i)) \delta(\Lambda_2 - \Lambda(M_2; \mu_i)) \left[ \mathcal{L}(d|M_1, M_2, \Lambda_1, \Lambda_2; \mathcal{H}) \right] .
\end{align}
Here, we marginalize over $\Lambda_{1,2}$ in addition to $M_{1,2}$ by introducing Dirac delta functions over $\Lambda(M;\mu)$. 
This accounts for the fact that the likelihood available from standard parameter estimation does not include intrinsic correlations among $\Lambda_{1,2}$ (cf.~Ref.~\cite{LVC_properties}); most stochastic samplers employed in GW parameter estimation sample from $\mathcal{L}(d|M_1, M_2, \Lambda_1, \Lambda_2; \mathcal{H})$, treating $\Lambda_{1,2}$ as completely independent parameters.\footnote{
Some parameter estimation analyses may include non-trivial priors for $M_{1,2}$ and $\Lambda_{1,2}$, but we can account for this by appropriately re-weighting their samples.} 

Thus, provided that $\mathcal{L}(d|M_1, M_2, \Lambda_1, \Lambda_2; \mathcal{H})$ is known, we need only compute the integral \eqref{distr} to obtain a posterior process for the EOS.
In practice, however, standard parameter estimation gives us access to a finite number of samples from the likelihood instead of the distribution itself.
Assuming that the exact marginal conditional likelihood can be accurately modeled from the available samples \cite{GW170817samples}, we simply approximate $\mathcal{L}(d|M_1, M_2, \Lambda_1, \Lambda_2; \mathcal{H})$ with a Gaussian KDE (see, e.g., Ref.~\cite{LVC_properties}). 
The integral is then evaluated with Monte Carlo techniques.
Although Monte Carlo integration is relatively inefficient, it provides the most straightforward way to compute the posterior process.
More advanced methods, such as Markov-Chain Monte Carlo, may be amenable to the inclusion of a GP prior, but the associated jump proposals are non-trivial (see, e.g., Refs.~\cite{Wang2013, Titsias2011}).
We therefore rely exclusively on direct Monte Carlo integration in this paper, implementing the following algorithm to calculate the posterior process:

\begin{enumerate}[label=\emph{\alph*})]
    \item{draw a synthetic \EOS~$\mu^{(a)}_i \sim P(\mu_i|\{\mu^{(\alpha)}_{j^\ast}\}, \vec{\sigma})$ as a realization of the GP prior;}
    \item{draw a pair of component masses $M_1^{(a)}, M_2^{(a)} \sim P(M_1, M_2|\mathcal{H}$) from a specified prior;}
    \item{compute tidal deformabilities $\Lambda_1^{(a)}$ and $\Lambda_2^{(a)}$ associated with $M_1^{(a)}$, $M_2^{(a)}$, and $\mu^{(a)}_i$ via the equations of stellar structure;}
    \item{evaluate the corresponding GW likelihood $\mathcal{L}^{(a)}=\mathcal{L}(d|M_1^{(a)}, M_2^{(a)}, \Lambda_1^{(a)}, \Lambda_2^{(a)}; \mathcal{H})$ via the KDE; and}
    \item{repeat until a sufficiently large number of samples have been collected.}
\end{enumerate}
Because our samples have unequal weights, we gauge the precision of the Monte Carlo integral through the \emph{effective} number of samples
\begin{equation}
    N_\mathrm{eff} = \exp\left(-\sum\limits_a w^{(a)} \log w^{(a)} \right) \quad \left| \quad w^{(a)} = \left( \mathcal{L}^{(a)} \left/ \sum\limits_b \mathcal{L}^{(b)} \right) \right.\right. 
\end{equation}
based on the entropy of the weights observed \textit{a posteriori}.
The larger $N_{\text{eff}}$, the better the precision.
If only a few samples have significant weights, then both the entropy and $N_\mathrm{eff}$ will be small.
However, if most samples have similar weights, the entropy and $N_\mathrm{eff}$ will be large, possibly approaching the upper bound $N_\mathrm{eff} = N$, achieved for equal weights.

Monte Carlo integration also provides a means of directly estimating the evidence $P(d|\{\mu^{(\alpha)}_{j^\ast}\}, \vec{\sigma}, \mathcal{H})$ for a given prior process as a functional integral over $\mu(p)$:
\begin{align} \label{evidence}
    P(d|\{\mu^{(\alpha)}_{j^\ast}\}, \vec{\sigma}, \mathcal{H}) & = \int \mathcal{D} \mu\, dM_1 dM_2\, P(\mu|\{\mu^{(\alpha)}_{j^\ast}\}, \vec{\sigma}) P(M_1, M_2|\mathcal{H}) \mathcal{L}\left(d|M_1, M_2, \Lambda\big(M_1; \mu\big), \Lambda\big(M_2; \mu\big); \mathcal{H} \right) \nonumber \\
                                                              & \approx \frac{1}{N} \sum\limits_a^N \mathcal{L}^{(a)} \quad \left| \quad \begin{matrix} \mu^{(a)} \sim P(\mu|\{\mu^{(\alpha)}_{j^\ast}\}, \vec{\sigma}) \\ M_1^{(a)}, M_2^{(a)} \sim P(M_1, M_2|\mathcal{H}) \end{matrix} \right. .
\end{align}
Unlike the posterior process, the sampling uncertainty in $P(d|\{\mu^{(\alpha)}_{j^\ast}\}, \vec{\sigma}, \mathcal{H})$ scales inversely with $\sqrt{N}$ rather than $\sqrt{N_\mathrm{eff}}$, although $N_\mathrm{eff} \propto N$ as $N\rightarrow\infty$.
Because our KDE approximation of the likelihood has an unknown normalization factor, only relative measures of evidence are meaningful, and we therefore compute the Bayes factor---the ratio of evidences---when comparing support for different priors.

\section{Gaussian process equation-of-state priors}\label{sec:method}

In our specific implementation, we condition our GP \EOS~prior on a fiducial set of candidate {\EOS}s selected from extant nuclear-theoretic models.
Our training set of candidate {\EOS}s is composed of seven well-established models \cite{EOScatalog}: \texttt{alf2} \cite{ALF2}, \texttt{eng} \cite{ENG}, \texttt{h4} \cite{H4}, \texttt{mpa1} \cite{MPA1}, \texttt{ms1}, \texttt{ms1b} \cite{MS1}, and \texttt{sly} \cite{SLy}. 
These models span a wide range of stiffnesses, and support a $1.93 \, M_{\odot}$ star, a conservative observational bound on the maximum NS mass \cite{Antoniadis}. 
Our candidate {\EOS}s are purely hadronic, except for \texttt{alf2} and \texttt{H4}, which contain color-flavor-locked quarks and hyperons, respectively. 

First off, we seek to build an uninformative \EOS~prior process that uniformly covers the entire range of physically viable {\EOS}s, from the stiffest possible (saturating the causality constraint) to the softest possible (marginally failing to support a $1.93 \, M_{\odot}$ NS).
Within this range, synthetic {\EOS}s~should vary as freely as possible subject to the physical constraints discussed in Sec.~\ref{sec:model}.
We therefore use the tabulated {\EOS}s to provide only weak, general guidance to the form of the synthetic {\EOS}s.
This is achieved by selecting a large white-noise variance $\sigma_n$.
Thus, synthetic {\EOS}s drawn from the prior can depart significantly from the mean of the tabulated {\EOS}s upon which they are conditioned (cf.~Eq.~\eqref{E}) and can exhibit different types of functional behavior (cf.~Eq.~\eqref{Cov}) while exploring a large swath of the pressure-density plane.
We refer to this fiducial \EOS~prior process as the \emph{model-agnostic prior}.

We also explore an alternate choice of prior.
Our second prior seeks to conform more closely to our tabulated {\EOS}s, producing much tighter \textit{a priori} bounds in the pressure-density plane.
In this case, hyperparameters are chosen to yield synthetic \EOS~behavior that reproduces the features of the input {\EOS}s: $\sigma_n$ is set to be quite small, and the synthetic EOSs consequently depart relatively little from the average of the candidate EOSs.
We call the resulting \EOS~prior process the \emph{model-informed prior}.

\begin{figure}
    \begin{minipage}{0.49\textwidth}
        \begin{center}
            prior processes \\
            \includegraphics[width=1.00\textwidth, clip=True, trim=0.70cm 0.55cm 0.60cm 0.20cm]{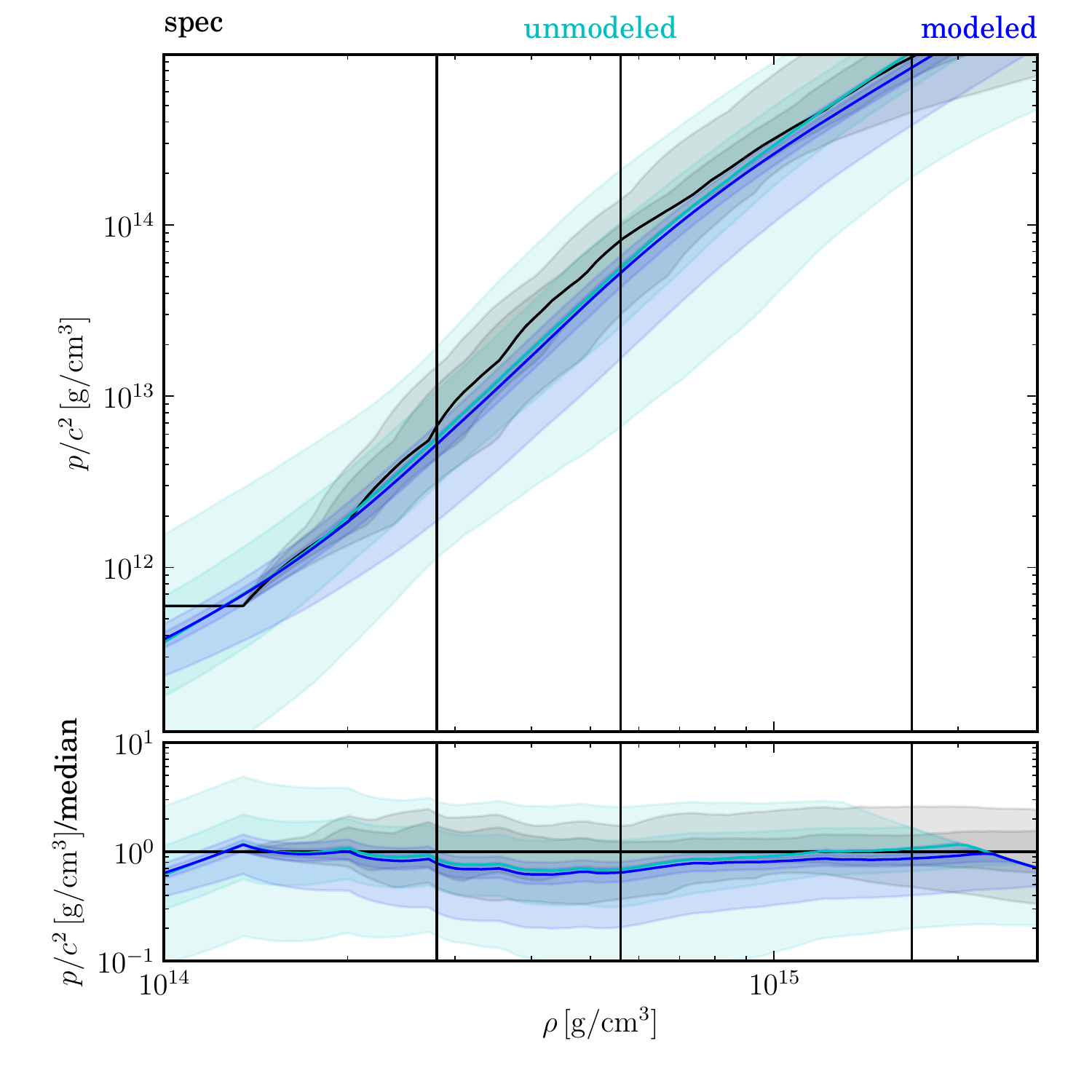}
        \end{center}
    \end{minipage}
    \begin{minipage}{0.49\textwidth}
        \begin{center}
            posterior processes \\
            \includegraphics[width=1.00\textwidth, clip=True, trim=0.70cm 0.55cm 0.60cm 0.20cm]{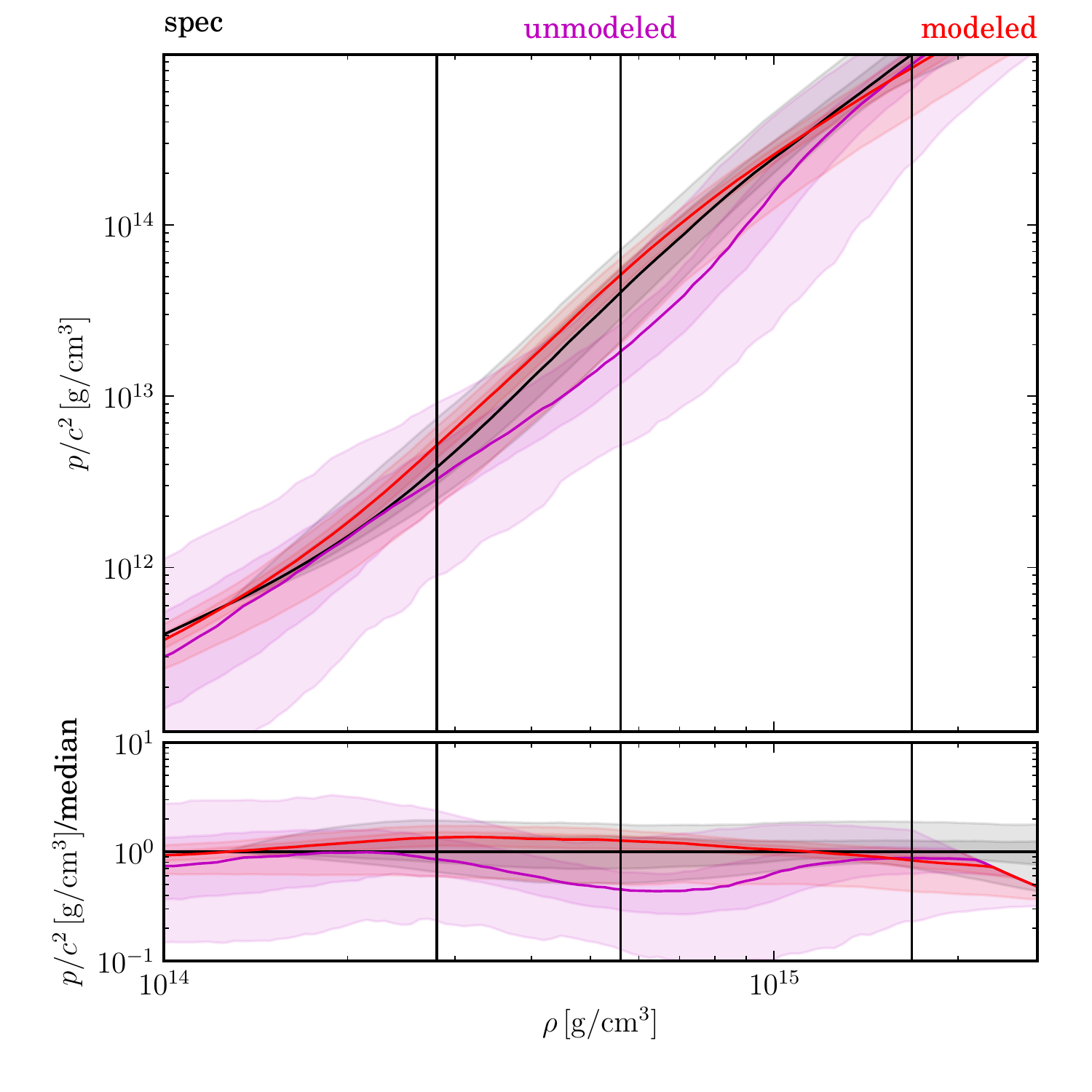}
        \end{center}
    \end{minipage}
    \caption{
        A comparison for GW170817 between the spectral prior from Ref.~\cite{LVC_eos} (\textit{black}) and our model-agnostic (\textit{light blue}) and model-informed (\textit{dark blue}) priors (\textit{left}), as well as between the spectral posteriors (\textit{black}) and our model-agnostic (\textit{light red}) and model-informed (\textit{dark red}) posteriors (\textit{right}).
       Solid lines show the medians and shaded regions correspond to 50\% and 90\% credible regions.
        We see that our model-agnostic posterior favors even softer {\EOS}s than the spectral approach, whereas the model-informed prior favors slightly stiffer {\EOS}s than the spectral approach.
        However, the associated uncertainties suggest all results are consistent.
    }
    \label{fig:prior_spec}
\end{figure}

Our GP priors are shown in Fig.~\ref{fig:gpr draw} along with our set of tabulated {\EOS}s.
We see that the model-agnostic prior supports a much broader range of possible {\EOS}s, while remaining centered---like the model-informed prior---on our representative set of candidate models.
We also compare our prior processes to the spectral method's \cite{LVC_eos} in Fig.~\ref{fig:prior_spec}.
One observes that the spectral prior is centered on a similar region of the pressure-density plane as both our model-agnostic and model-informed priors, but that our model-agnostic prior covers a somewhat broader region.
Because our synthetic {\EOS}s are not restricted to a single parametric form, they also encapsulate a wider range of functional behaviors than the spectral {\EOS}s, though this may not be apparent from the two-dimensional projection represented in the figures.

We first evaluate the non-parametric method's performance by analyzing simulated GW signals with both the model-agnostic and model-informed GP prior processes~(Sec.~\ref{sec:injections}).
We then place constraints on the true NS \EOS~with GW170817 data (Sec.~\ref{sec:gw170817}).
When performing the Monte Carlo integration, we draw component masses from a uniform prior consistent with the low-spin prior of Ref.~\cite{LVC_GW170817}, and we iterate until $O(10^3)$ effective samples have been returned, typically corresponding to $O(10^5)$ draws from the prior.
The two parallel inferences demonstrate how the non-parametric method can be tuned to assume different levels of \textit{a priori} confidence in the fiducial candidate {\EOS}s.
We remark, however, that neither of the priors constructed here should be construed as an optimal representation of the variability in the NS \EOS.
Our selection of hyperparameters and training {\EOS}s simply provides a qualitative illustration of possible designs for the GP prior process; the optimization of the \EOS~prior will be pursued in future work.
For example, the fiducial set of candidate {\EOS}s could be expanded to include more recently developed models with broader phenomenology, including first-order phase transitions or strange quark matter.
The hyperparameters could also be selected more judiciously to produce an \EOS~prior process that covers an even larger region of the pressure-density plane; this is particularly relevant for the model-informed prior, as excessively stiff or soft {\EOS}s tend to fall toward its edge.
Nonetheless, the GP priors introduced here are sufficiently broad to demonstrate the non-parametric method's efficacy, and we expect the model-agnostic prior in particular to resemble more carefully designed \EOS~prior processes.

\section{Simulated gravitational wave signals}\label{sec:injections}

We perform an injection campaign to demonstrate the non-parametric method's ability to infer a known EOS from GW data.
This proof of principle study is conducted with the intention of applying the same analysis to GW170817, and, accordingly, our simulated signals are designed to resemble that event.
We define three different sources with parameters similar to those inferred for GW170817, and inject the corresponding GW signals into real detector noise surrounding the event~\cite{LVC_GW170817,GW170817hoft}.
In particular, the injections have component masses and spins that are compatible with the inferred source properties for GW170817; the sources are placed at a comparable distance and orientation relative to the detectors; and the signals' network signal-to-noise ratios are approximately the same ($\text{SNR} \approx 35$) \cite{LVC_properties}.
For each of these sources, we further select one of three candidate {\EOS}s---\texttt{ms1}, \texttt{h4}, \texttt{sly}---ranging from relatively stiff to soft.
The choice of injected masses and \EOS~determines the tidal deformabilities in the simulated GW signal. 
The $M_{1,2}$ and $\Lambda_{1,2}$ values for each mock event, as well as the associated EOS, are listed in Table~\ref{tb:inj}.
Injections are performed with the \texttt{TaylorF2} waveform (see, e.g., Ref.~\cite{Buonanno2009}) and the $M_{1,2}$-$\Lambda_{1,2}$ marginal likelihoods are sampled with \texttt{lalinference}~\cite{Veitch} using the same.

\begin{figure}
    \begin{minipage}{0.49\textwidth}
        \begin{center}
            model-agnostic \\
            \includegraphics[width=1.00\textwidth, clip=True, trim=0.50cm 1.95cm 0.3cm 0.3cm]{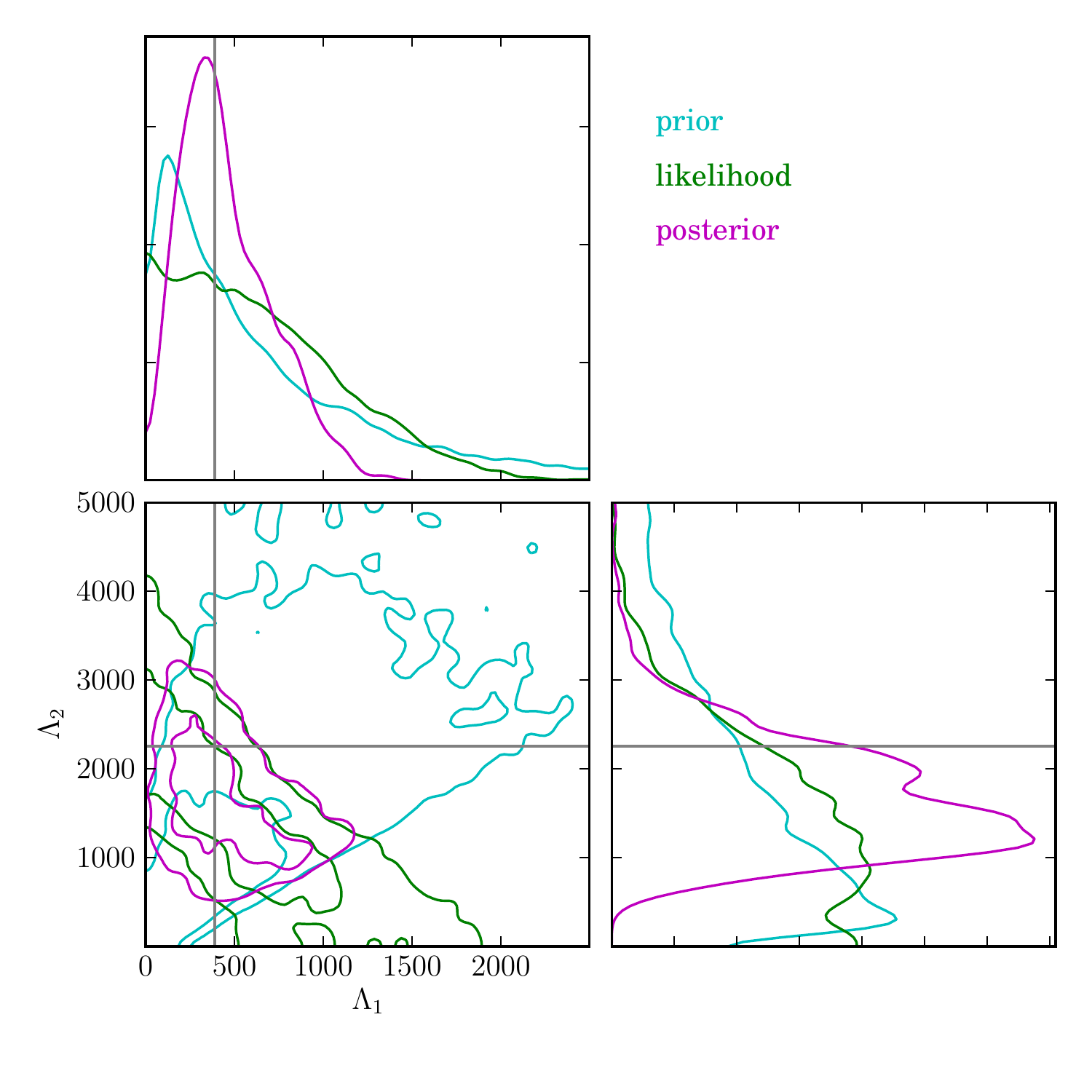}
            \includegraphics[width=1.00\textwidth, clip=True, trim=0.50cm 1.10cm 0.3cm 6.8cm]{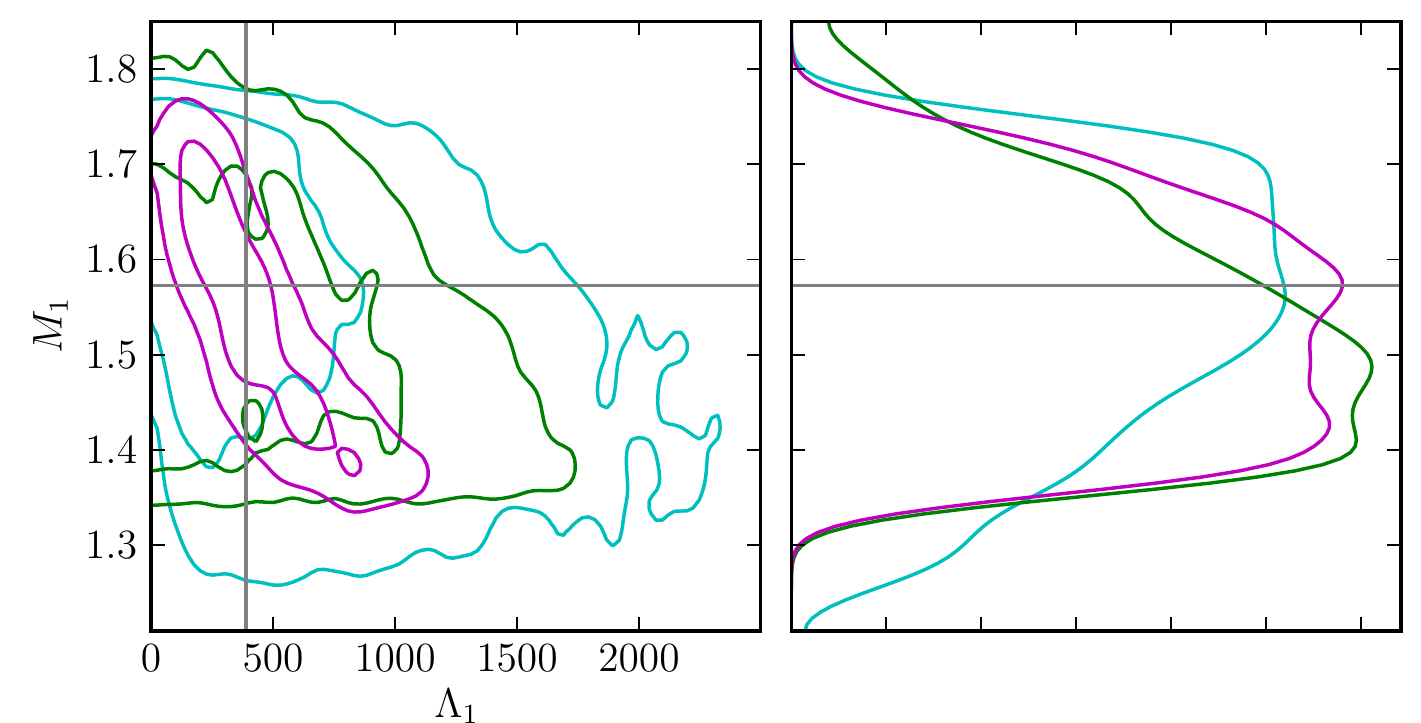}
        \end{center}
    \end{minipage}
    \begin{minipage}{0.49\textwidth}
        \begin{center}
            model-informed \\
            \includegraphics[width=1.00\textwidth, clip=True, trim=0.50cm 1.95cm 0.3cm 0.3cm]{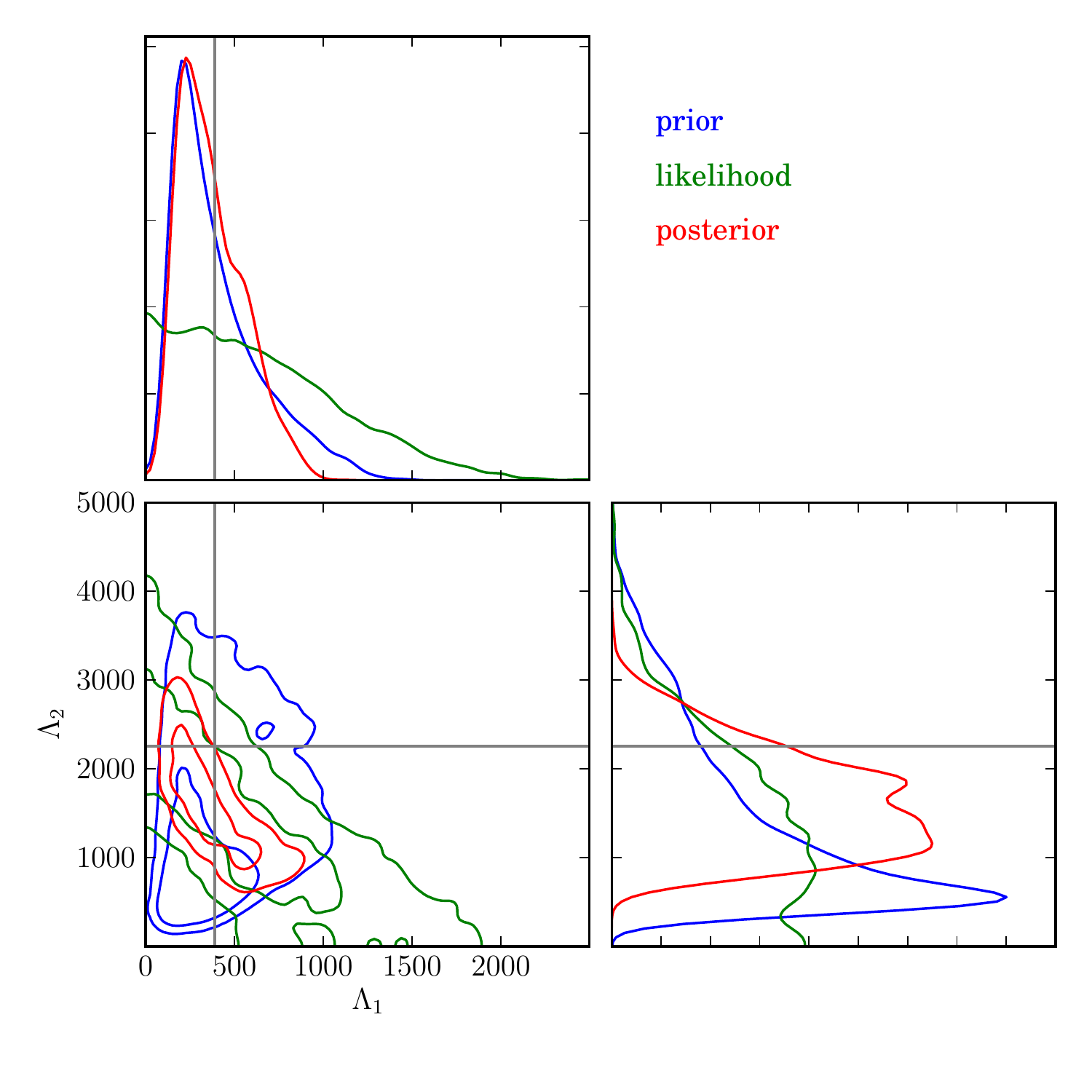}
            \includegraphics[width=1.00\textwidth, clip=True, trim=0.50cm 1.10cm 0.3cm 6.8cm]{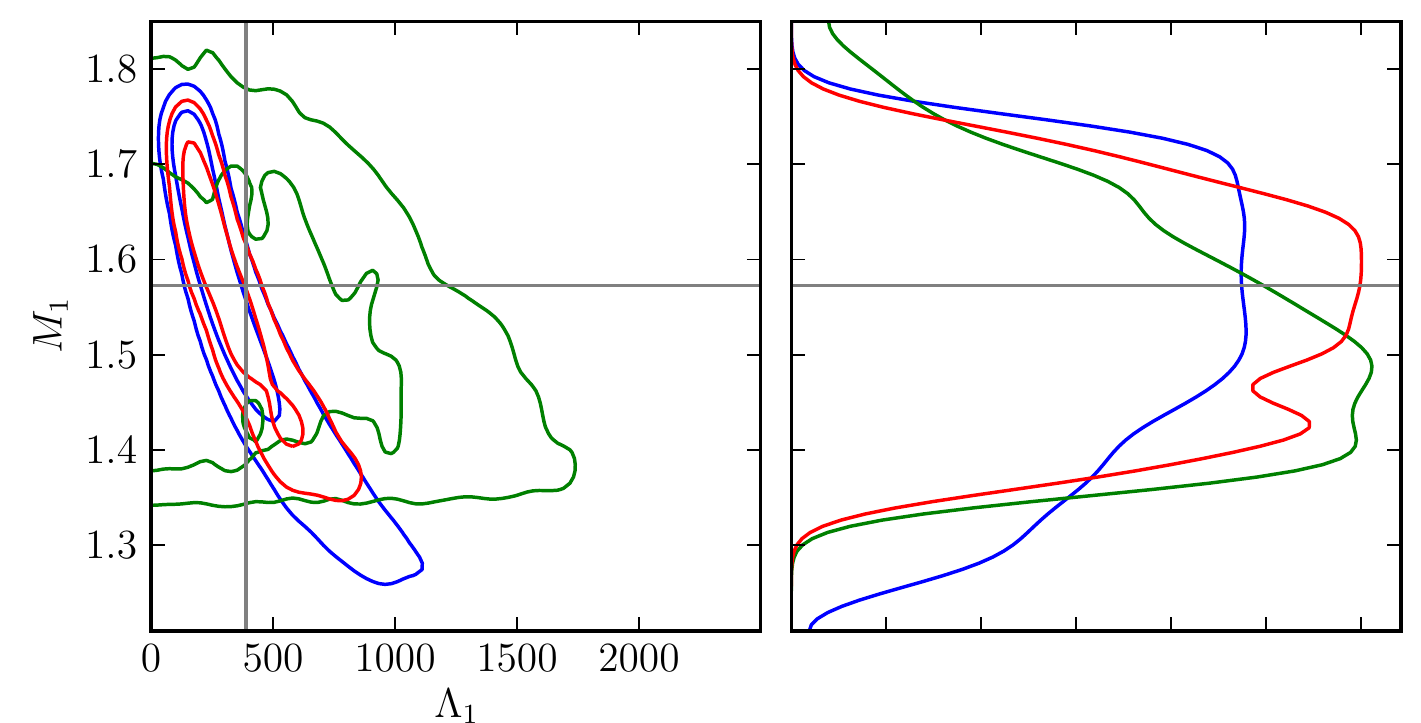}
        \end{center}
    \end{minipage}
    \caption{
        Prior (\textit{blue}), likelihood (\textit{green}), and posterior (\textit{red}) distributions for our model-agnostic (\textit{left}) and model-informed (\textit{right}) GP priors for a single simulated injection, event I with \texttt{h4} from Table~\ref{tb:inj}.
        Contours for the joint distributions correspond to 50\% and 90\% credible regions.
        Because of the broader support in the model-agnostic prior, the injected value (\textit{grey cross-hairs}) is comfortably contained in the associated posterior's support.
        We note that the model-informed prior appears to contain the injected value in some marginal distributions, but it actually lives near the edge of the allowed $M$-$\Lambda$ region.
        This accounts for the fact that the $\Lambda_{1,2}$ posterior barely contains the injection even though it appears to be contained in the $\Lambda_{1,2}$ prior. Since we perform inference with a non-trivial prior in a four-dimensional space, the resulting constraints may not be obvious from two-dimensional projections.
    }
    \label{fig:corner inj1 h4 prior post}
\end{figure}

\begin{figure}
    \begin{minipage}{0.49\textwidth}
        \begin{center}
            model-agnostic \\
            \includegraphics[width=1.00\textwidth, clip=True, trim=0.70cm 0.55cm 0.60cm 0.20cm]{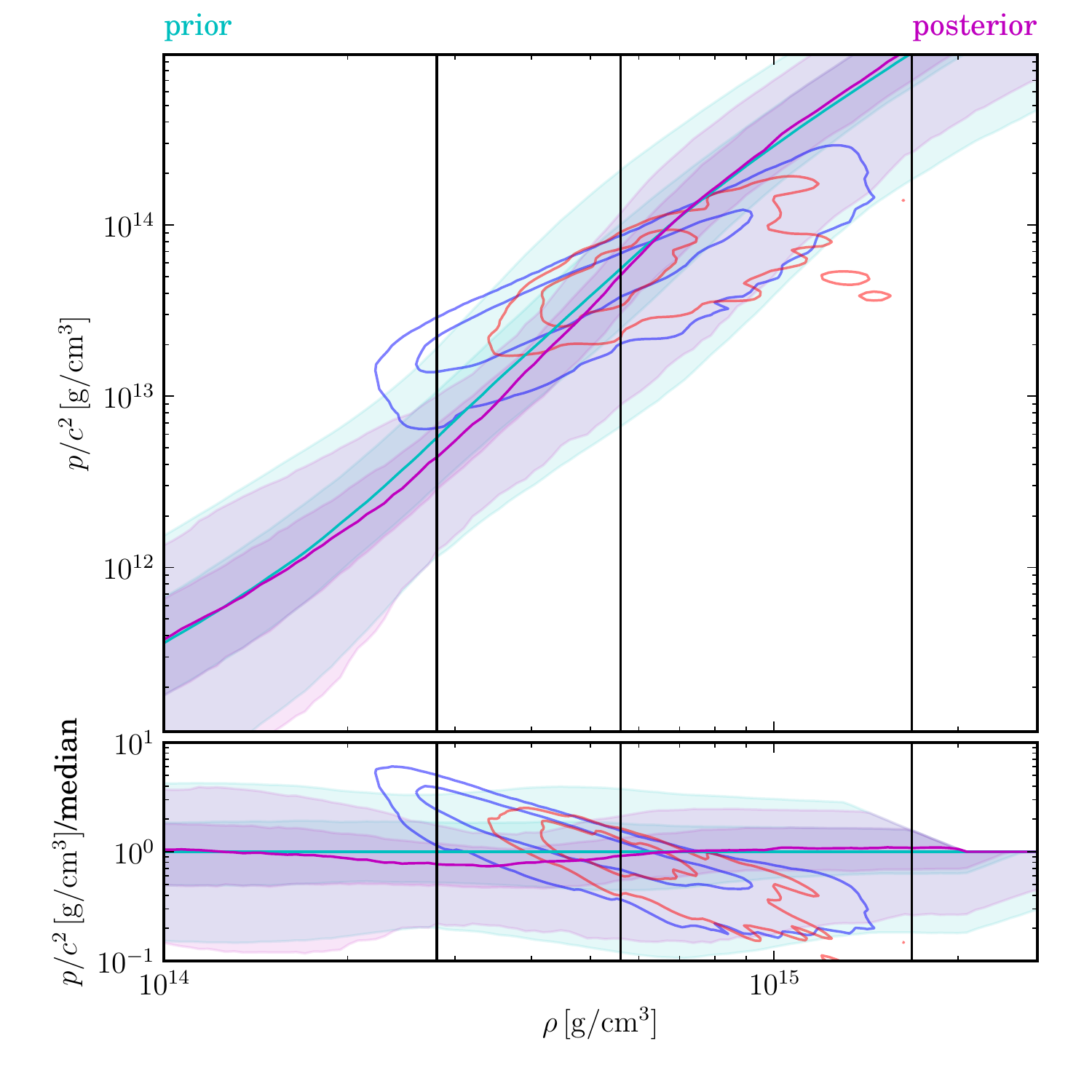}
        \end{center}
    \end{minipage}
    \begin{minipage}{0.49\textwidth}
        \begin{center}
            model-informed \\
            \includegraphics[width=1.00\textwidth, clip=True, trim=0.70cm 0.55cm 0.60cm 0.20cm]{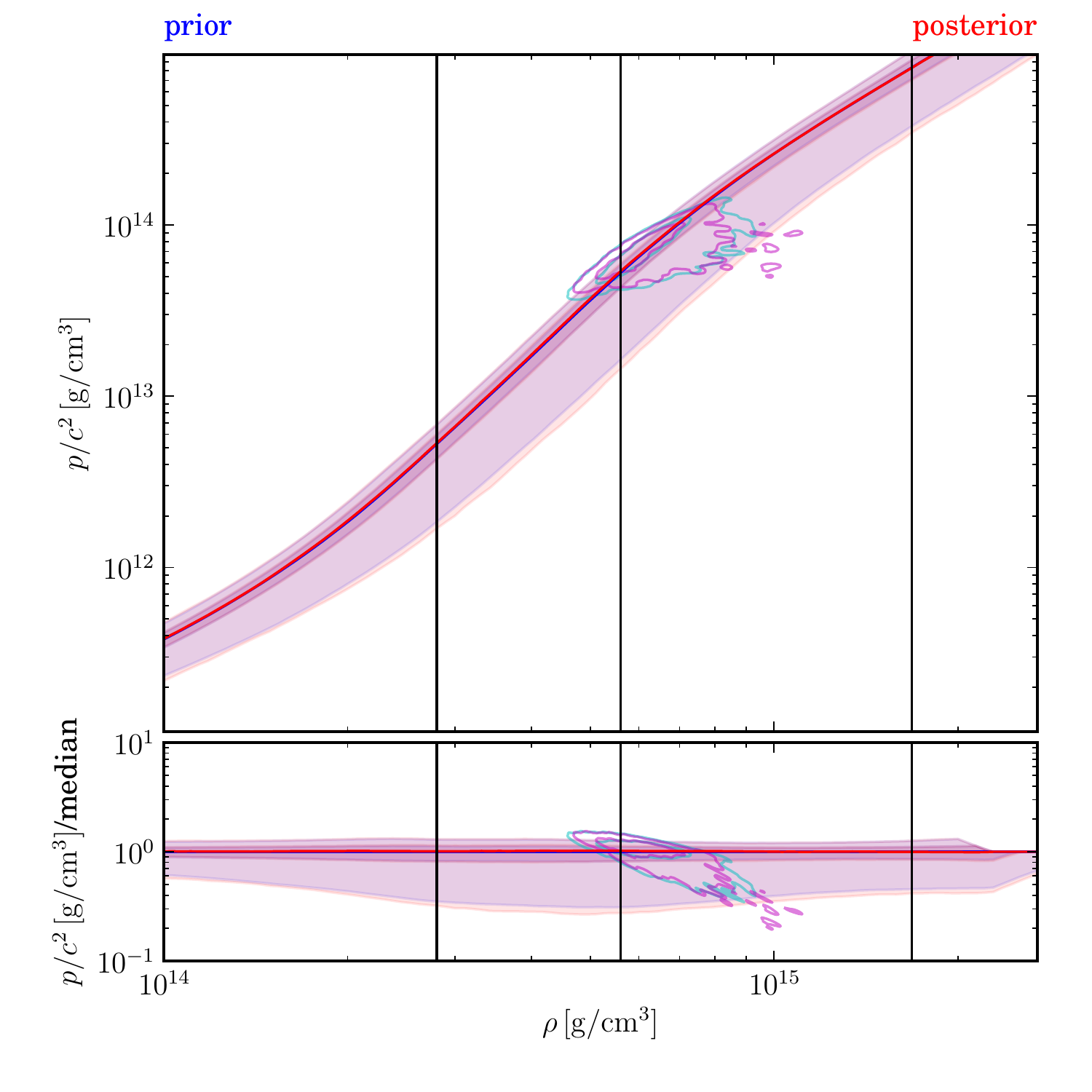}
        \end{center}
    \end{minipage}
    \caption{
        Processes corresponding to the injection in Fig.~\ref{fig:corner inj1 h4 prior post} (event I with \texttt{h4} from Table~\ref{tb:inj}), showing the prior (\textit{blue}) and posterior (\textit{red}) for our model-agnostic (\textit{left}) and model-informed (\textit{right}) priors.
        Solid lines show the medians and shaded regions correspond to 50\% and 90\% credible regions for the one-dimensional marginal distributions for $p(\rho)$.
        Contours correspond to the 50\% and 90\% credible regions for the central density and pressure of $M_1$ and vertical black bars denote $\rhonuc$, $2\rhonuc$, and $6\rhonuc$.
        We see stronger relative \textit{a posteriori} constraints with the model-agnostic prior, whereas the posterior obtained with the model-informed prior is nearly indistinguishable from the prior itself.
    }
    \label{fig:process inj1 h4 prior post}
\end{figure}

The posteriors obtained for one simulated event, injection I with \texttt{h4}, are shown by way of example in Figs.~\ref{fig:corner inj1 h4 prior post} and \ref{fig:process inj1 h4 prior post} along with the associated priors. 
Fig.~\ref{fig:corner inj1 h4 prior post} shows projections of the joint probability distribution over $M_{1,2}$ and $\Lambda_{1,2}$ into the $\Lambda_1$-$\Lambda_2$ and $\Lambda_1$-$M_1$ subspaces, as well as marginal distributions for $M_1$, $\Lambda_1$ and $\Lambda_2$. 
On the left, one can see that the model-agnostic prior supports a wide range of possible $M$-$\Lambda$ relations, and the inferred posterior comfortably includes the injected value.
This is in keeping with our expectations, since the model-agnostic prior is designed to contain all feasible {\EOS}s.
Nonetheless, the inference produced informative constraints on the tidal deformabilities, as the $90\%$ confidence region of the posterior is much reduced relative to the prior.

Fig.~\ref{fig:process inj1 h4 prior post} shows the corresponding \EOS~prior and posterior processes in the pressure-density plane.
In the lefthand panel, the broad model-agnostic prior is significantly narrowed by the inference, especially around $2\rho_{\mathrm{nuc}}$, producing a posterior process that favors an \EOS~somewhat softer than the prior's median.
The inference also yields information about the central density $\rho_c$ and pressure $p_c$ of each NS, since the $M$-$\rho_c$ relation for each synthetic EOS is known.
The overplotted contours indicate the posterior for the more massive component's central values.
Constraints on the \EOS~above $\rho_{c,1}$, the central density associated with $M_1$, arise exclusively through correlations within the synthetic {\EOS}s; the GW likelihood directly constrains the \EOS~only below $\rho_{c,1}$.

The righthand panels of Figs.~\ref{fig:corner inj1 h4 prior post} and \ref{fig:process inj1 h4 prior post} show our inference with the model-informed prior.
In terms of the joint $M_{1,2}$-$\Lambda_{1,2}$ probability distribution, the model-informed prior is much more restrictive, leading to a tighter posterior than in the model-agnostic case.
Specifically, the model-informed posterior has limited support for very soft {\EOS}s (small tidal deformabilities).
This is consistent with the fact that the model-informed prior is closely tied to the training set of candidate {\EOS}s, which includes models no softer than \texttt{sly} ($\Lambda_{1.4} \approx 290$), whereas the less constrained synthetic {\EOS}s from the model-agnostic prior explore softer regions of the pressure-density plane. 
Despite the model-informed posterior's smaller support, the injected parameters $M_{1,2}$ and $\Lambda_{1,2}$ are still recovered within $90\%$ confidence, although $\tilde{\Lambda}$ happens to lie marginally outside the $90\%$ confidence region for this event.

In the model-informed case, we note that the posterior in the pressure-density plane is nearly identical to the prior. 
This is because the model-informed prior is already quite narrow, and a single noisy GW event conveys only a limited amount of information.
We expect tighter constraints to be achievable with combined observations from multiple events \cite{Lackey,Agathos}.

The full set of injected signals is presented in Appendix~\ref{sec:complete inj}, and we summarize the results here, reporting our method's performance for each of the injections in Table~\ref{tb:inj}. 
We quote the injected $M_{1,2}$, $\Lambda_{1,2}$ and $\tilde{\Lambda}$ values along with one-dimensional 90\% credible regions around the maxima \textit{a posteriori}.
In general, we find that the component masses are always well recovered with both priors, while $\Lambda_{1,2}$ are typically well recovered with the model-agnostic prior, although the maxima \textit{a posteriori} are not always centered on the injected values.
This is due to a combination of noise fluctuations (the likelihood does not peak at the injected values) and the influence of our priors (the injected values lie near the edge of the prior).
Unsurprisingly, $\tilde{\Lambda}$ is generally less sensitive to the priors and is consequently recovered more robustly.
The GW data constrain $\Lambda_{1,2}$ primarily through $\tilde{\Lambda}$, so \textit{a posteriori} relative degrees of belief for combinations of $\Lambda_{1,2}$ that produce the same $\tilde{\Lambda}$ are mostly informed by the prior.

In contrast to the model-agnostic prior, $\Lambda_{1,2}$ are not always well recovered with the model-informed prior.
While several two-dimensional projections, such as the marginal $\Lambda_1$-$\Lambda_2$ distribution, appear to comfortably contain the injected signal, in reality it lies near the edge of the allowed $M$-$\Lambda$ relations in some cases (see, e.g., Fig.~\ref{fig:corner inj1 h4 prior post}).
Hence, the injection is found near the edge of the posterior because of the constraints applied in the full four-dimensional space.
This indicates that our model-informed prior could be too tight to reliably infer the true \EOS~if it is significantly softer or stiffer than the candidate EOSs included in our training set.
Our model-agnostic prior is more trustworthy in this regard, though it generally produces looser constraints because of its broader support.
What's more, the model-informed prior tends to more strongly influence the inferred $\Lambda_{1,2}$, resulting at times in injected values recovered outside the 90\% credible regions.
However, this is to be expected for injected {\EOS}s lying near or beyond the edge of our prior.
We only expect our confidence regions to have the correct coverage if the injected signals were drawn directly from our priors, which is not the case here. Although \texttt{ms1}, \texttt{h4} and \texttt{sly} are among the candidate EOSs upon which the GP was conditioned, they are not themselves synthetic EOSs drawn from the EOS prior process.

Since the recovered posteriors for the injections depend on the GP prior, we may quantify the degree to which the GW data prefer one prior or the other via Bayesian model selection.
We estimate the Bayes factor
\begin{equation}
    \Bayes = \frac{P(d|\{\mu^{(\alpha)}_{j^\ast}\}, \vec{\sigma}_\mathrm{un}, \mathcal{H})}{P(d|\{\mu^{(\alpha)}_{j^\ast}\}, \vec{\sigma}_\mathrm{mod}, \mathcal{H})}
\end{equation}
between our model-agnostic and model-informed priors for all events listed in Table~\ref{tb:inj}, calculating the evidences according to Eq.~\eqref{evidence}.
For most simulated signals, neither prior is strongly preferred by the data, with the exception of injection I with \texttt{ms1}.
As \texttt{ms1} is particularly stiff, it lies at the very edge of our model-informed prior's support, and therefore there is much more evidence in favor of the model-agnostic prior.
This appears to be more prevalent for particularly stiff {\EOS}s than for particularly soft ones (compare \Bayes~for \texttt{ms1} and \texttt{sly}), but we typically find that the GW data do not disfavor our model-informed prior at high confidence. One could imagine performing a similar model selection scheme on different GP priors conditioned on specific features of the NS EOS, to test, e.g., for the existence of a first-order phase transition in the NS EOS.

\begin{table}
    \begin{tabular}{lcccrrrcccccc}
    \hline \hline
   \multirow{2}{*}{ }  & \multicolumn{6}{c}{Injected} & \multicolumn{5}{c}{Recovered with model-agnostic (model-informed) prior} & \multirow{2}{*}{\Bayes} \\
    & \multicolumn{1}{c}{$m_1 \, [M_{\odot}]$}  & \multicolumn{1}{c}{$m_2 \, [M_{\odot}]$}  & \EOS & \multicolumn{1}{c}{$\Lambda_1$} & \multicolumn{1}{c}{$\Lambda_2$} & \multicolumn{1}{c}{$\tilde{\Lambda}$} & \multicolumn{1}{c}{$m_1 \, [M_{\odot}]$} & \multicolumn{1}{c}{$m_2 \, [M_{\odot}]$} & \multicolumn{1}{c}{$\Lambda_1$} & \multicolumn{1}{c}{$\Lambda_2$} & \multicolumn{1}{c}{$\tilde{\Lambda}$} & \\
    \hline 
 &                          &                       & \multirow{2}{*}{MS1} & \multirow{2}{*}{ 714} & \multirow{2}{*}{3320} & \multirow{2}{*}{1550} &   \injoneunmsonePOSTmone  &   \injoneunmsonePOSTmtwo  &   \injoneunmsonePOSTlone  &   \injoneunmsonePOSTltwo  &   \injoneunmsonePOSTlamtilde  & \multirow{2}{*}{\bayesinjonemsone} \\
&                          &                       &                      &                       &                       &                       & (\injonemodmsonePOSTmone) & (\injonemodmsonePOSTmtwo) & (\injonemodmsonePOSTlone) & (\injonemodmsonePOSTltwo) & (\injonemodmsonePOSTlamtilde) & \\
    \cline{4-13}
  \multirow{2}{*}{(I)} & \multirow{2}{*}{1.57} & \multirow{2}{*}{1.19} & \multirow{2}{*}{ H4} & \multirow{2}{*}{ 390} & \multirow{2}{*}{2250} & \multirow{2}{*}{ 983} &   \injoneunhfourPOSTmone  &   \injoneunhfourPOSTmtwo  &   \injoneunhfourPOSTlone  &   \injoneunhfourPOSTltwo  &   \injoneunhfourPOSTlamtilde  & \multirow{2}{*}{\bayesinjonehfour} \\
 &                         &                       &                      &                       &                       &                       & (\injonemodhfourPOSTmone) & (\injonemodhfourPOSTmtwo) & (\injonemodhfourPOSTlone) & (\injonemodhfourPOSTltwo) & (\injonemodhfourPOSTlamtilde) & \\
    \cline{4-13}
 &                         &                       & \multirow{2}{*}{SLy} & \multirow{2}{*}{ 130} & \multirow{2}{*}{ 834} & \multirow{2}{*}{ 354} &     \injoneunslyPOSTmone  &     \injoneunslyPOSTmtwo  &     \injoneunslyPOSTlone  &     \injoneunslyPOSTltwo  &     \injoneunslyPOSTlamtilde  & \multirow{2}{*}{\bayesinjonesly} \\
  &                        &                       &                      &                       &                       &                       &   (\injonemodslyPOSTmone) &   (\injonemodslyPOSTmtwo) &   (\injonemodslyPOSTlone) &   (\injonemodslyPOSTltwo) &   (\injonemodslyPOSTlamtilde) & \\
    \hline 
  &                        &                       & \multirow{2}{*}{MS1} & \multirow{2}{*}{1580} & \multirow{2}{*}{1630} & \multirow{2}{*}{1600} &   \injfourunmsonePOSTmone  &   \injfourunmsonePOSTmtwo  &   \injfourunmsonePOSTlone  &   \injfourunmsonePOSTltwo  &   \injfourunmsonePOSTlamtilde  & \multirow{2}{*}{\bayesinjfourmsone} \\
 &                         &                       &                      &                       &                       &                       & (\injfourmodmsonePOSTmone) & (\injfourmodmsonePOSTmtwo) & (\injfourmodmsonePOSTlone) & (\injfourmodmsonePOSTltwo) & (\injfourmodmsonePOSTlamtilde) & \\
    \cline{4-13}
  \multirow{2}{*}{(II)} &  \multirow{2}{*}{1.37} & \multirow{2}{*}{1.36} & \multirow{2}{*}{ H4} & \multirow{2}{*}{1000} & \multirow{2}{*}{1034} & \multirow{2}{*}{1020} &   \injfourunhfourPOSTmone  &   \injfourunhfourPOSTmtwo  &   \injfourunhfourPOSTlone  &   \injfourunhfourPOSTltwo  &   \injfourunhfourPOSTlamtilde  & \multirow{2}{*}{\bayesinjfourhfour} \\
     &                     &                       &                      &                       &                       &                       & (\injfourmodhfourPOSTmone) & (\injfourmodhfourPOSTmtwo) & (\injfourmodhfourPOSTlone) & (\injfourmodhfourPOSTltwo) & (\injfourmodhfourPOSTlamtilde) & \\
    \cline{4-13}
      &                    &                       & \multirow{2}{*}{SLy} & \multirow{2}{*}{ 349} & \multirow{2}{*}{ 362} & \multirow{2}{*}{ 355} &     \injfourunslyPOSTmone  &     \injfourunslyPOSTmtwo  &     \injfourunslyPOSTlone  &     \injfourunslyPOSTltwo  &     \injfourunslyPOSTlamtilde  & \multirow{2}{*}{\bayesinjfoursly} \\
       &                   &                       &                      &                       &                       &                       &   (\injfourmodslyPOSTmone) &   (\injfourmodslyPOSTmtwo) &   (\injfourmodslyPOSTlone) &   (\injfourmodslyPOSTltwo) &   (\injfourmodslyPOSTlamtilde) & \\
    \hline 
  &                        &                       & \multirow{2}{*}{MS1} & \multirow{2}{*}{1270} & \multirow{2}{*}{1850} & \multirow{2}{*}{1530} &   \injfiveunmsonePOSTmone  &   \injfiveunmsonePOSTmtwo  &   \injfiveunmsonePOSTlone  &   \injfiveunmsonePOSTltwo  &   \injfiveunmsonePOSTlamtilde  & \multirow{2}{*}{\bayesinjfivemsone} \\
   &                       &                       &                      &                       &                       &                       & (\injfivemodmsonePOSTmone) & (\injfivemodmsonePOSTmtwo) & (\injfivemodmsonePOSTlone) & (\injfivemodmsonePOSTltwo) & (\injfivemodmsonePOSTlamtilde) & \\
    \cline{4-13}
   \multirow{2}{*}{(III)} & \multirow{2}{*}{1.42} & \multirow{2}{*}{1.33} & \multirow{2}{*}{ H4} & \multirow{2}{*}{ 782} & \multirow{2}{*}{1190} & \multirow{2}{*}{ 968} &   \injfiveunhfourPOSTmone  &   \injfiveunhfourPOSTmtwo  &   \injfiveunhfourPOSTlone  &   \injfiveunhfourPOSTltwo  &   \injfiveunhfourPOSTlamtilde  & \multirow{2}{*}{\bayesinjfivehfour} \\
    &                      &                       &                      &                       &                       &                       & (\injfivemodhfourPOSTmone) & (\injfivemodhfourPOSTmtwo) & (\injfivemodhfourPOSTlone) & (\injfivemodhfourPOSTltwo) & (\injfivemodhfourPOSTlamtilde) & \\
    \cline{4-13}
     &                     &                       & \multirow{2}{*}{SLy} & \multirow{2}{*}{ 269} & \multirow{2}{*}{ 420} & \multirow{2}{*}{ 338} &     \injfiveunslyPOSTmone  &     \injfiveunslyPOSTmtwo  &     \injfiveunslyPOSTlone  &     \injfiveunslyPOSTltwo  &     \injfiveunslyPOSTlamtilde  & \multirow{2}{*}{\bayesinjfivesly} \\
      &                    &                       &                      &                       &                       &                       &   (\injfivemodslyPOSTmone) &   (\injfivemodslyPOSTmtwo) &   (\injfivemodslyPOSTlone) &   (\injfivemodslyPOSTltwo) &   (\injfivemodslyPOSTlamtilde) & \\ 
    \hline \hline
                                                                                                &   \multicolumn{6}{c}{\multirow{2}{*}{GW170817}} &   \gwphenomunloPOSTmone  &   \gwphenomunloPOSTmtwo  &   \gwphenomunloPOSTlone  &   \gwphenomunloPOSTltwo  &   \gwphenomunloPOSTlamtilde  & \multirow{2}{*}{\bayesgwphenomlo} \\ 
     &                     &                       &                      &                       &                       &                       & (\gwphenommodloPOSTmone) & (\gwphenommodloPOSTmtwo) & (\gwphenommodloPOSTlone) & (\gwphenommodloPOSTltwo) & (\gwphenommodloPOSTlamtilde) & \\
    \hline \hline
    \end{tabular}
    \caption{
        Injected parameters used to test \EOS~recovery and the associated one-dimensional posterior credible regions.
        Three mass ratios and three injected equations of state are considered.
        We report the 90\% credible regions for the model-agnostic (model-informed) prior along with the maximum \textit{a posteriori} for each parameter separately.
        Additionally, we report the Bayes factor between our priors (\Bayes), generally showing no strong preference for either prior with a single event.
        Error estimates correspond to $1$-$\sigma$ uncertainty for Bayes factors.
        Analogous results for GW170817 are shown as well, with similar conclusions.
    }
    \label{tb:inj}
\end{table}

\section{GW170817}\label{sec:gw170817}

\begin{figure}
    \begin{minipage}{0.49\textwidth}
        \begin{center}
            model-agnostic \\
            \includegraphics[width=1.00\textwidth, clip=True, trim=0.50cm 1.95cm 0.3cm 0.3cm]{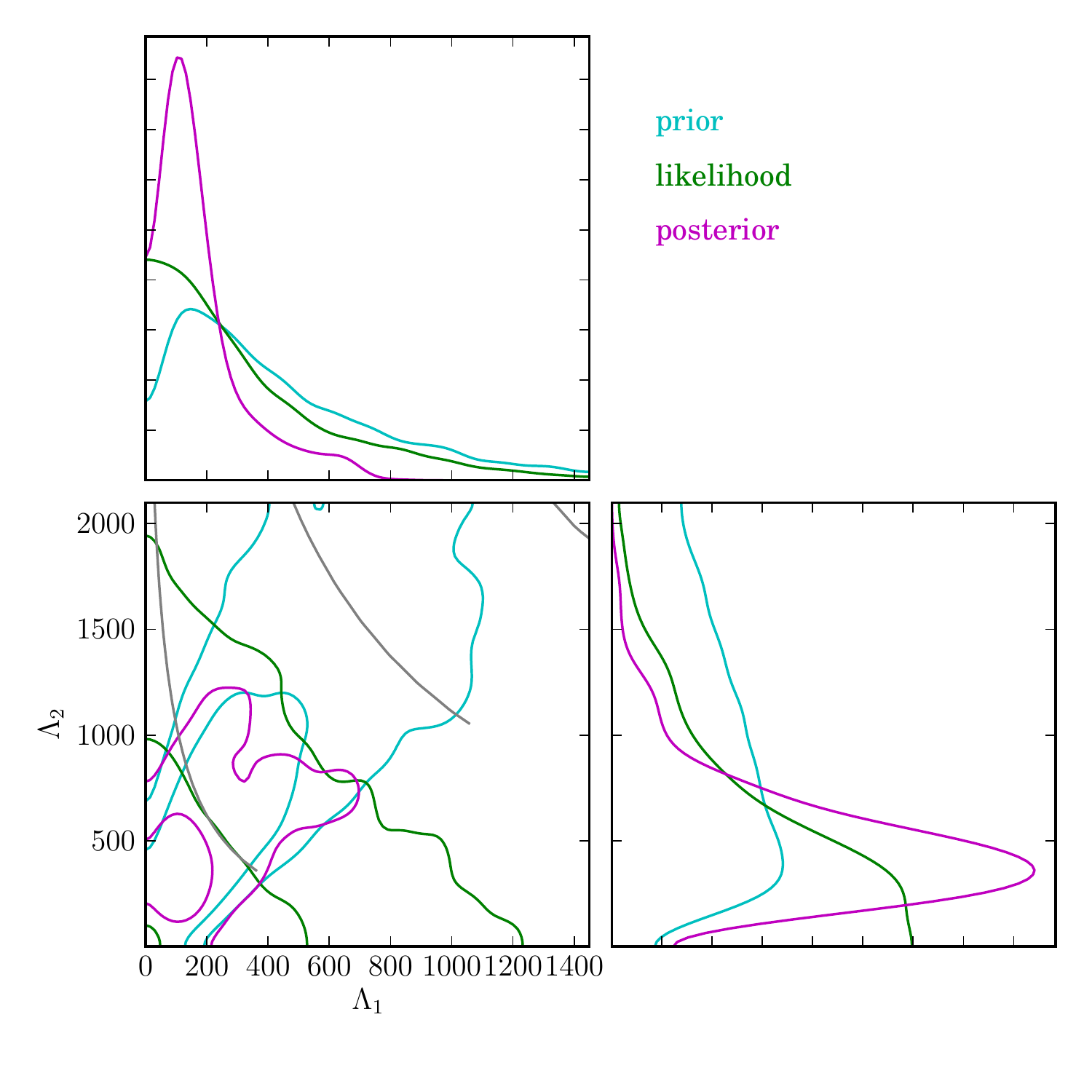}
            \includegraphics[width=1.00\textwidth, clip=True, trim=0.50cm 1.10cm 0.3cm 6.8cm]{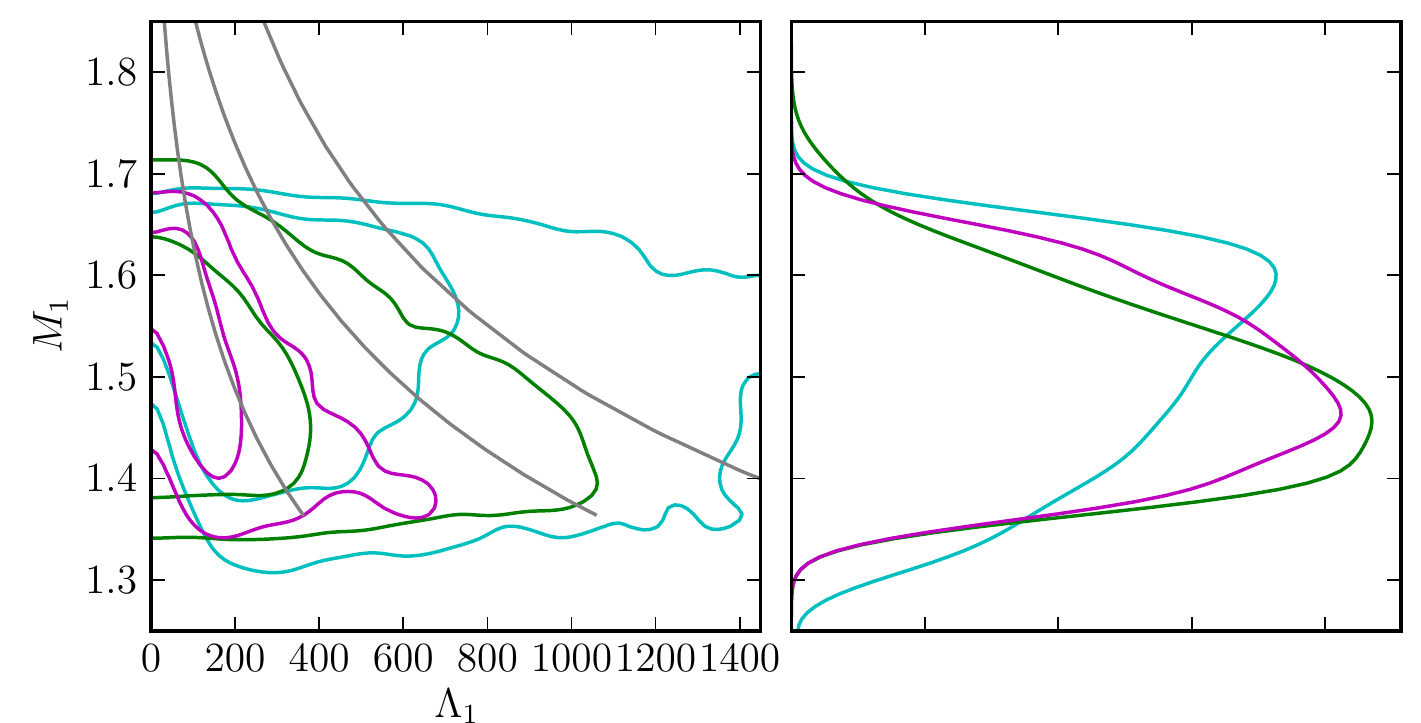}
        \end{center}
    \end{minipage}
    \begin{minipage}{0.49\textwidth}
        \begin{center}
            model-informed \\
            \includegraphics[width=1.00\textwidth, clip=True, trim=0.50cm 1.95cm 0.3cm 0.3cm]{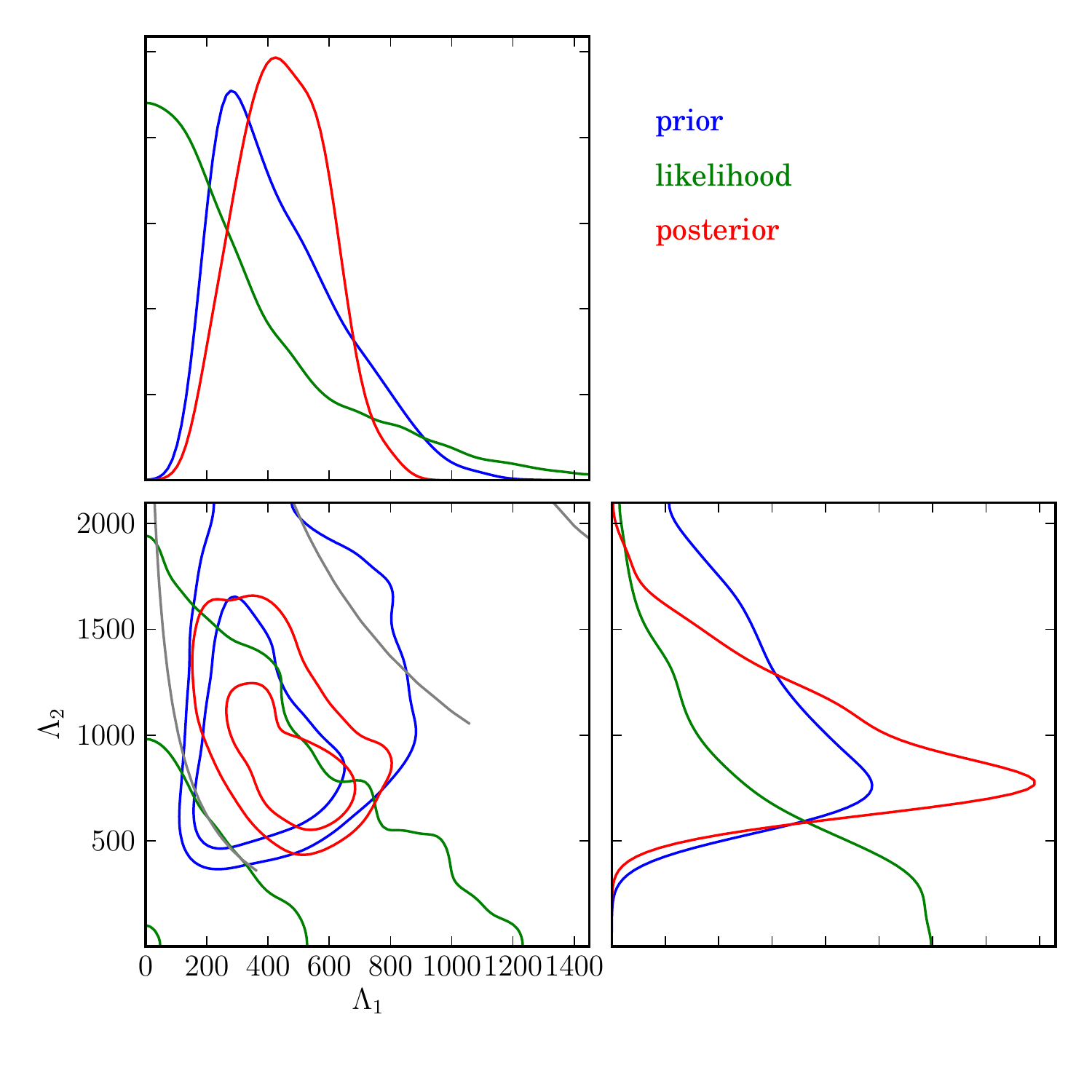}
            \includegraphics[width=1.00\textwidth, clip=True, trim=0.50cm 1.10cm 0.3cm 6.8cm]{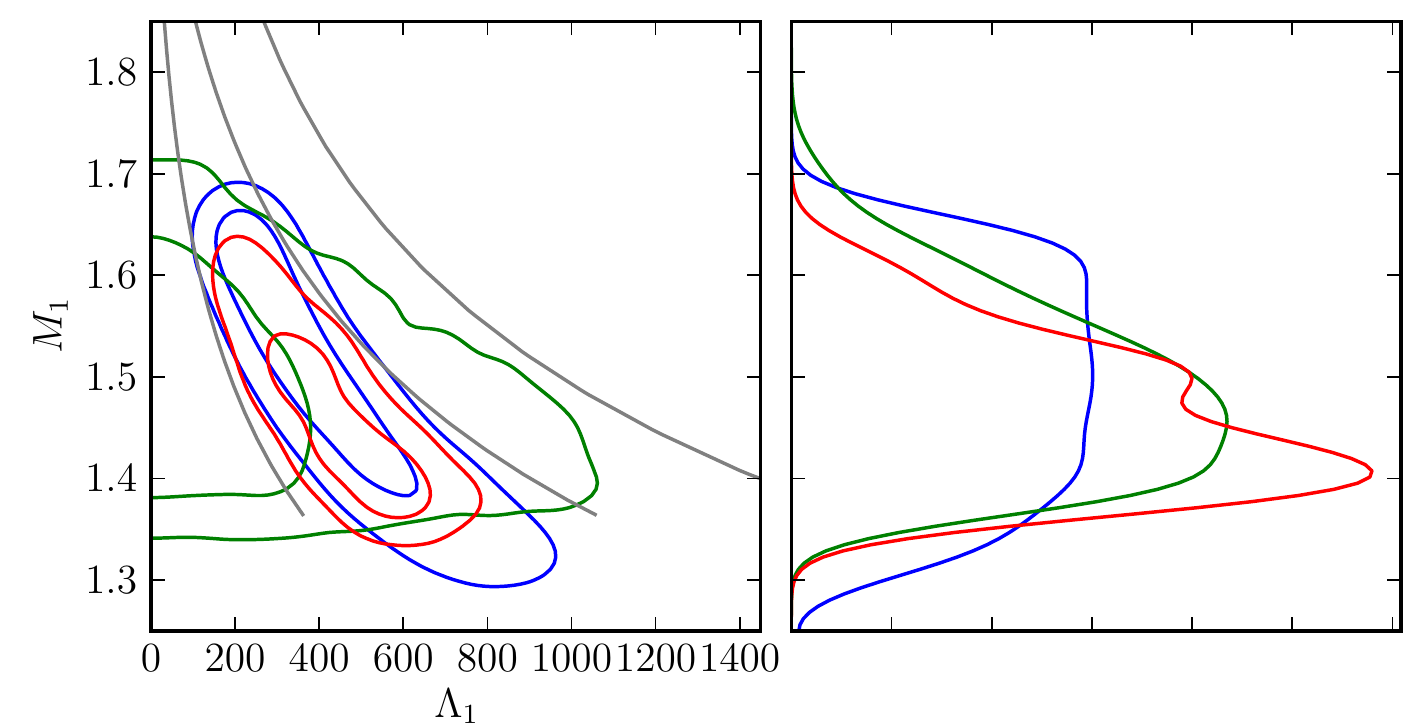}
        \end{center}
    \end{minipage}
    \caption{
        Marginal distributions for GW170817, analogous to Fig.~\ref{fig:corner inj1 h4 prior post}.
        The priors (\textit{blue}), likelihood (green), and posteriors (\textit{red}) are shown for both the model-agnostic (\textit{left}) and model-informed (\textit{right}) priors.
        Reference curves (\textit{grey}) are shown for (from left to right) \texttt{sly}, \texttt{h4}, and \texttt{ms1}.
        Contours in the joint distributions correspond to the 50\% and 90\% credible regions, and we see that the model-agnostic prior's \textit{a posteriori} constraints are similar to those reported in Ref.~\cite{LVC_eos}, whereas the model-informed prior favors significantly larger $\Lambda_{1,2}$ but similar masses.
    }
    \label{fig:corner gw170817 prior post}
\end{figure}

\begin{figure}
    \begin{minipage}{0.49\textwidth}
        \begin{center}
            model-agnostic \\
            \includegraphics[width=1.00\textwidth, clip=True, trim=0.70cm 0.55cm 0.60cm 0.20cm]{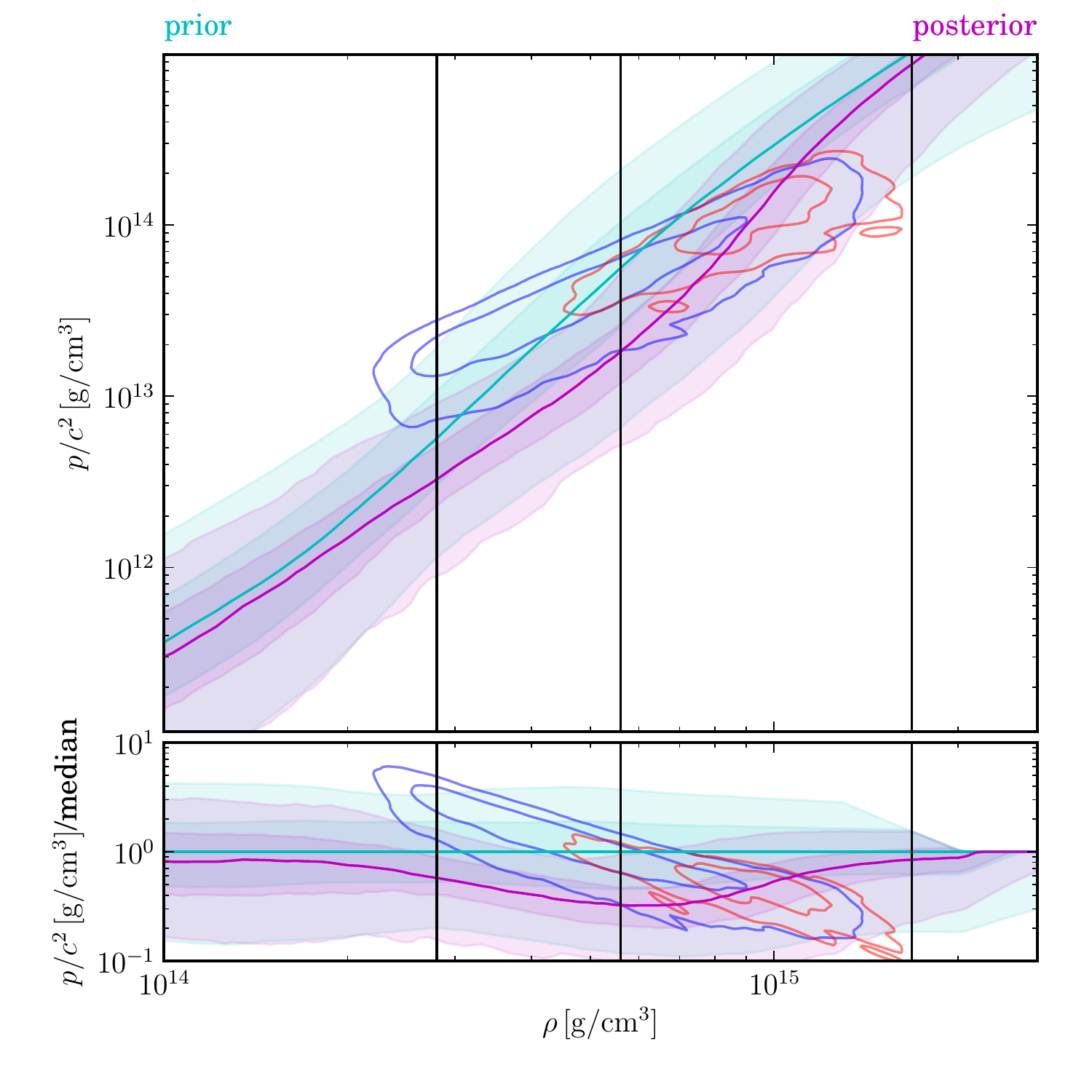}
        \end{center}
    \end{minipage}
    \begin{minipage}{0.49\textwidth}
        \begin{center}
            model-informed \\
            \includegraphics[width=1.00\textwidth, clip=True, trim=0.70cm 0.55cm 0.60cm 0.20cm]{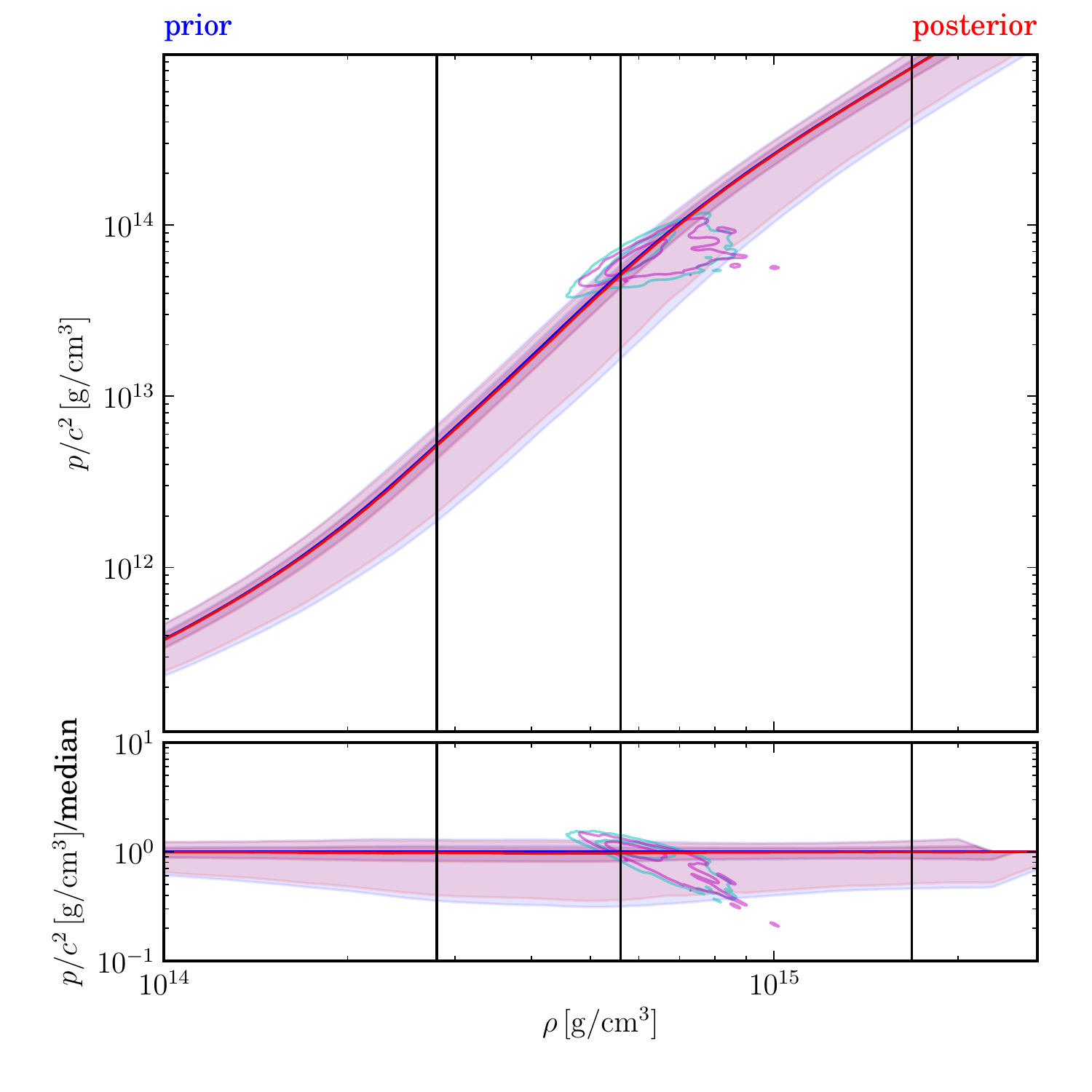}
        \end{center}
    \end{minipage}
    \caption{
        Processes for the model-agnostic (\textit{left}) and model-informed (\textit{right}) priors with GW170817, both \textit{a posteriori} (\textit{red}) and \textit{a priori} (\textit{blue}).
        We find only marginally tightened posterior constraints with the model-informed prior, whereas the model-agnostic posterior clearly favors softer {\EOS}s.
        As in Fig.~\ref{fig:process inj1 h4 prior post}, solid lines show medians, shaded regions represent 50\% and 90\% credible regions for $p(\rho)$, contours represent the 50\% and 90\% credible regions for the central densities and pressures of $M_1$, and vertical bars denote $\rhonuc$, $2\rhonuc$, and $6\rhonuc$.
    }
    \label{fig:process gw170817 prior post}    
\end{figure}

Having tested our non-parametric inference on simulated GW events, we now apply it to real data from GW170817.
We repeat the analysis of the previous section using the same model-agnostic and model-informed priors. 
Our recovered $M_{1,2}$-$\Lambda_{1,2}$ posteriors and \EOS~posterior processes are shown in Figs.~\ref{fig:corner gw170817 prior post} and \ref{fig:process gw170817 prior post}, respectively.
One-dimensional credible regions for the component masses and tidal deformabilities are reported in Table~\ref{tb:inj}, which also presents the Bayes factor between the two priors.

With the model-agnostic prior, we find that GW170817 places significant \textit{a posteriori} constraints on the tidal deformabilities, favoring a limited region corresponding to relatively soft {\EOS}s.
Indeed, we infer a chirp deformability of $\tilde{\Lambda} = \gwphenomunloPOSTlamtilde$ (maximum \textit{a posteriori} and 90\% credible region).
Using $\Lambda(M;\mu)$ for every synthetic EOS $\mu^{(a)}(p)$, we find $\Lambda_{1.4}=\gwphenomunloPOSTlamonepointfour$.
This suggests some of the smallest tidal deformabilities of any study to date.
Our findings agree with the results of previous analyses, namely $\tilde{\Lambda} = 222^{+420}_{-138}$ \cite{De} and $\Lambda_{1.4} = 190^{+390}_{-120}$ \cite{LVC_eos}.
Moreover, the $90\%$ confidence region of our $\Lambda_1$-$\Lambda_2$ posterior, shown in Fig.~\ref{fig:corner gw170817 prior post}, is markedly similar to the corresponding region of the posterior obtained with the spectral method (cf.~Fig.~1 of Ref.~\cite{LVC_eos}).

From Fig.~\ref{fig:process gw170817 prior post}, we observe a preference for a slightly softer-than-average pressure-density relation at mid-range densities.
In particular, we obtain the $90\%$-credible regions $p(2\rhonuc)=\gwphenomunloPOSTpressureattworhonuc$ \pressureunits~and \linebreak \mbox{$p(6\rhonuc)=\gwphenomunloPOSTpressureatsixrhonuc$ \pressureunits}~on the pressure at twice and six times nuclear density. 
These values agree with the constraints $p(2\rhonuc)=3.5^{+2.7}_{-1.7}\times 10^{34}$ \pressureunits~and $p(6\rhonuc)=9.0^{+7.9}_{-2.6}\times 10^{35}$ \pressureunits~obtained in Ref.~\cite{LVC_eos}.
Furthermore, Fig.~\ref{fig:process gw170817 prior post} displays our inference for the central density $\rho_{c,1}$ and pressure $p_{c,1}$ of the more massive component of the binary; we find that the posterior peaks at $\rho_{c,1} \approx 3 \rho_{\mathrm{nuc}}$.
Since $M(\rho_c;\mu)$ is known for each synthetic EOS, the posterior process can also be used to make an inference for the maximum mass supported by the NS EOS.
We find that  $M_\mathrm{max}=\mmaxPOSTun$ $M_\odot$ with the model-agnostic prior.

Repeating the inference with the model-informed prior, we recover a smaller \textit{a posteriori} $90\%$ confidence region that has less support for soft {\EOS}s, as can be seen in the right-hand panel of Fig.~\ref{fig:corner gw170817 prior post}. 
Our inferred tidal deformabilities are thus somewhat larger: $\tilde{\Lambda} = \gwphenommodloPOSTlamtilde$ and $\Lambda_{1.4}=\gwphenommodloPOSTlamonepointfour$.
The upper bounds on the tidal deformabilities are similar to the results of the original analyses of GW170817 \cite{LVC_GW170817,LVC_properties}, but the lower bounds are much larger.
This is because the tabulated {\EOS}s used to construct the model-informed prior cannot produce $\Lambda_{1.4} \lesssim 290$ (the \texttt{sly} value), while the GW data appear to support deformabilities $\Lambda_{1.4} \lesssim 200$.
Electromagnetic observations of the kilonova associated with GW170817 have found similarly large lower bounds for $\tilde{\Lambda}$ \cite{Radice,Coughlin2018,RadiceDai}.
The numerical relativity simulations of Ref.~\cite{RadiceDai}, in particular, suggest $\tilde{\Lambda} \gtrsim 300$, in mild tension with our model-agnostic result but more compatible with our model-informed analysis.
Curiously, the $\Lambda_{1.4}$ 90\% confidence region for the model-agnostic prior includes a long tail to large values and contains the maximum \textit{a posteriori} estimate from the model-informed prior, although the 50\% confidence region is much tighter.

The \EOS~posterior process for the model-informed prior is plotted in the right-hand panel of Fig.~\ref{fig:process gw170817 prior post}.
The pressure constraints extracted from the posterior are $p(2\rhonuc)=\gwphenommodloPOSTpressureattworhonuc$ \pressureunits~and \linebreak \mbox{$p(6\rhonuc)=\gwphenommodloPOSTpressureatsixrhonuc$ \pressureunits}, consistent with Ref.~\cite{LVC_eos} (see Fig.~\ref{fig:prior_spec}).
The posterior for the central density $\rho_{c,1}$ peaks around $2\rho_{\text{nuc}}$, and the constraint on the maximum mass supported is tightened to $M_\mathrm{max}=\mmaxPOSTmod$ $M_\odot$.

Overall, our results for GW170817 are consistent with the literature~\cite{LVC_GW170817,LVC_properties,LVC_eos,De}, favoring relatively low $\tilde{\Lambda}$ and a correspondingly soft \EOS.
In fact, Fig.~\ref{fig:prior_spec} shows that our inferred model-agnostic posterior process for $\mu(p)$ favors a slightly softer {\EOS} than that found with the spectral method \cite{LVC_eos}, though the uncertainties are quite broad.
On the other hand, the model-informed posterior prefers slightly larger pressures than the spectral method's.
Nonetheless, compared to our priors---i.e.~$p_\mathrm{prior}(2\rhonuc)=\gwphenomunloPRIORpressureattworhonuc$ (\gwphenommodloPRIORpressureattworhonuc) \pressureunits~for the model-agnostic (model-informed) prior---we generally see smaller 90\% confidence regions and shifted modes in the pressure-density plane \textit{a posteriori}.

Although the differences between prior and posterior are more pronounced in the model-agnostic case, neither prior is strongly favored by the data.
We find $\Bayes = \bayesgwphenomlo$ (point estimate and 1-$\sigma$ uncertainty), implying that our inference of the tidal deformabilities and the \EOS~itself is relatively prior-dominated.
In other words, the data from GW170817 alone cannot strongly differentiate between a model that yields $p(2\rhonuc)=\gwphenomunloPOSTpressureattworhonuc\,\pressureunits$ and one that yields $p(2\rhonuc)=\gwphenommodloPOSTpressureattworhonuc\,\pressureunits$, suggesting that all such constraints should be interpreted with care.
We expect that combined data from multiple GW signals will be needed to discriminate between different models for the EOS.

\section{Conclusions}\label{sec:conclusion}

GW measurements of the macroscopic properties of NSs offer a promising means of deducing information about the nuclear microphysics that governs their internal structure.
The non-parametric framework for NS \EOS~inference developed here provides direct constraints on the pressure-density relation inside the star from the standard GW likelihood.
The GP formulation of the \EOS~can compactly and faithfully represent a more diverse set of possible {\EOS}s than parametric models, and since its systematic uncertainties are independent of those from previous analyses, non-parametric inference can be used to complement and corroborate existing approaches.
The full EOS posterior process produced by our approach allows pressures and densities within the NS to be inferred, derived properties like the maximum NS mass to be constrained, and  model selection between various \EOS~priors to be performed.

We carried out a study of simulated GW170817-like signals using real detector noise as a demonstration of our method's efficacy at recovering a known EOS.
The inference produced informative constraints on the tidal deformabilities and the pressure-density relation, in the sense that the 90\% confidence region of the posterior was reduced.
Overall, we find that injections with stiffer EOSs favor stiffer EOSs \textit{a posteriori}, and softer injected EOSs favor softer posteriors, as expected.
However, the constraints depend somewhat strongly on the prior assumed, and the data typically do not favor the model-agnostic prior over the model-informed one, or vice-versa.

Using the same model-informed and model-agnostic priors, we analyzed GW170817 to infer tidal deformabilities, pressures and the maximum mass supported by the EOS. We summarize the results below and discuss their significance.

\emph{Tidal deformabilities.} Analyzing GW170817 with the model-agnostic prior, we infer a canonical deformability of $\Lambda_{1.4} = 160^{+448}_{-113}$ and a chirp deformability of $\tilde{\Lambda} = 210^{+383}_{-113}$, in comfortable agreement with previous analyses \cite{De,LVC_eos}.
The maximum \textit{a posteriori} values we obtain for the canonical and chirp deformabilities are the smallest claimed to date, but come with broad uncertainties.
Such small tidal deformabilities favor an EOS that resembles the softest nuclear-theoretic models available.
Repeating the analysis with the model-informed prior, the inferred tidal deformabilities are $\Lambda_{1.4} = 556^{+163}_{-172}$ and $\tilde{\Lambda} = 631^{+164}_{-122}$.
The 90\% confidence regions from both priors overlap, although the maxima \textit{a posteriori} are notably different.

\emph{Pressures.} The posterior process places direct constraints on the EOS in the pressure-density plane, enabling us to extract the constraints  $p(2\rhonuc)=\gwphenomunloPOSTpressureattworhonuc$ \pressureunits~and $p(6\rhonuc)=\gwphenomunloPOSTpressureatsixrhonuc$ \pressureunits~on the pressure at twice and six times the nuclear saturation density with the model-agnostic prior.
These figures are consistent with the pressure constraints calculated with the spectral method.
Qualitatively, we observe a similar preference for lower-than-average pressures at mid-range densities as was noted in Ref.~\cite{LVC_eos}.
The pressures inferred with the model-informed prior are $p(2\rho_{\text{nuc}}) = \gwphenommodloPOSTpressureattworhonuc$ \pressureunits~and $p(6\rho_{\text{nuc}}) = \gwphenommodloPOSTpressureatsixrhonuc$ \pressureunits, favoring  a slightly stiffer {\EOS}.
Nonetheless, the results of the inference are still consistent with those of previous analyses \cite{LVC_eos} within their uncertainties.

\emph{Maximum mass.} The EOS posterior process gives us access to maximum-mass information through correlations between the high densities relevant for $M_{\text{max}}$ and the mid-range densities probed by GW170817.
With the model-agnostic prior, we constrain the maximum NS mass to be $M_\mathrm{max}=\mmaxPOSTun\, M_\odot$, disfavoring especially stiff candidate EOSs that produce $M_{\text{max}} \gtrsim 2.5 \, M_{\odot}$.
Ours is the first reported constraint derived exclusively from GW data.
This is compatible with multimessenger studies of GW170817 via EOS model selection \cite{Margalit} and approximate universal relations \cite{Rezzolla2018}, numerical relativity \cite{Shibata}, and general-relativistic magnetohydrodynamic simulations \cite{Ruiz}, which account for the observed spectrum of the electromagnetic counterpart, the lifetime of the merger remnant, and the GW-inferred source properties. Taken together, the multimessenger studies suggest $2.15  \, M_{\odot} \lesssim M_{\text{max}} \lesssim 2.28 \, M_{\odot}$.
The constraint on the maximum NS mass is tightened to $M_{\text{max}} = 2.04^{+0.22}_{-0.002} \,  M_{\odot}$ with the model-informed prior.

\emph{Bayes factor.} Despite producing noticeably dissimilar \textit{a posteriori} credible regions for the tidal deformabilities and the pressure-density relation, the Bayes factors between the model-agnostic and model-informed priors are $O(1)$ for virtually all of the simulated events considered.
For GW170817, we find $\Bayes=\bayesgwphenomlo$ (point estimate and 1-$\sigma$ uncertainty).
This implies that data from a single event are not sufficiently informative to strongly favor either model.
We conclude that it is likely that all \EOS~constraints derived from GW170817 alone are relatively prior-dominated.
Thus, the \EOS~priors must be carefully motivated on physical grounds and associated constraints interpreted scrupulously.

Given the prior-dependence of the results, revisiting the choice of candidate EOSs used to condition the GP priors may be worthwhile.
One could expand the training set to include candidate EOSs softer than \texttt{sly} and stiffer than \texttt{ms1}.
Since early indications are that the NS EOS is relatively soft, extending the prior at its soft edge is particularly important for obtaining reliable model-informed inferences from GW data.
Adding candidate EOSs with, e.g., first order phase transitions or strange quark matter could also make the training set more fully representative of the theoretical possibilities for the EOS.
Despite already covering a large area of the pressure-density plane, the model-agnostic prior could also benefit from conditioning on an expanded and more representative set of input EOSs, as it would assimilate an even greater degree of variation in the functional behavior of the EOS.
Coupled with improved hyperparameter selection via automated marginalization over many possible settings (cf.~Eq.~\eqref{mixture}), it may be possible to deliver tighter constraints on the tidal deformabilities and the pressure-density relation with priors better grounded in nuclear theory.

Although we found that posterior constraints from a single GW event are rather broad and prior-dependent, we expect that the impact of prior assumptions will eventually be overcome by the combined data from many GW signals.
Our Monte Carlo integration scheme can be straightforwardly extended to incorporate an arbitrary number of events: we simply draw masses (e.g., $M^{(a)}_{1,A}$) consistent with each event $A$ separately for each synthetic \EOS, and then combine the likelihoods $\mathcal{L}_A$ from each event so that the overall weight for the $a^{\text{th}}$ synthetic \EOS~is $\mathcal{L}^{(a)} = \prod_A \mathcal{L}_A(d_A|M_{1,A}^{(a)}, M_{2,A}^{(a)}, \Lambda_{1,A}^{(a)}, \Lambda_{2,A}^{(a)}; \mathcal{H}_A)$.
Electromagnetic observations of NS radii from NICER and moment of inertia measurements from radio observations can also be incorporated as additional likelihoods when available.
A precise quantification of the non-parametric method's ability to constrain the \EOS~with multiple GW signals will be investigated in future work. 

As illustrated by our Bayes factors between the model-agnostic and model-informed priors, our method naturally provides evidence estimates which can be used for Bayesian hypothesis ranking.
Leveraging the flexibility that the non-parametric representation offers when selecting the \EOS~prior, one could perform diverse kinds of model selection beyond the simple example presented here. For instance, evidence for different NS matter compositions (e.g.~purely hadronic vs.~hyperonic) or phenomenological features of the \EOS~(e.g.~strong phase transitions) could be compared.
This can be done by changing the set of tabulated {\EOS}s used to condition the underlying GP, so as to construct separate priors reflecting the typical behavior of different classes of {\EOS}.
In the context of GW170817, a hypothesis ranking scheme could also be applied to determine whether the GW data identify the event as a binary NS, a binary black hole, or a NS-black hole merger.
Model selection with a GP representation for the EOS has the potential to shed light on a variety of interesting questions about the macroscopic properties and internal structure of NSs.
More broadly, the versatility of non-parametric EOS inference will help maximize the information obtained from future observations, serving as a key component in the effort to constrain the supranuclear EOS.

\acknowledgments

The authors thank Daniel Holz, Shaon Ghosh and the members of the LVC Extreme Matter group for useful discussions about this work. Computational resources provided by the LIGO Laboratory and supported by National Science Foundation Grants PHY-0757058 and PHY-0823459 are gratefully acknowledged.
P.~L. is supported in part by the Natural Sciences and Engineering Research Council of Canada, and by NSF grants PHY 15-05124 and PHY 17-08081 to the University of Chicago.
R.~E. is supported at the University of Chicago by the Kavli Institute for Cosmological Physics through an endowment from the Kavli Foundation and its founder Fred Kavli.

\bibliography{gpr_eos-refs}

\appendix

\section{Complete results for simulated GW signals}\label{sec:complete inj}

\begin{figure}
    \begin{center}
        model-agnostic
    \end{center}
    \begin{minipage}{0.32\textwidth}
        \begin{center}
            \hspace{1.2cm} \texttt{ms1} \\
            \includegraphics[width=1.00\textwidth, clip=True, trim=0.50cm 1.10cm 6.80cm 6.80cm]{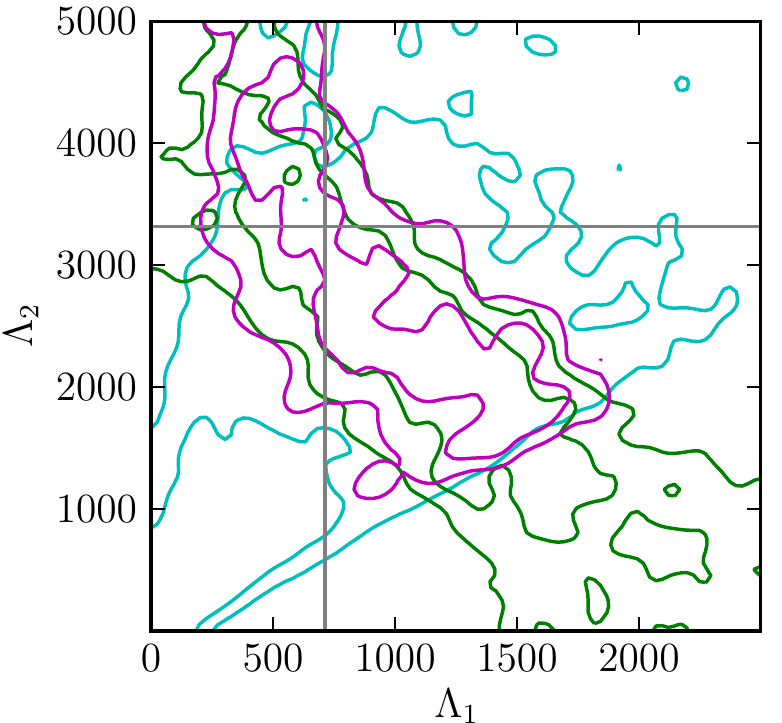}
            \includegraphics[width=1.00\textwidth, clip=True, trim=0.50cm 1.10cm 6.80cm 6.80cm]{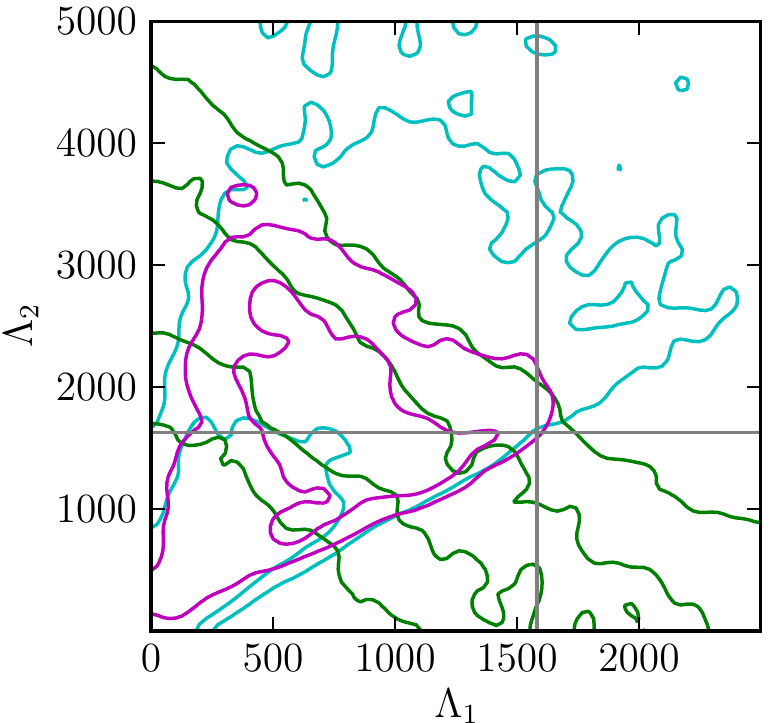}
            \includegraphics[width=1.00\textwidth, clip=True, trim=0.50cm 1.10cm 6.80cm 6.80cm]{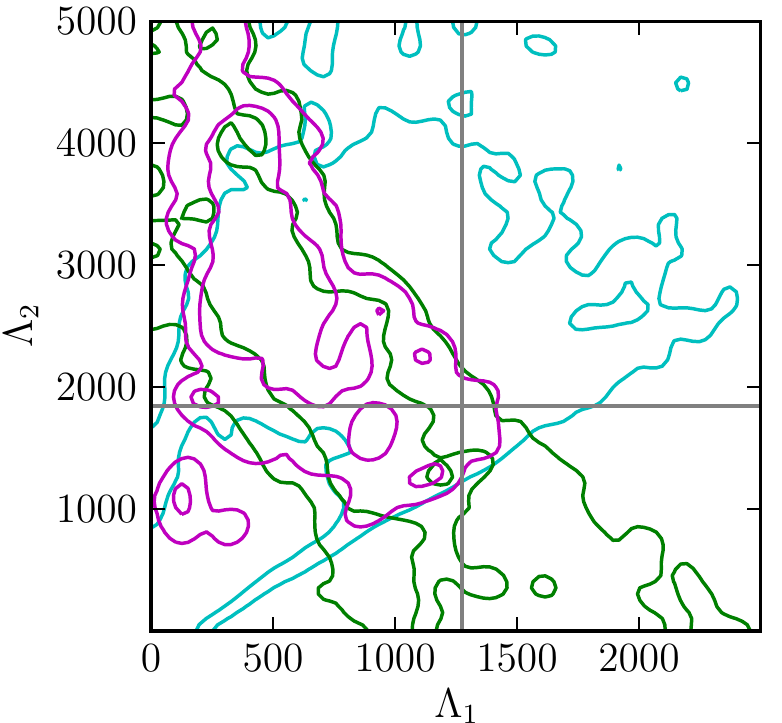}
        \end{center}
    \end{minipage}
    \begin{minipage}{0.32\textwidth}
        \begin{center}
            \hspace{1.2cm} \texttt{h4} \\
            \includegraphics[width=1.00\textwidth, clip=True, trim=0.50cm 1.10cm 6.80cm 6.80cm]{magenta_cyan-figures/kde-corner-samples_lambda1-lambda2_inj1unh4_prior-likelihood-posterior_noref}
            \includegraphics[width=1.00\textwidth, clip=True, trim=0.50cm 1.10cm 6.80cm 6.80cm]{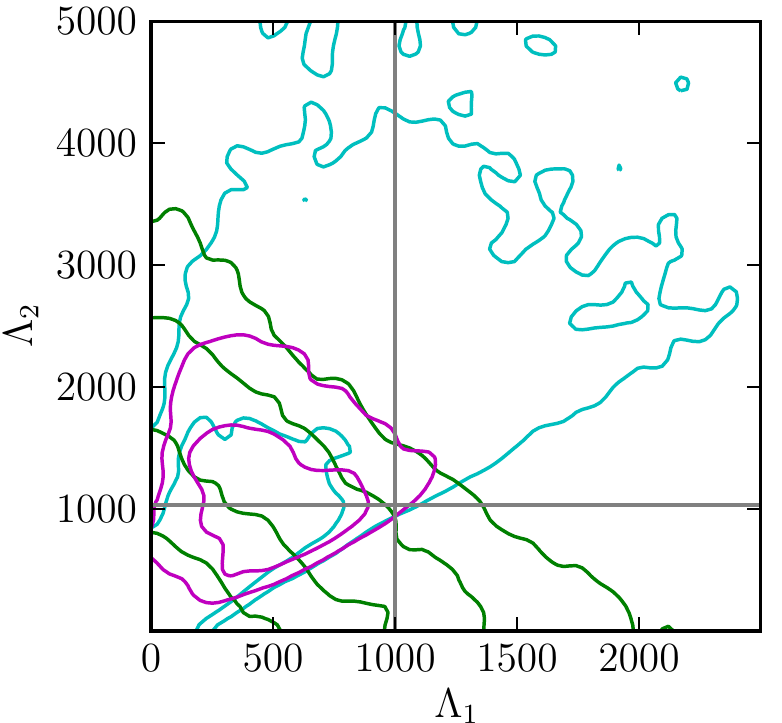}
            \includegraphics[width=1.00\textwidth, clip=True, trim=0.50cm 1.10cm 6.80cm 6.80cm]{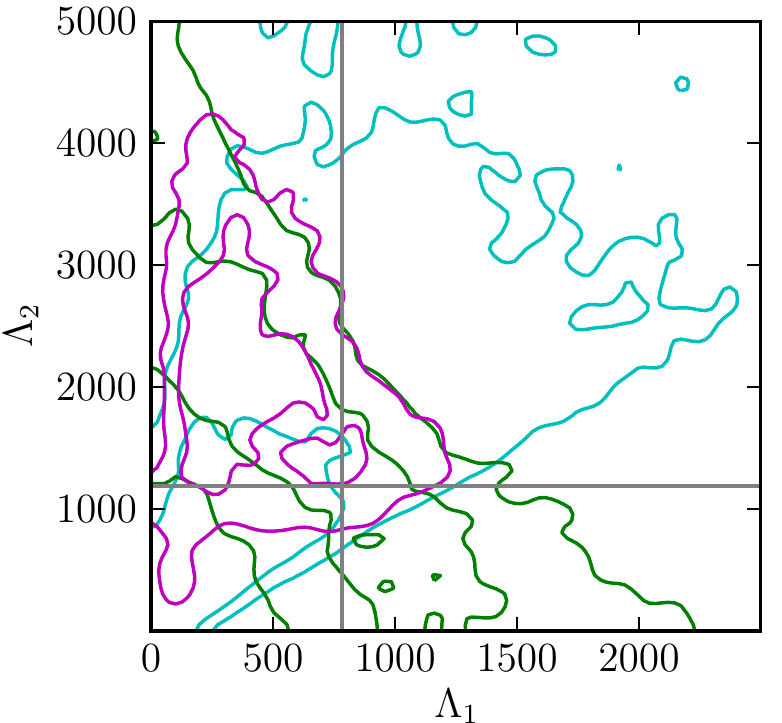}
        \end{center}
    \end{minipage}
    \begin{minipage}{0.32\textwidth}
        \begin{center}
            \hspace{1.2cm} \texttt{sly} \\
            \includegraphics[width=1.00\textwidth, clip=True, trim=0.50cm 1.10cm 6.80cm 6.80cm]{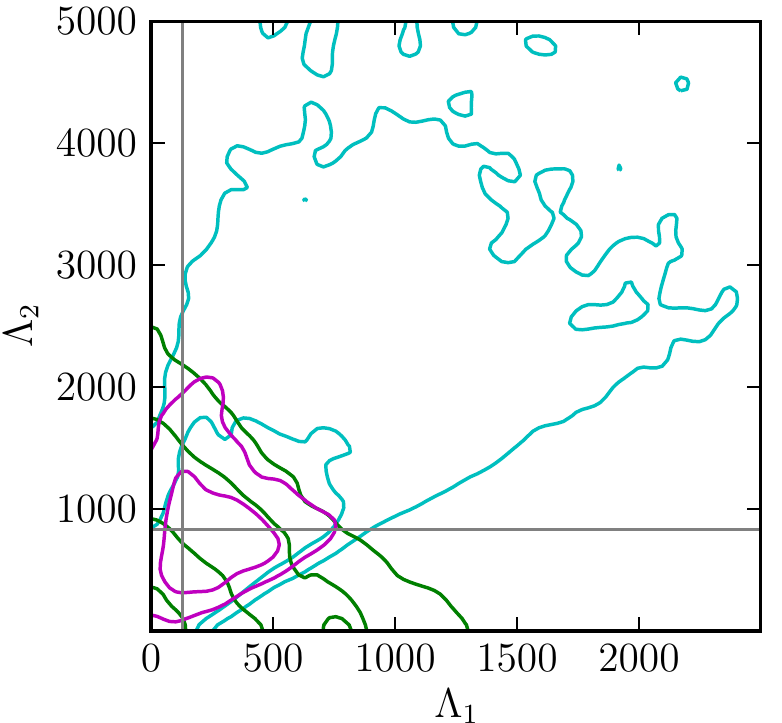}
            \includegraphics[width=1.00\textwidth, clip=True, trim=0.50cm 1.10cm 6.80cm 6.80cm]{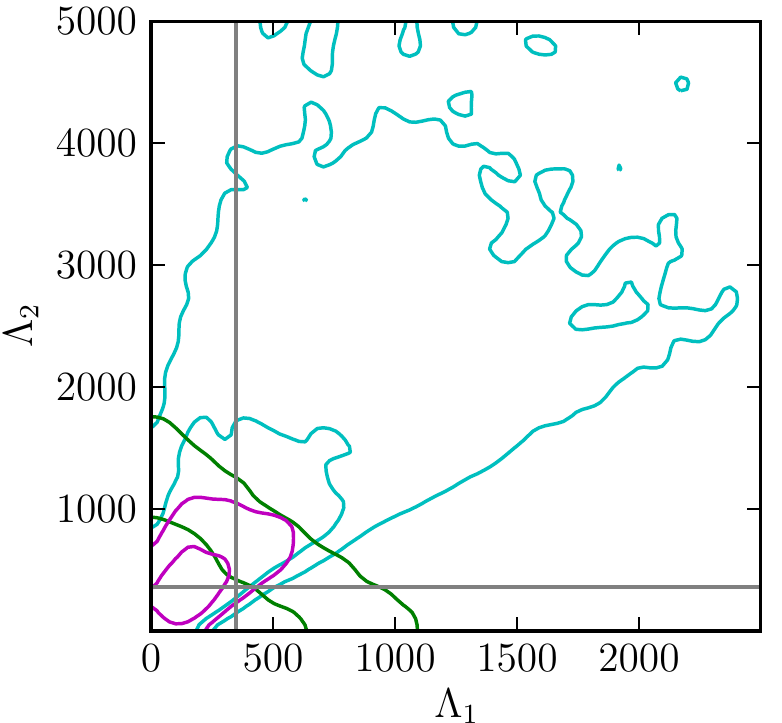}
            \includegraphics[width=1.00\textwidth, clip=True, trim=0.50cm 1.10cm 6.80cm 6.80cm]{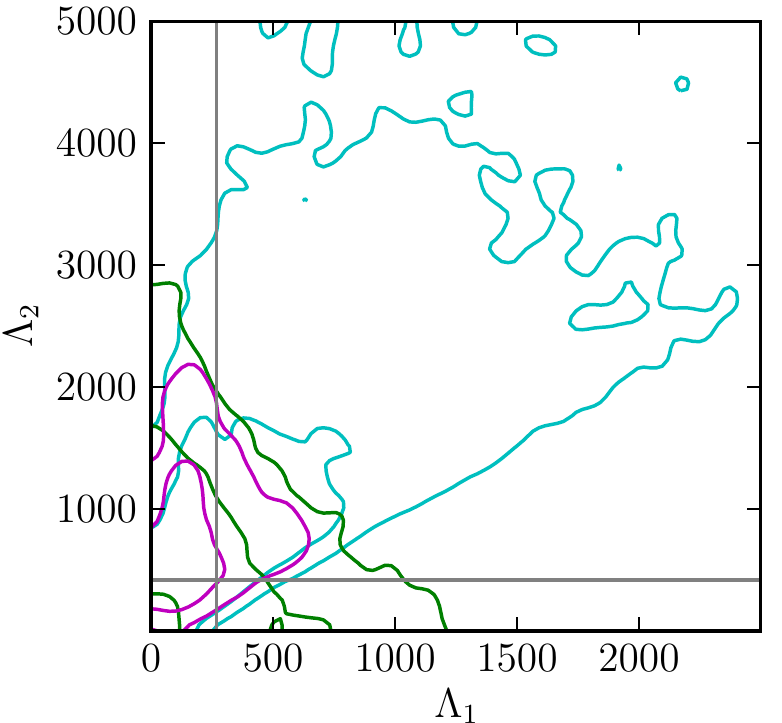}
        \end{center}
    \end{minipage}
    \caption{
        $\Lambda_1$--$\Lambda_2$ joint prior (\textit{blue}), likelihood (\textit{green}), and posterior (\textit{red}) distributions using the model-agnostic prior.
        Injected values are shown as grey cross-hairs and the colored contours correspond to 50\% and 90\% confidence regions.
        Rows correspond to events I-III from Table~\ref{tb:inj} and columns are labeled according to the injected \EOS.
    }
    \label{fig:inj un lambda1-lambda2}
\end{figure}

\begin{figure}
    \begin{center}
        model-agnostic
    \end{center}
    \begin{minipage}{0.32\textwidth}
        \begin{center}
            \hspace{1.2cm} \texttt{ms1} \\
            \includegraphics[width=1.00\textwidth, clip=True, trim=0.50cm 1.10cm 6.80cm 6.80cm]{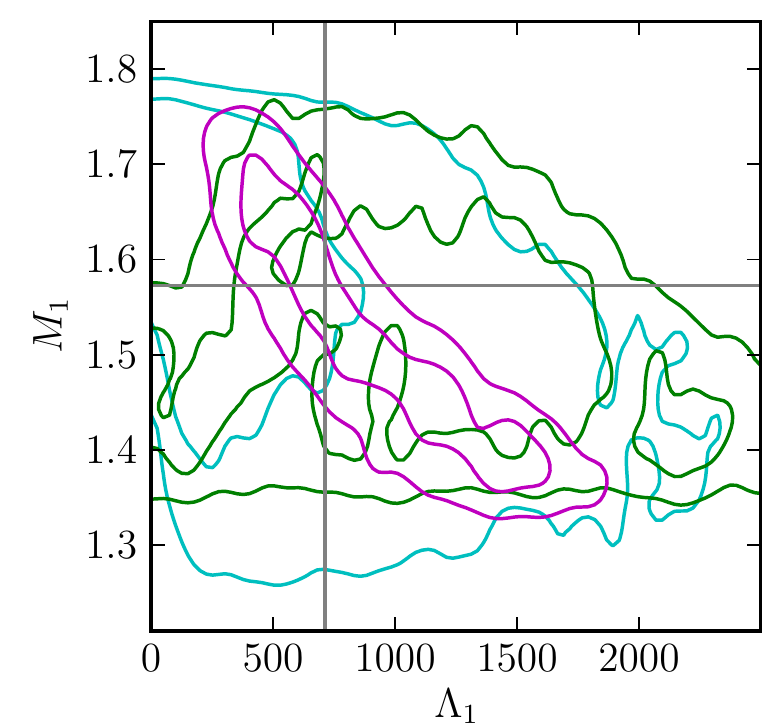}
            \includegraphics[width=1.00\textwidth, clip=True, trim=0.50cm 1.10cm 6.80cm 6.80cm]{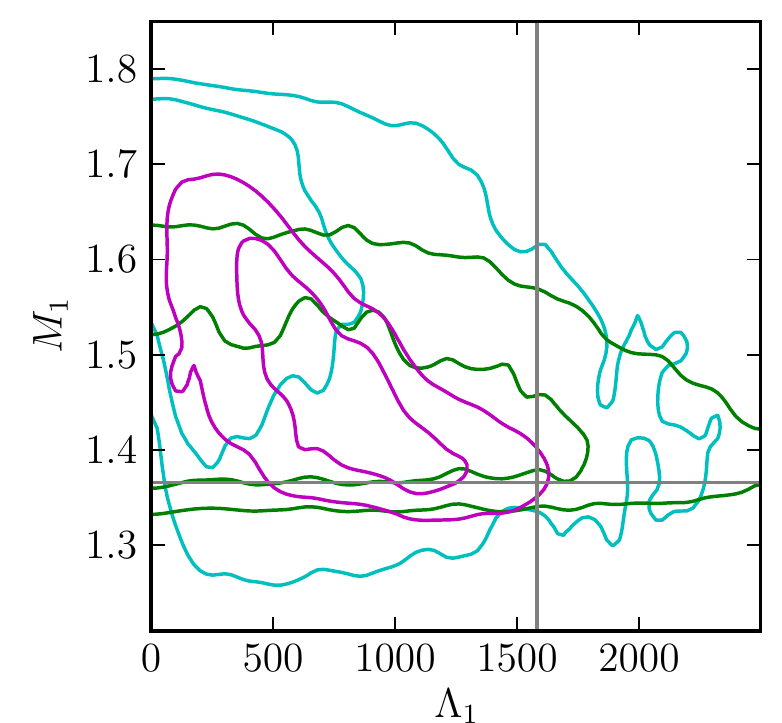}
            \includegraphics[width=1.00\textwidth, clip=True, trim=0.50cm 1.10cm 6.80cm 6.80cm]{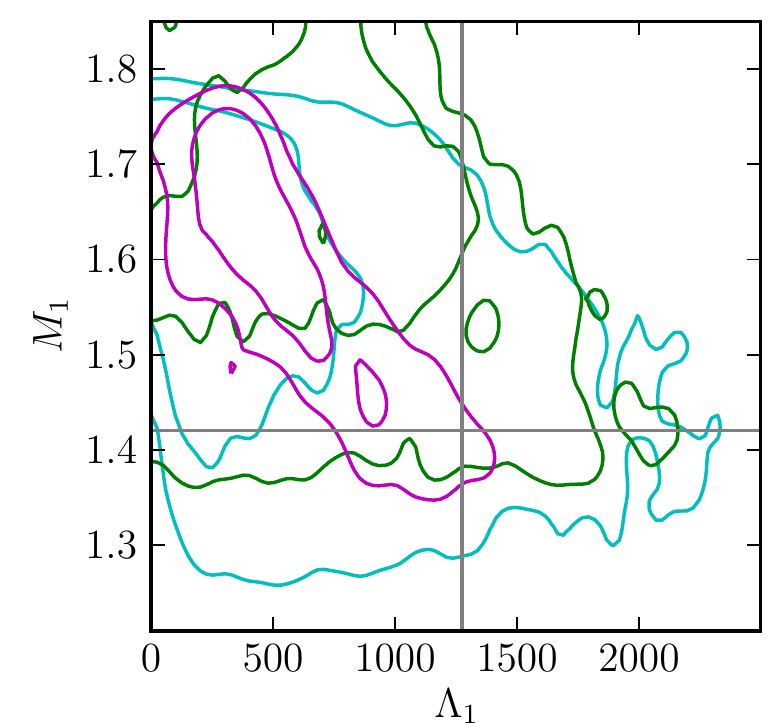}
        \end{center}
    \end{minipage}
    \begin{minipage}{0.32\textwidth}
        \begin{center}
            \hspace{1.2cm} \texttt{h4} \\
            \includegraphics[width=1.00\textwidth, clip=True, trim=0.50cm 1.10cm 6.80cm 6.80cm]{magenta_cyan-figures/kde-corner-samples_lambda1-m1_source_inj1unh4_prior-likelihood-posterior_noref}
            \includegraphics[width=1.00\textwidth, clip=True, trim=0.50cm 1.10cm 6.80cm 6.80cm]{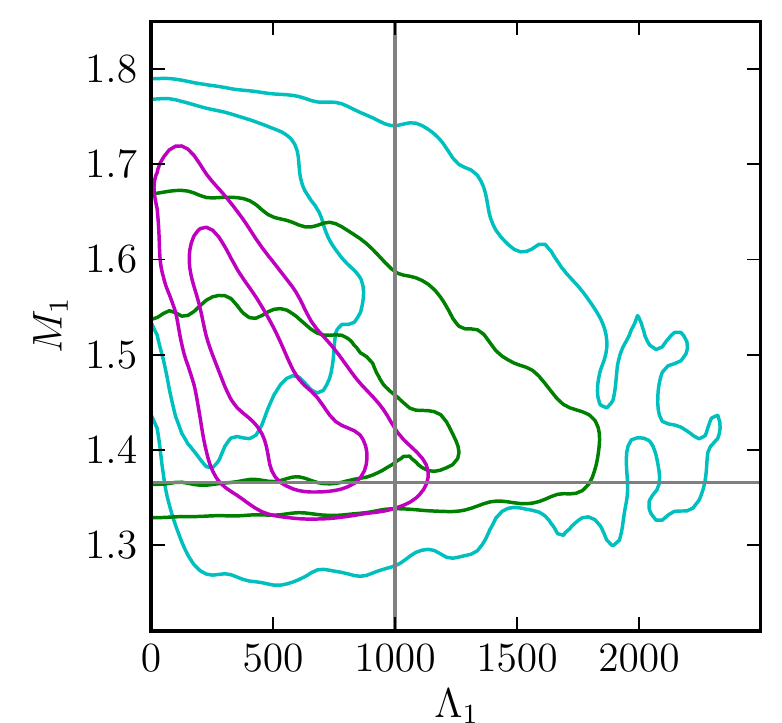}
            \includegraphics[width=1.00\textwidth, clip=True, trim=0.50cm 1.10cm 6.80cm 6.80cm]{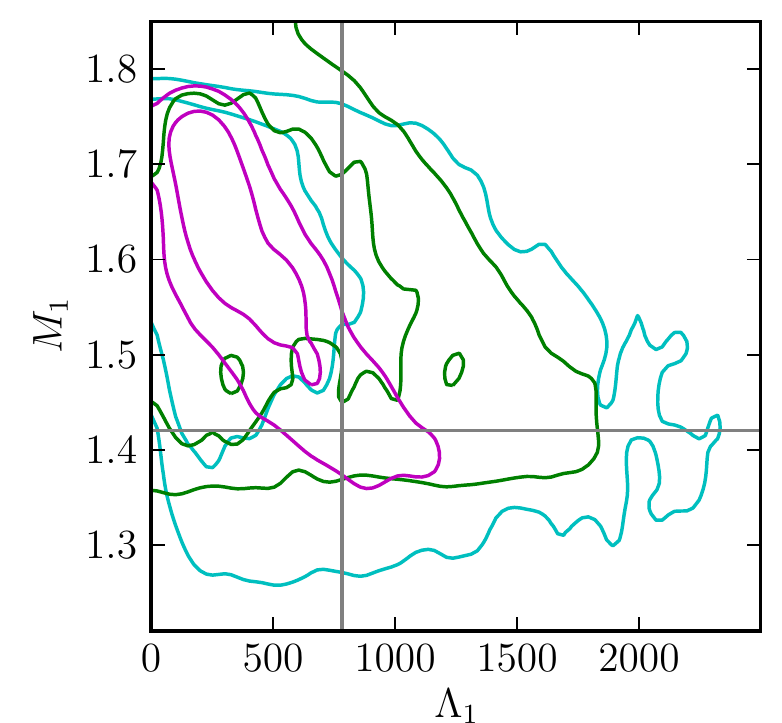}
        \end{center}
    \end{minipage}
    \begin{minipage}{0.32\textwidth}
        \begin{center}
            \hspace{1.2cm} \texttt{sly} \\
            \includegraphics[width=1.00\textwidth, clip=True, trim=0.50cm 1.10cm 6.80cm 6.80cm]{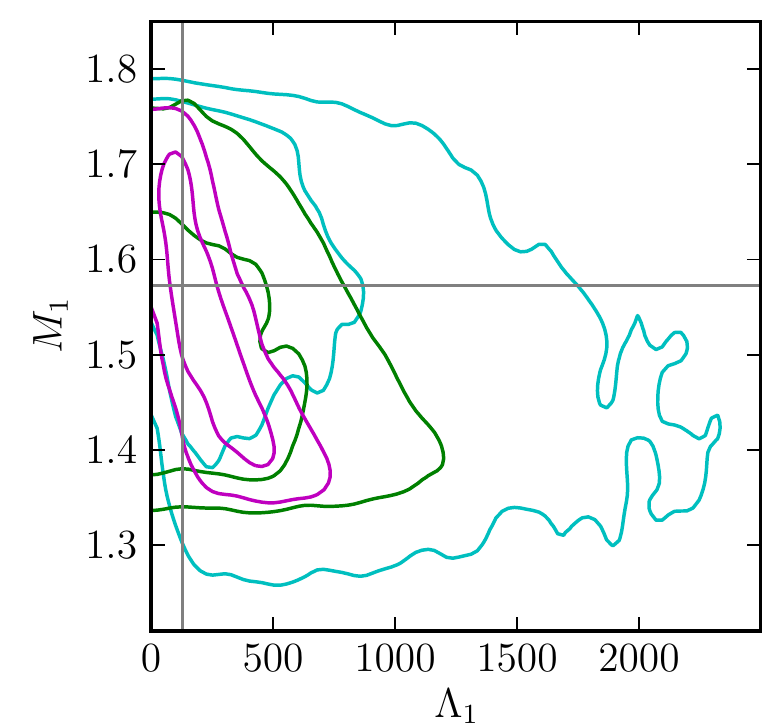}
            \includegraphics[width=1.00\textwidth, clip=True, trim=0.50cm 1.10cm 6.80cm 6.80cm]{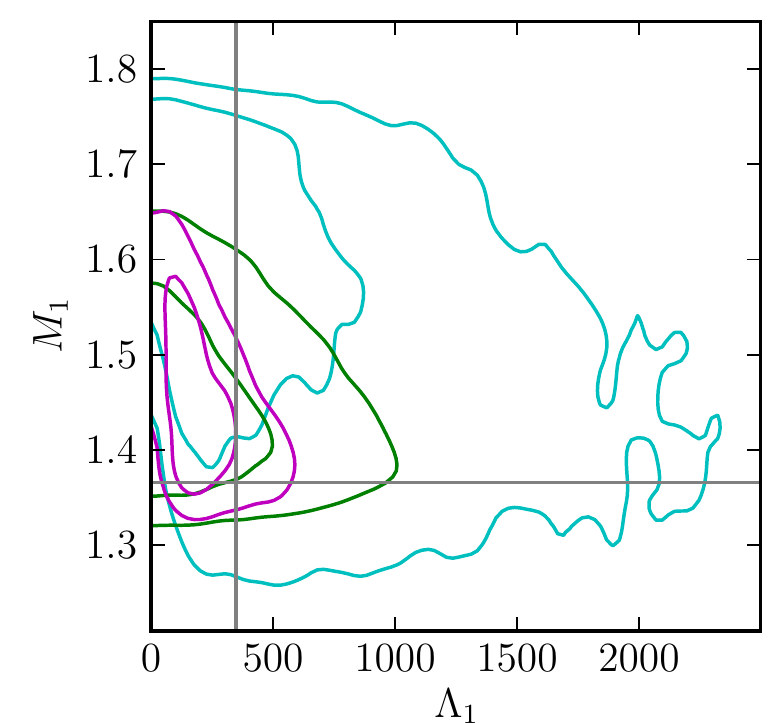}
            \includegraphics[width=1.00\textwidth, clip=True, trim=0.50cm 1.10cm 6.80cm 6.80cm]{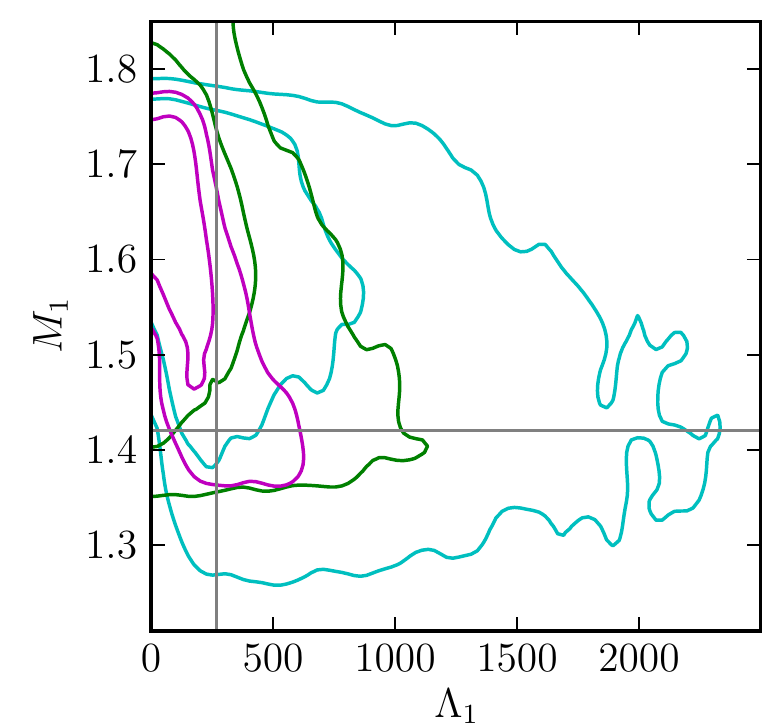}
        \end{center}
    \end{minipage}
    \caption{
        $\Lambda_1$--$M_1$ joint prior (\textit{blue}), likelihood (\textit{green}), and posterior (\textit{red}) distributions using the model-agnostic prior.
        Injected values are shown as grey cross-hairs and the colored contours correspond to 50\% and 90\% confidence regions.
        Rows correspond to events I-III from Table~\ref{tb:inj} and columns are labeled according to the injected \EOS.
    }
    \label{fig:inj un lambda1-m1_source}
\end{figure}

\begin{figure}
    \begin{center}
        model-agnostic
    \end{center}
    \begin{minipage}{0.31\textwidth}
        \begin{center}
            \texttt{ms1} \\
            \includegraphics[width=1.00\textwidth, clip=True, trim=0.05cm 0.12cm 0.45cm 0.12cm]{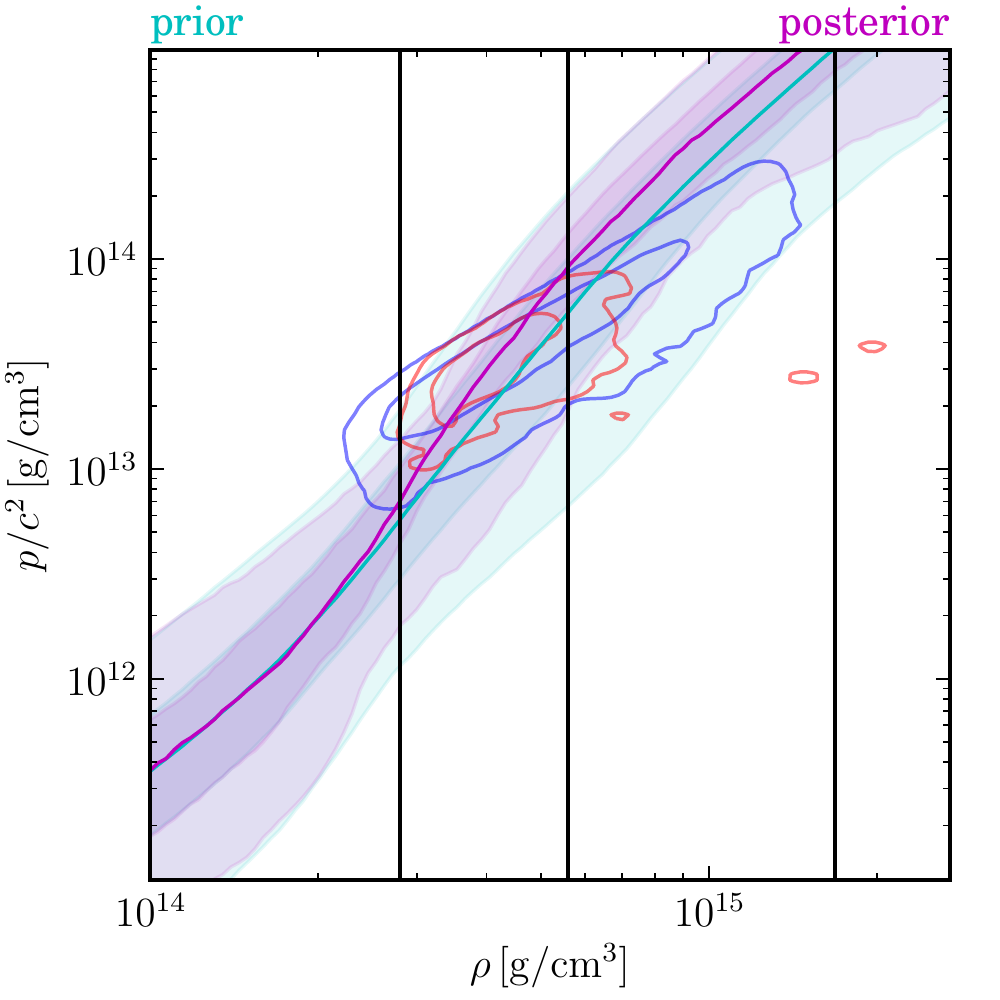}
            \includegraphics[width=1.00\textwidth, clip=True, trim=0.05cm 0.12cm 0.45cm 0.12cm]{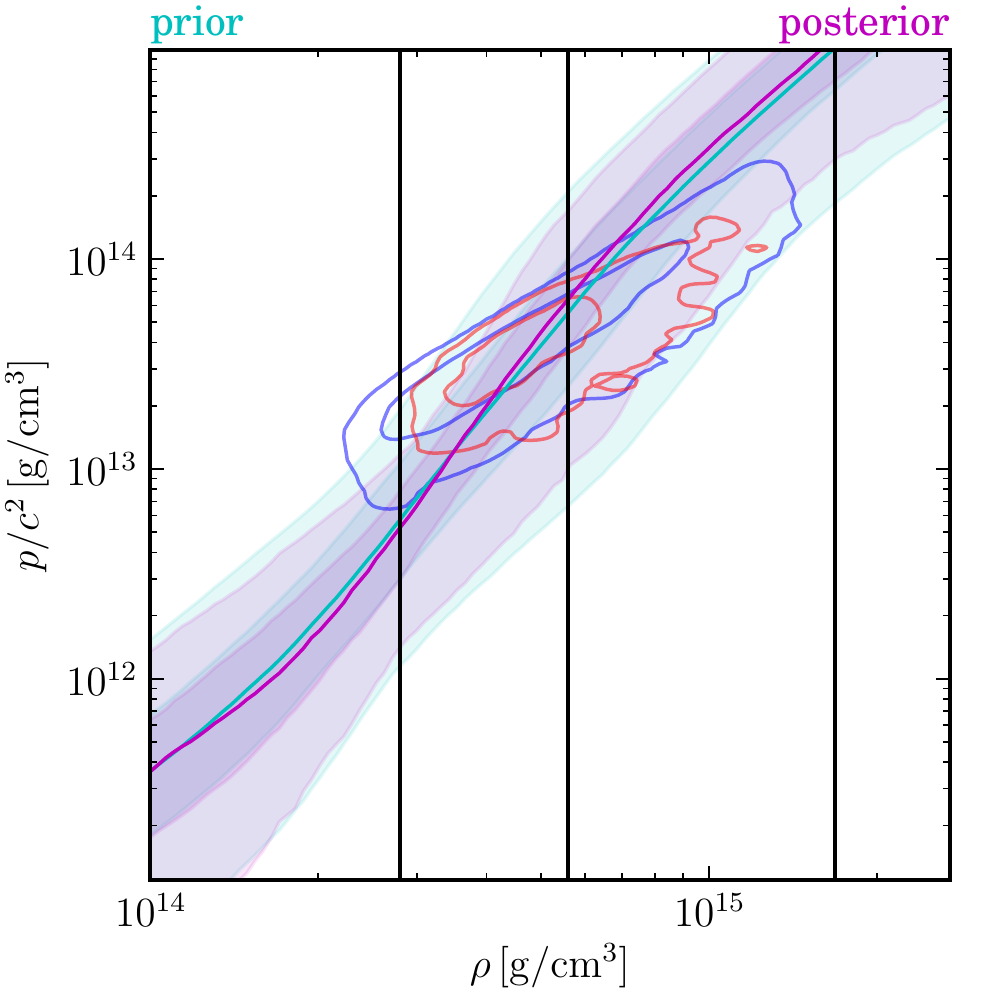}
            \includegraphics[width=1.00\textwidth, clip=True, trim=0.05cm 0.12cm 0.45cm 0.12cm]{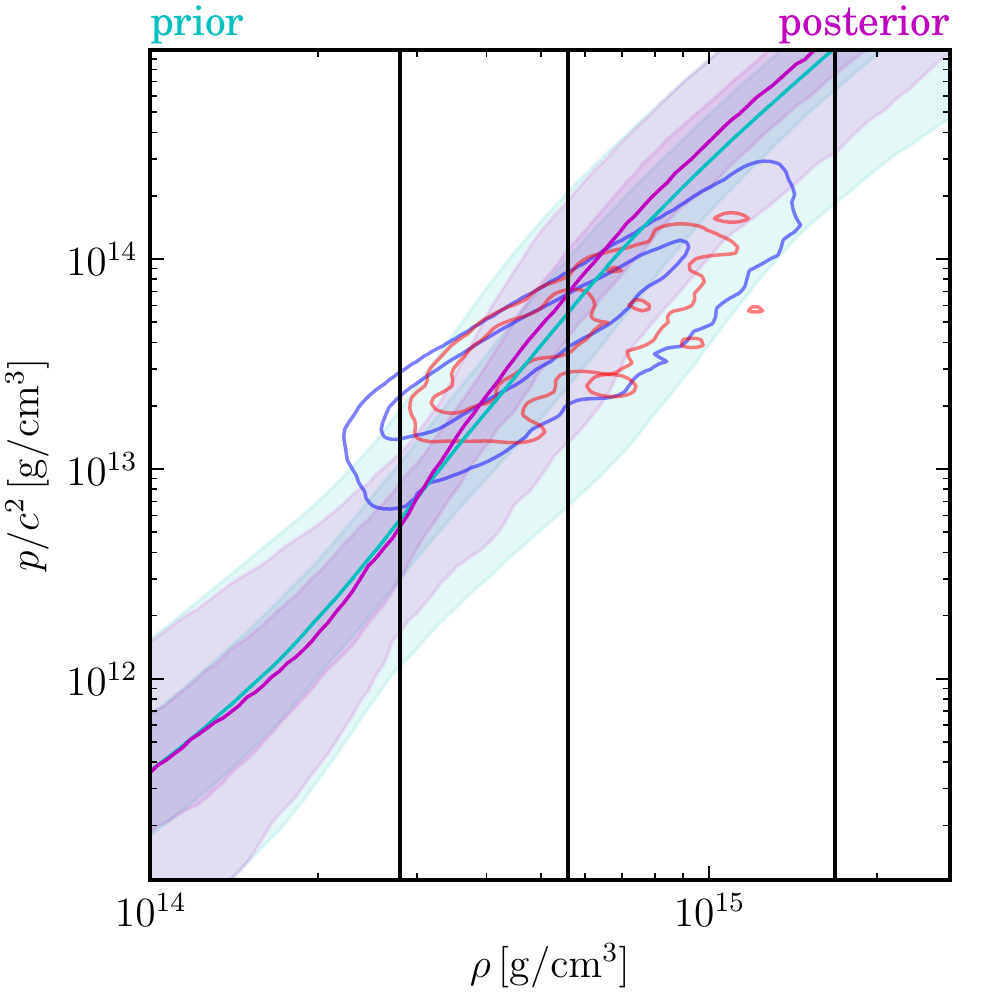}
        \end{center}
    \end{minipage}
    \begin{minipage}{0.31\textwidth}
        \begin{center}
            \texttt{h4} \\
            \includegraphics[width=1.00\textwidth, clip=True, trim=0.05cm 0.12cm 0.45cm 0.12cm]{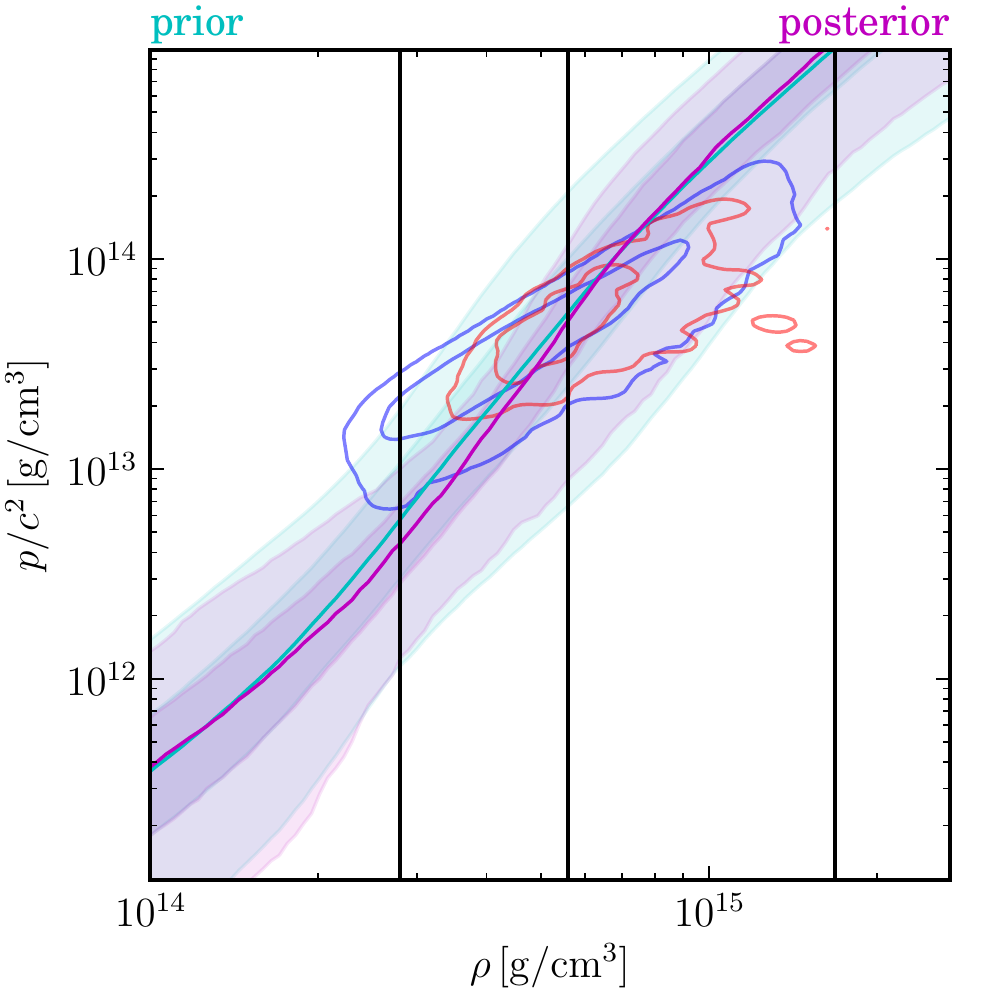}
            \includegraphics[width=1.00\textwidth, clip=True, trim=0.05cm 0.12cm 0.45cm 0.12cm]{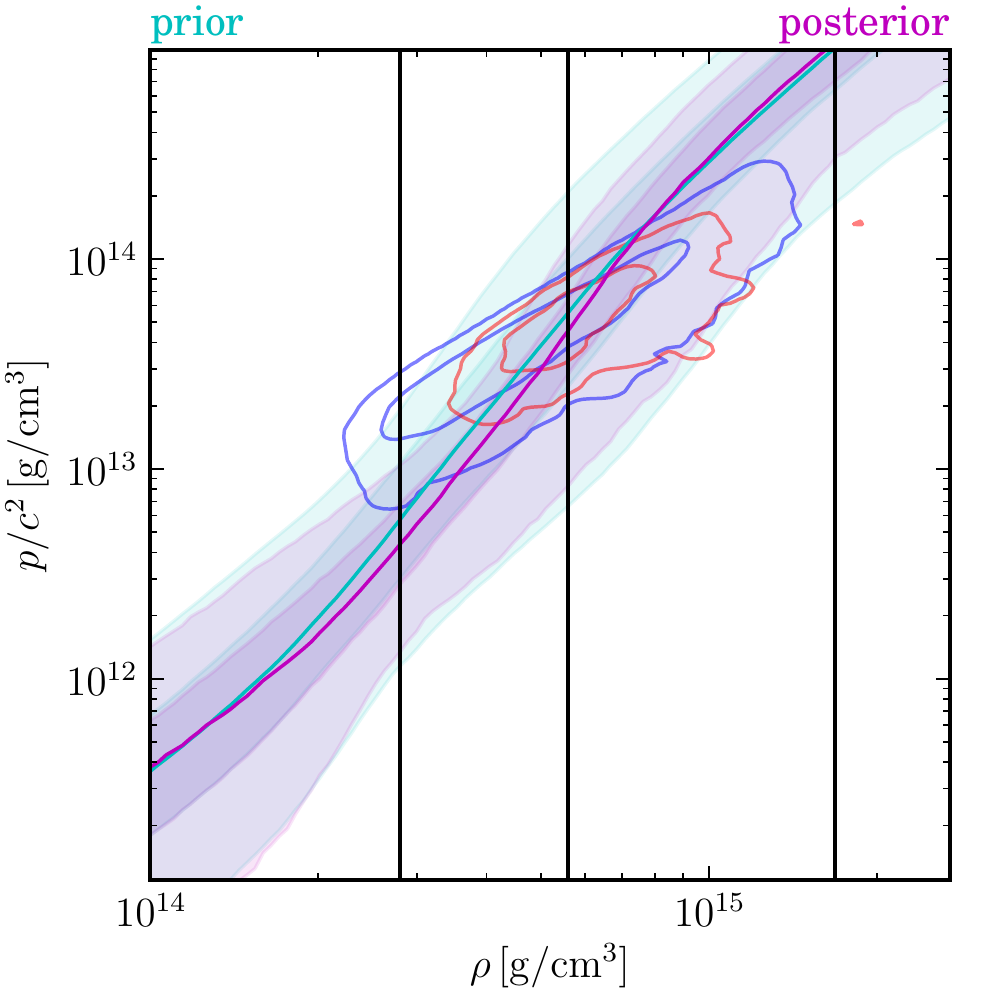}
            \includegraphics[width=1.00\textwidth, clip=True, trim=0.05cm 0.12cm 0.45cm 0.12cm]{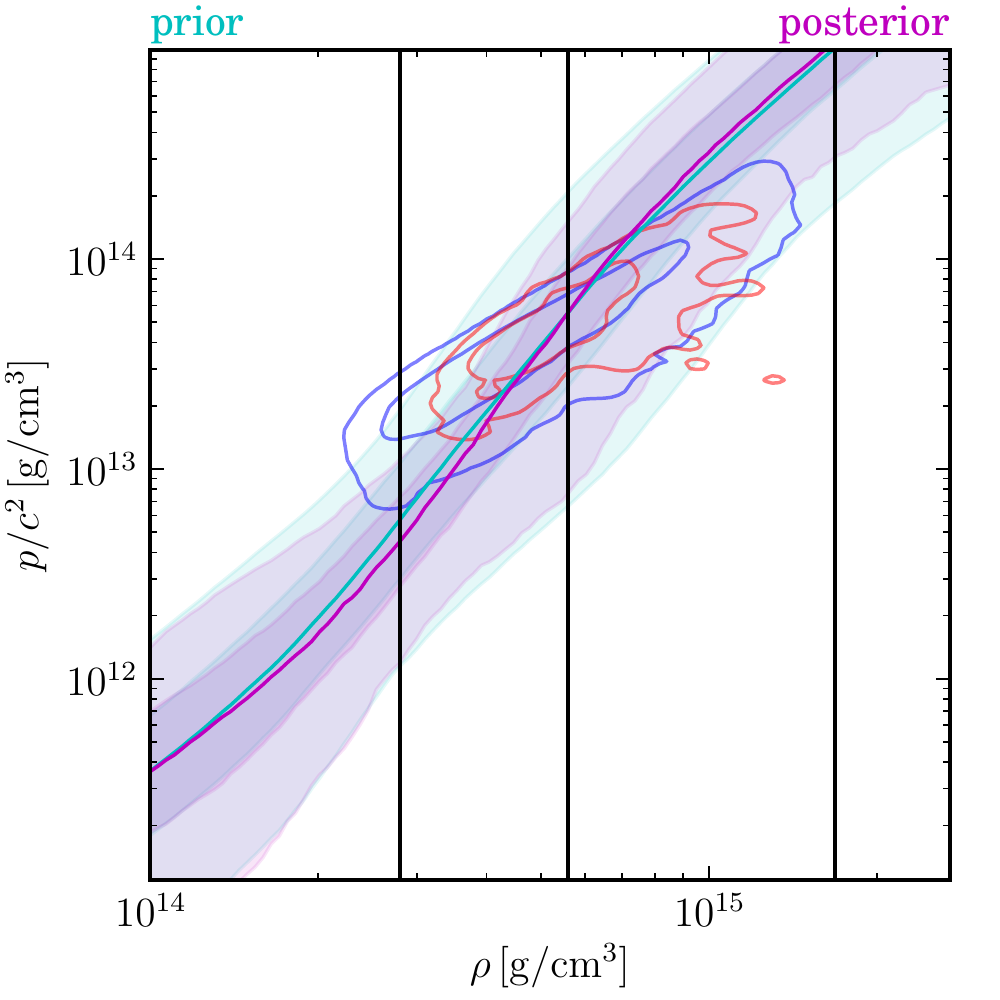}
        \end{center}
    \end{minipage}
    \begin{minipage}{0.31\textwidth}
        \begin{center}
            \texttt{sly} \\
            \includegraphics[width=1.00\textwidth, clip=True, trim=0.05cm 0.12cm 0.45cm 0.12cm]{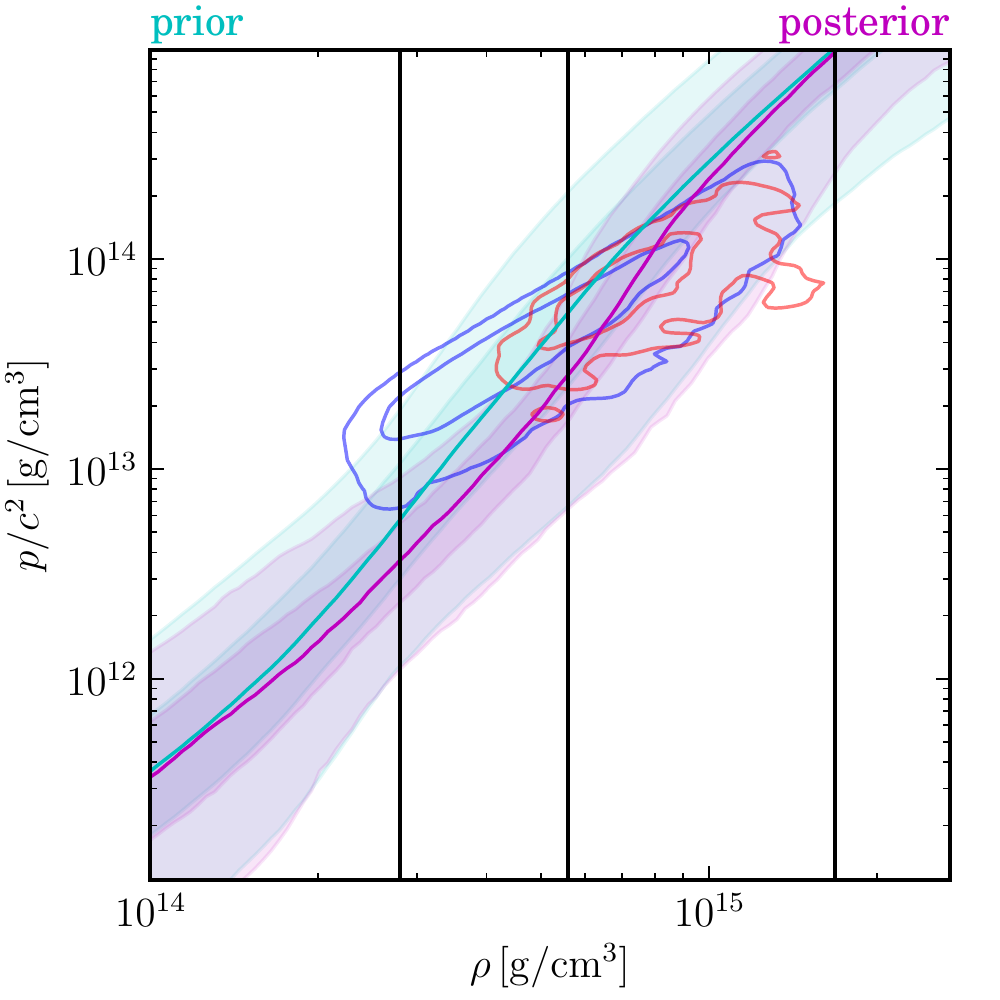}
            \includegraphics[width=1.00\textwidth, clip=True, trim=0.05cm 0.12cm 0.45cm 0.12cm]{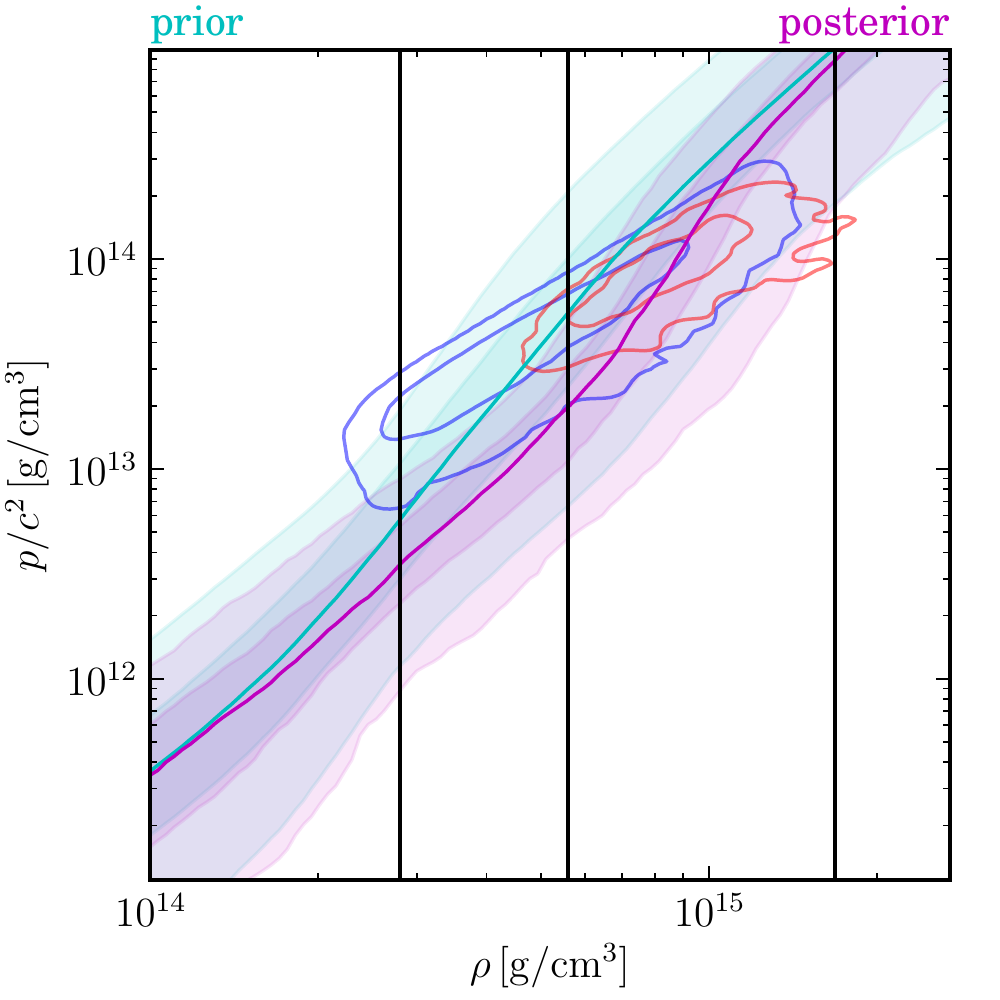}
            \includegraphics[width=1.00\textwidth, clip=True, trim=0.05cm 0.12cm 0.45cm 0.12cm]{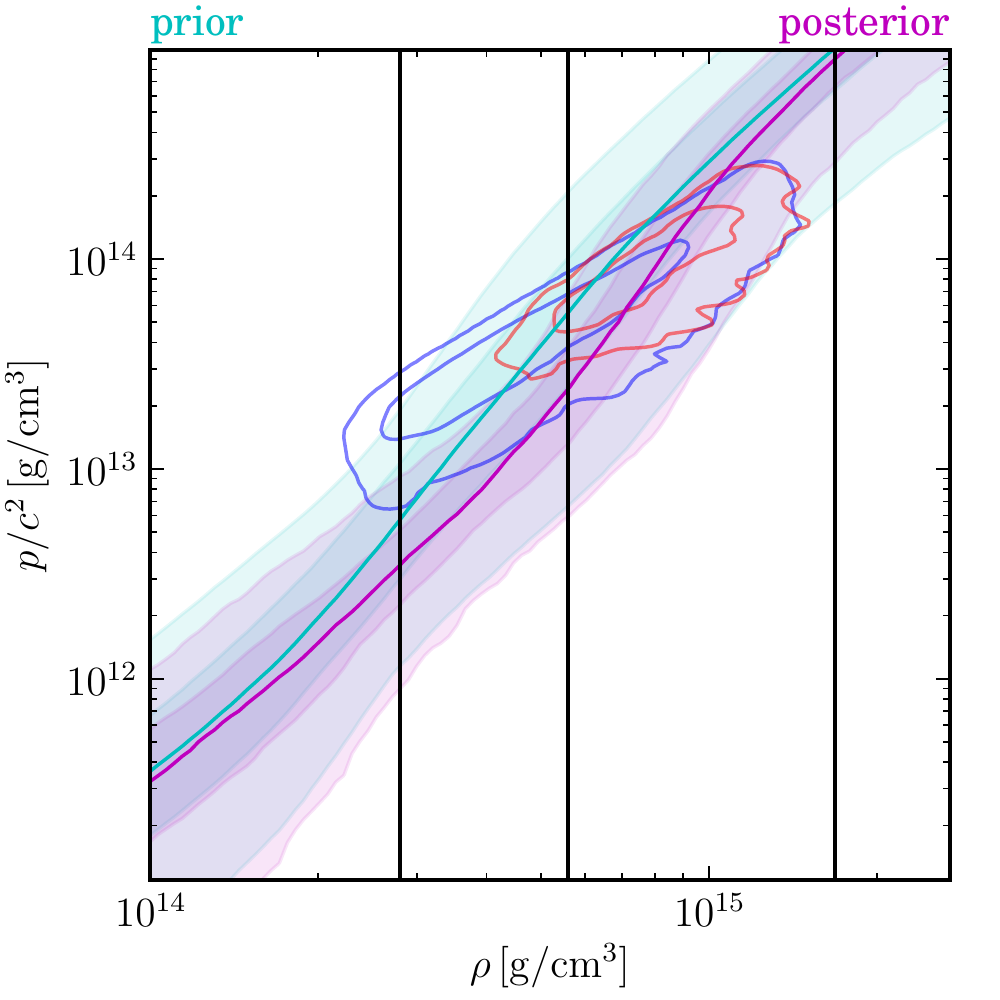}
        \end{center}
    \end{minipage}
    \caption{
        Prior (\textit{blue}) and posterior (\textit{red}) processes for injections for the model-agnostic prior.
        Solid lines show the medians, shaded regions represent the 50\% and 90\% credible regions for $p(\rho)$, and the contours represent the 50\% and 90\% credible regions for the central densities and pressures of $M_1$.
        Rows correspond to events I-III from Table~\ref{tb:inj} and columns are labeled according to injected \EOS.
    }
    \label{fig:inj un process}
\end{figure}

\begin{figure}
    \begin{center}
        model-informed
    \end{center}
    \begin{minipage}{0.32\textwidth}
        \begin{center}
            \hspace{1.2cm} \texttt{ms1} \\
            \includegraphics[width=1.00\textwidth, clip=True, trim=0.50cm 1.10cm 6.80cm 6.80cm]{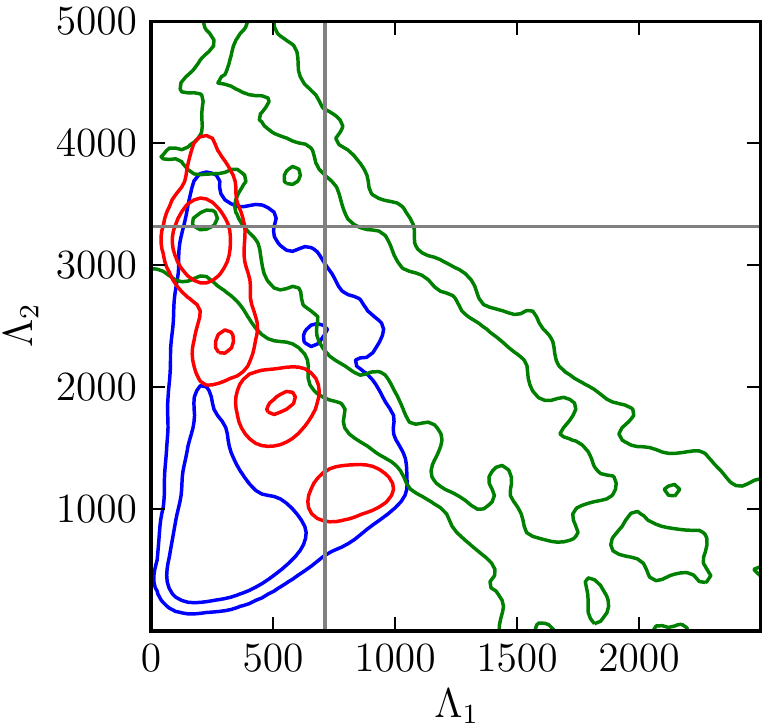}
            \includegraphics[width=1.00\textwidth, clip=True, trim=0.50cm 1.10cm 6.80cm 6.80cm]{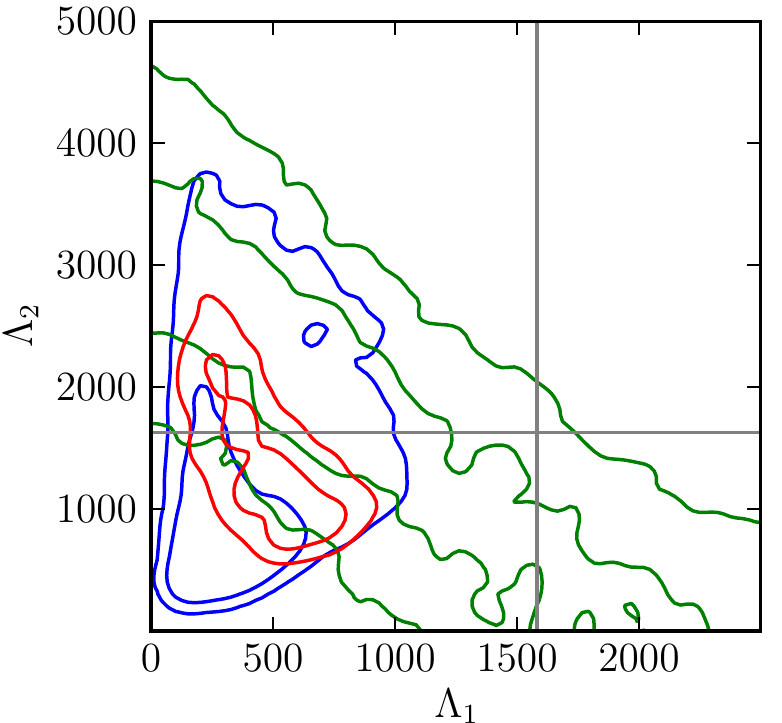}
            \includegraphics[width=1.00\textwidth, clip=True, trim=0.50cm 1.10cm 6.80cm 6.80cm]{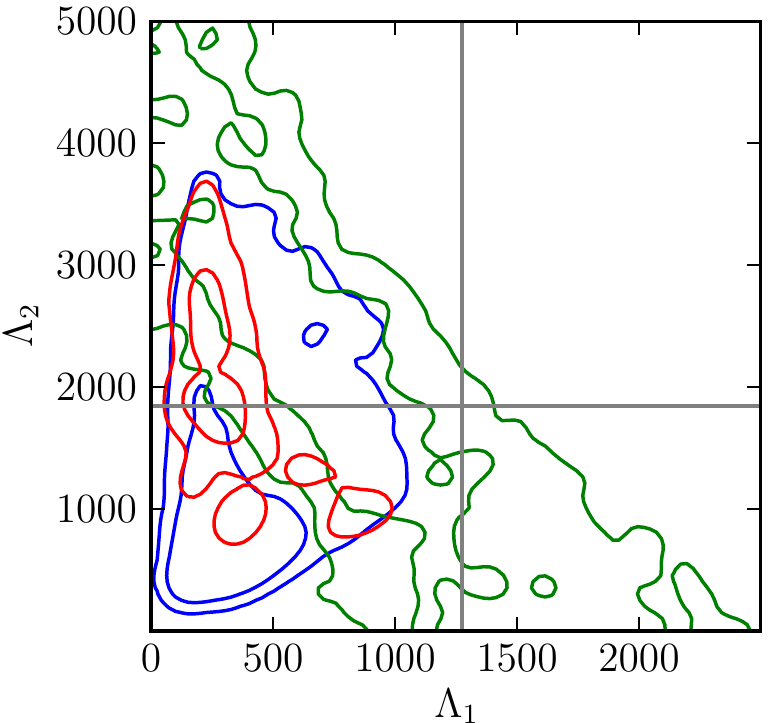}
        \end{center}
    \end{minipage}
    \begin{minipage}{0.32\textwidth}
        \begin{center}
            \hspace{1.2cm} \texttt{h4} \\
            \includegraphics[width=1.00\textwidth, clip=True, trim=0.50cm 1.10cm 6.80cm 6.80cm]{red_blue-figures/kde-corner-samples_lambda1-lambda2_inj1modh4_prior-likelihood-posterior_noref}
            \includegraphics[width=1.00\textwidth, clip=True, trim=0.50cm 1.10cm 6.80cm 6.80cm]{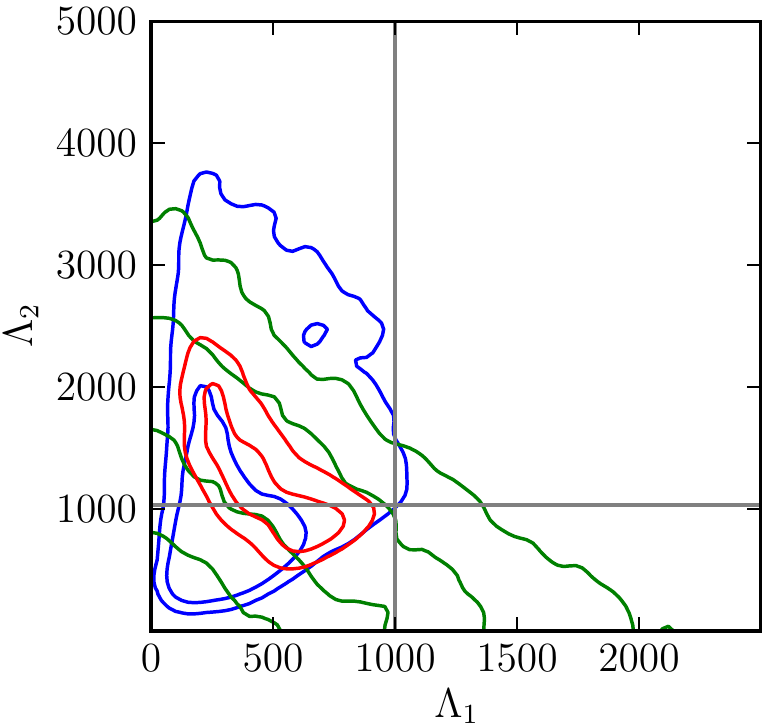}
            \includegraphics[width=1.00\textwidth, clip=True, trim=0.50cm 1.10cm 6.80cm 6.80cm]{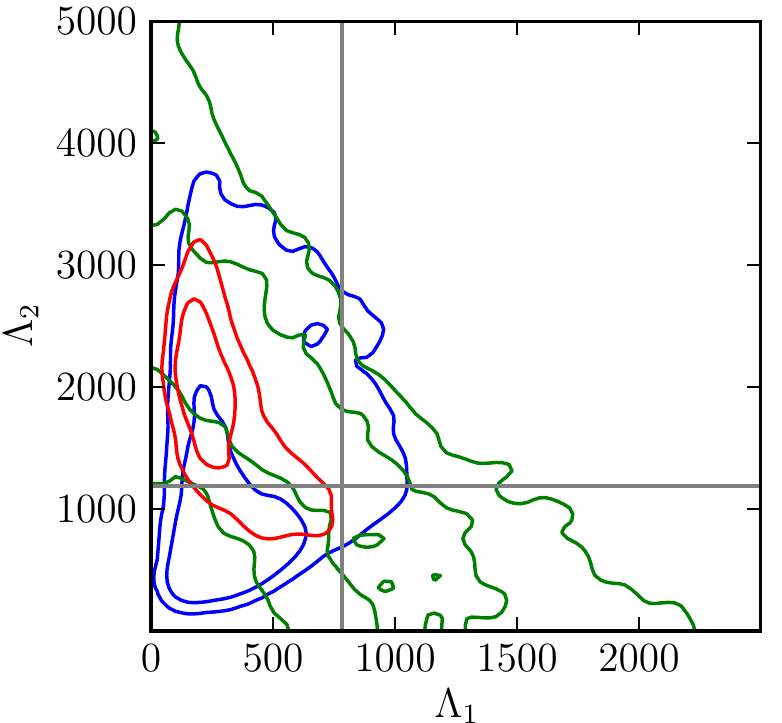}
        \end{center}
    \end{minipage}
    \begin{minipage}{0.32\textwidth}
        \begin{center}
            \hspace{1.2cm} \texttt{sly} \\
            \includegraphics[width=1.00\textwidth, clip=True, trim=0.50cm 1.10cm 6.80cm 6.80cm]{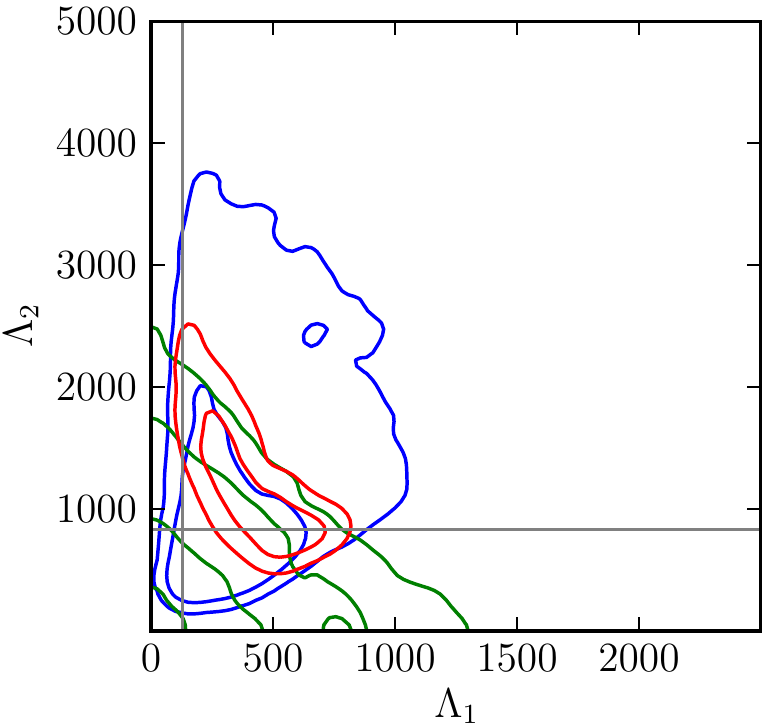}
            \includegraphics[width=1.00\textwidth, clip=True, trim=0.50cm 1.10cm 6.80cm 6.80cm]{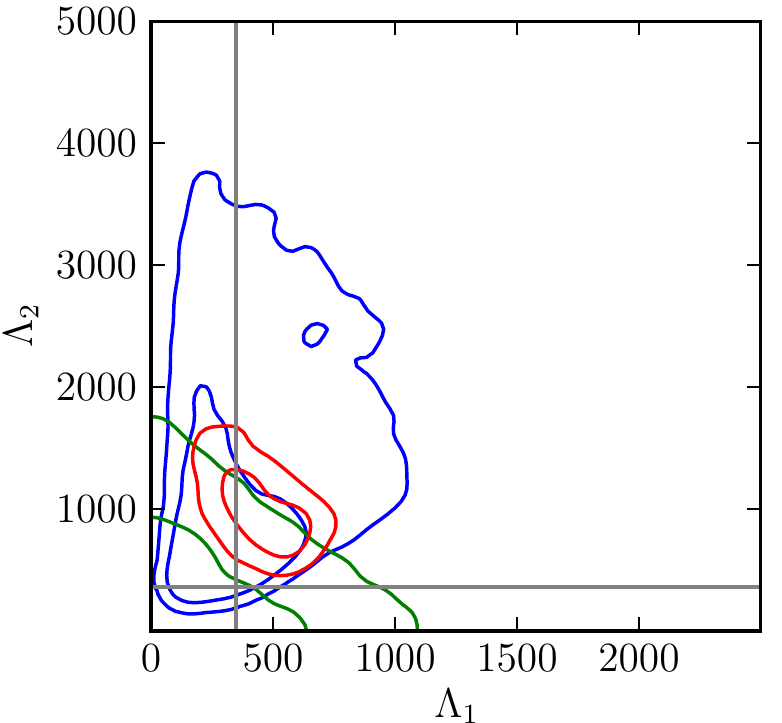}
            \includegraphics[width=1.00\textwidth, clip=True, trim=0.50cm 1.10cm 6.80cm 6.80cm]{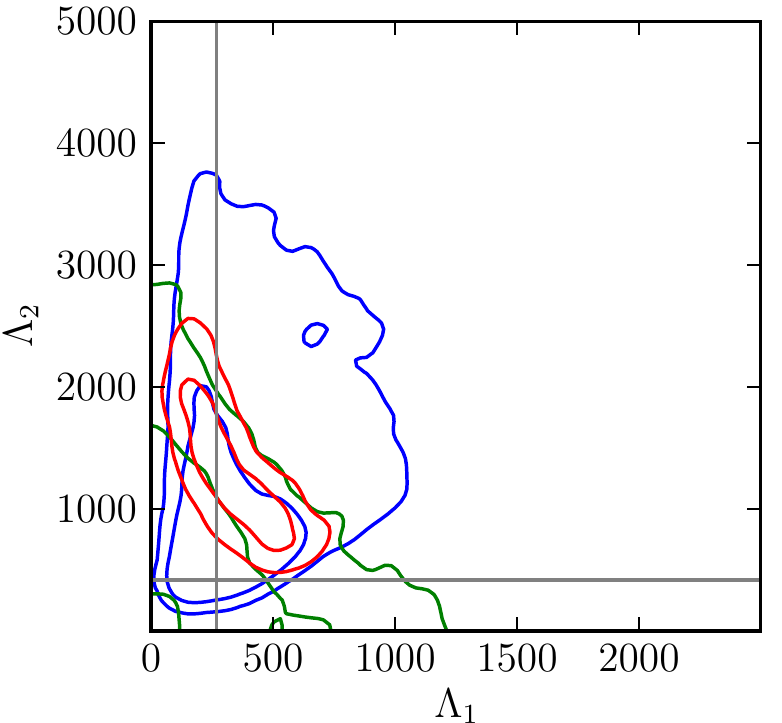}
        \end{center}
    \end{minipage}
    \caption{
        $\Lambda_1$--$\Lambda_2$ joint prior (\textit{blue}), likelihood (\textit{green}), and posterior (\textit{red}) distributions using the model-informed prior.
        Injected values are shown as grey cross-hairs and the colored contours correspond to 50\% and 90\% confidence regions.
        Rows correspond to events I-III from Table~\ref{tb:inj} and columns are labeled according to the injected \EOS.
        We note that \texttt{ms1}'s model-informed posterior for injection I is relatively under-sampled, with $N_\mathrm{eff} \sim \mathrm{a few}\times10^2$ even though $N>10^6$.
        What's more, this is the only injection that strongly favors one prior over the other, with $\Bayes = \bayesinjonemsone$, as evidenced by the dearth of of posterior support for the model-informed prior.
        Our evidence estimation, therefore, remains relatively precise even though errors in the posterior's shape are dominated by the relatively small number of samples with large weights.
    }
    \label{fig:inj mod lambda1-lambda2}
\end{figure}

\begin{figure}
    \begin{center}
        model-informed
    \end{center}
    \begin{minipage}{0.32\textwidth}
        \begin{center}
            \hspace{1.2cm} \texttt{ms1} \\
            \includegraphics[width=1.00\textwidth, clip=True, trim=0.50cm 1.10cm 6.80cm 6.80cm]{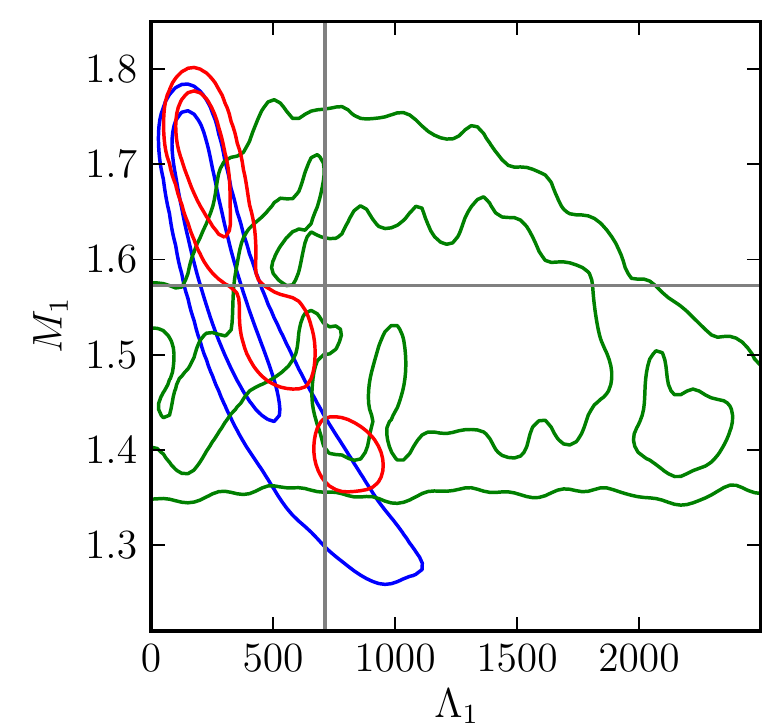}
            \includegraphics[width=1.00\textwidth, clip=True, trim=0.50cm 1.10cm 6.80cm 6.80cm]{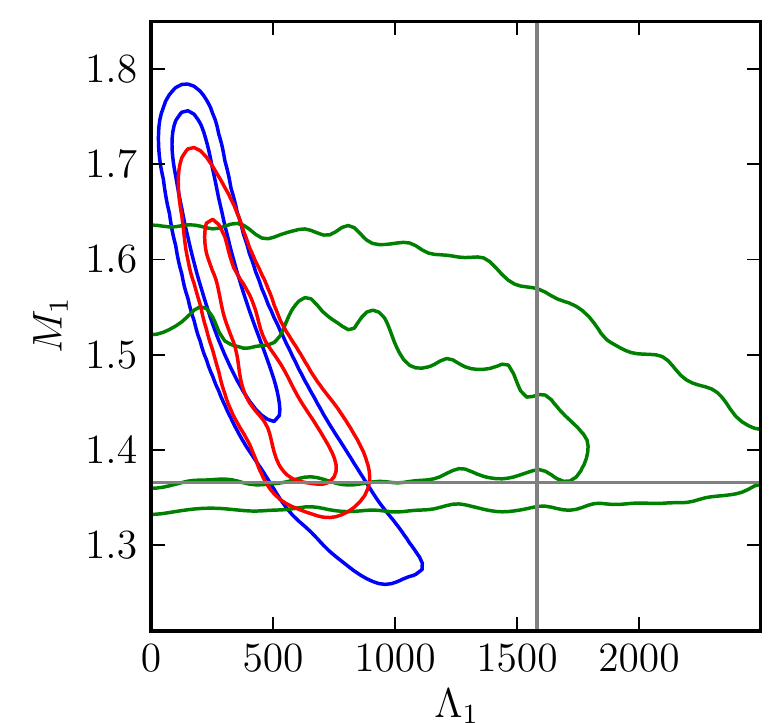}
            \includegraphics[width=1.00\textwidth, clip=True, trim=0.50cm 1.10cm 6.80cm 6.80cm]{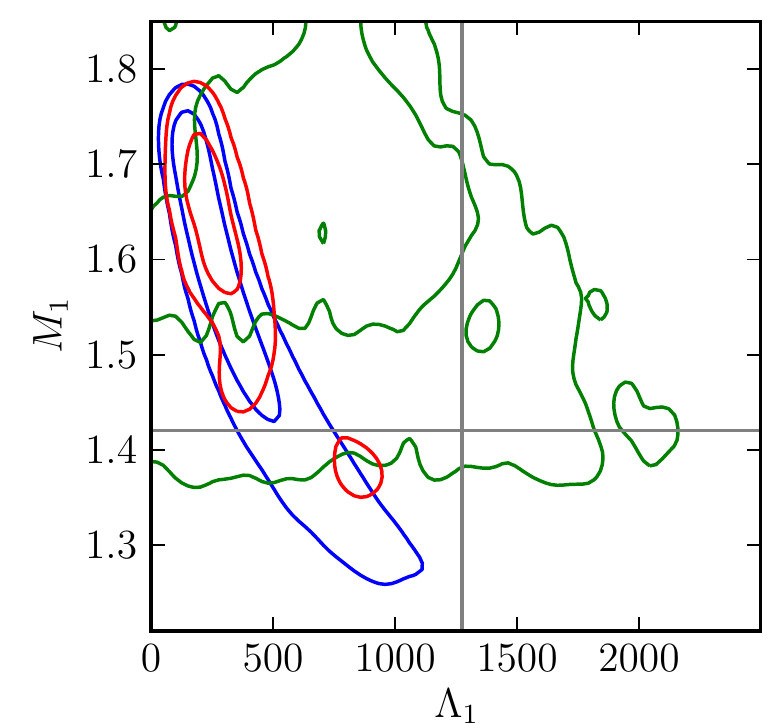}
        \end{center}
    \end{minipage}
    \begin{minipage}{0.32\textwidth}
        \begin{center}
            \hspace{1.2cm} \texttt{h4} \\
            \includegraphics[width=1.00\textwidth, clip=True, trim=0.50cm 1.10cm 6.80cm 6.80cm]{red_blue-figures/kde-corner-samples_lambda1-m1_source_inj1modh4_prior-likelihood-posterior_noref}
            \includegraphics[width=1.00\textwidth, clip=True, trim=0.50cm 1.10cm 6.80cm 6.80cm]{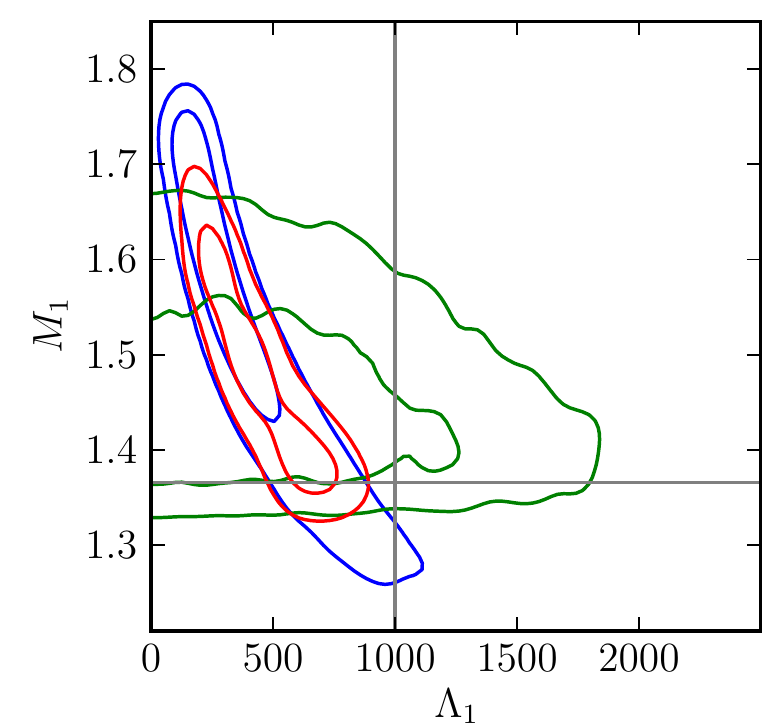}
            \includegraphics[width=1.00\textwidth, clip=True, trim=0.50cm 1.10cm 6.80cm 6.80cm]{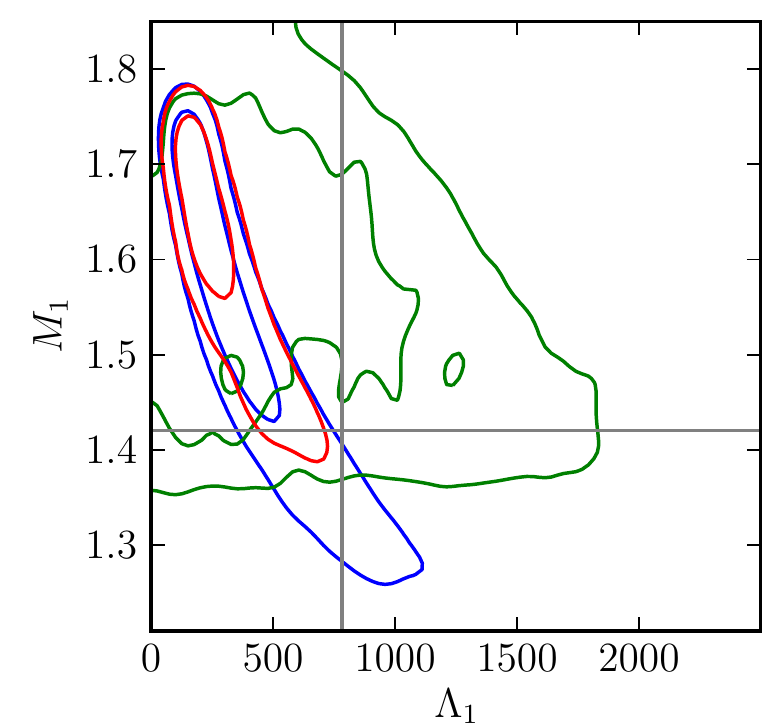}
        \end{center}
    \end{minipage}
    \begin{minipage}{0.32\textwidth}
        \begin{center}
            \hspace{1.2cm} \texttt{sly} \\
            \includegraphics[width=1.00\textwidth, clip=True, trim=0.50cm 1.10cm 6.80cm 6.80cm]{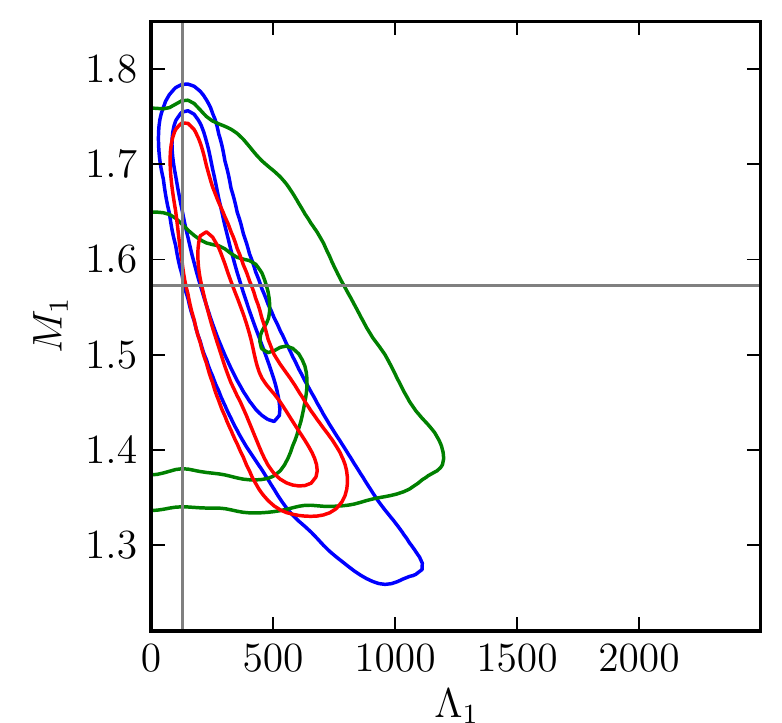}
            \includegraphics[width=1.00\textwidth, clip=True, trim=0.50cm 1.10cm 6.80cm 6.80cm]{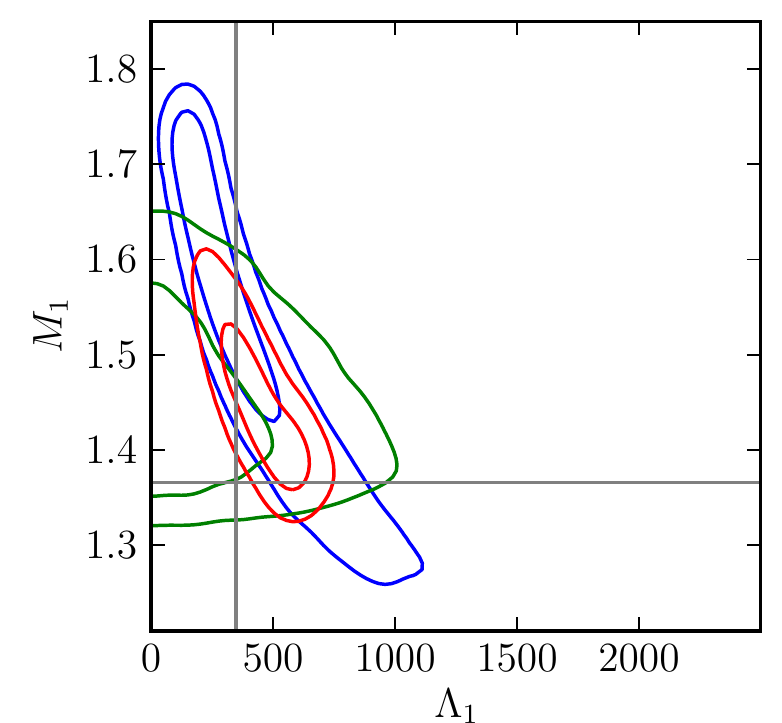}
            \includegraphics[width=1.00\textwidth, clip=True, trim=0.50cm 1.10cm 6.80cm 6.80cm]{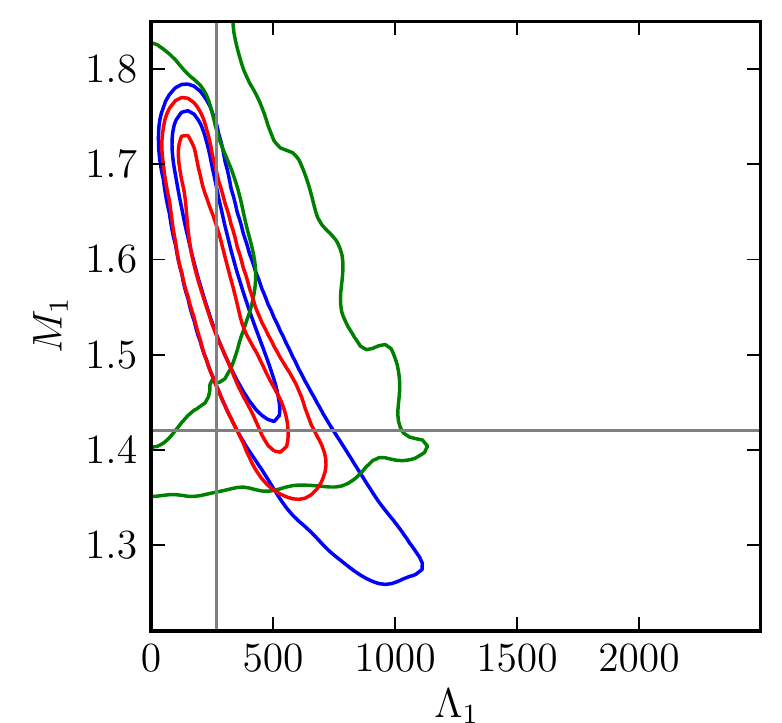}
        \end{center}
    \end{minipage}
    \caption{
        $\Lambda_1$--$M_1$ joint prior (\textit{blue}), likelihood (\textit{green}), and posterior (\textit{red}) distributions using the model-informed prior.
        Injected values are shown as grey cross-hairs and the colored contours correspond to 50\% and 90\% confidence regions.
        Rows correspond to events I-III from Table~\ref{tb:inj} and columns are labeled according to the injected \EOS.
    }
    \label{fig:inj mod lambda1-m1_source}
\end{figure}

In this appendix, we present complete results for the injection study described in Sec.~\ref{sec:injections}.
In Figs.~\ref{fig:inj un lambda1-lambda2}-\ref{fig:inj mod lambda1-m1_source}, we show the GW likelihood, non-parametric priors, and posteriors for each of the injections listed in Table~\ref{tb:inj}, finding the expected behavior in our posteriors.
We omit process plots for the model-informed prior for all injections because they closely resemble Fig.~\ref{fig:process inj1 h4 prior post}.
Injected signals with stiffer {\EOS}s (\texttt{ms1}) shift the posterior to stiffer pressures relative to the prior, whereas softer {\EOS}s (\texttt{sly}) produce lower pressures \textit{a posteriori}.
{\EOS}s that fall near the middle of the prior (\texttt{h4}) mainly produce tightened confidence regions.
We also see that the central densities for $M_1$ are shifted to smaller values for stiff {\EOS}s and larger values for soft {\EOS}s.
This is because stiffer (softer) {\EOS}s have larger (smaller) radii for the same mass, and $\rho_\mathrm{c} \sim M/R^3$ scales inversely with the NS size.
All these trends are readily apparent with the model-agnostic prior, whereas the model-informed prior often produces posteriors nearly identical to the prior.
However, as Table~\ref{tb:inj} shows, the data typically do not strongly favor one prior over the other.

Often, \texttt{ms1} is ``too stiff'' for our model-informed prior but only disfavored by our model-agnostic prior.
Although we attempt to construct the model-agnostic prior to support an extremely broad range of possible {\EOS}s, there are simply many more {\EOS}s that support relatively compact stars compared to the few that are at least as stiff as \texttt{ms1}.
This is because causality places a strict upper limit on how stiff an \EOS~can be, evidenced by the asymmetric one-dimensional marginal distributions in, e.g., Fig.~\ref{fig:process inj1 h4 prior post}, and our model-informed prior assigns relatively small weight to such {\EOS}s.
The majority of tabulated {\EOS}s used are not that stiff.
Similarly, \texttt{sly} is almost ``too soft'' for our model-informed prior, although it still appears to be better fit than \texttt{ms1} in most scenarios.
\texttt{h4} is ``just right,'' in that the model-informed prior captures its behavior well and is sometimes preferred over the model-agnostic one by as much as a factor of $\simeq 4$.

Generally, the model-agnostic prior's broader support means that the \textit{a posteriori} credible regions can stretch to include the injected values.
However, we find that this does not provide a better fit to the data than the model-informed prior, with the exception of \texttt{ms1} for injection I (first row of Table~\ref{tb:inj}, Figs.~\ref{fig:inj un lambda1-lambda2} and \ref{fig:inj mod lambda1-lambda2}).
There are a few other injections for which the injected value is near the edge of the credible regions, but these are typically near the edge of the support for both the model-informed and model-agnostic priors, resulting in no strong preference for one prior over the other.
We stress that this occurs because of a combination of factors.
Primarily, noise fluctuations in the data shift the likelihood relative to the injected value, meaning that the maximum likelihood parameters are not the injected ones.
Additionally, our priors are non-trivial in ($M_{1,2}$, $\Lambda_{1,2}$) and will prefer certain values over others.
One should only expect credible regions to correspond to correct coverage (e.g. 50\% of signals are recovered within the 50\% credible regions) if the signals are actually drawn from the prior.
Our injections are performed using a subset of the {\EOS}s used to construct our priors, but they are not actually drawn from our priors.
What's more, we chose relatively extreme {\EOS}s that live near the edge of our fiducial set, and, compounding the impact of noise fluctuations, it should consequently not be terribly surprising that some of the injected values are recovered near the edge of our 90\% credible regions.

\end{document}